\documentclass[aps,prx,10pt,twocolumn,superscriptaddress]{revtex4}

\usepackage{graphicx}
\usepackage{amssymb, amsmath}
\usepackage{color}
\usepackage{ulem}
\usepackage{bm}
\usepackage{bbm}
\usepackage{graphicx}
\usepackage{subfigure}
\usepackage{dcolumn}
\usepackage{mathtools}
\usepackage{pifont}
\usepackage{diagbox}
\usepackage{braket}
\usepackage{verbatim}
\usepackage{diagbox}

\usepackage{fancyhdr}
\def\I{\uppercase\expandafter{\romannumeral 1}}
\def\II{\uppercase\expandafter{\romannumeral 2}}
\def\III{{\uppercase\expandafter{\romannumeral 3}}}
\def\IV{{\uppercase\expandafter{\romannumeral 4}}}
\def\V{{\uppercase\expandafter{\romannumeral 5}}}
\def\VI{{\uppercase\expandafter{\romannumeral 6}}}
\def\VII{{\uppercase\expandafter{\romannumeral 7}}}
\def\i{\lowercase\expandafter{\romannumeral 1}}
\def\ii{\lowercase\expandafter{\romannumeral 2}}
\def\iii{{\lowercase\expandafter{\romannumeral 3}}}
\def\iv{{\lowercase\expandafter{\romannumeral 4}}}
\def\v{{\lowercase\expandafter{\romannumeral 5}}}
\def\vi{{\lowercase\expandafter{\romannumeral 6}}}
\def\vii{{\lowercase\expandafter{\romannumeral 7}}}

\def\angstrom{\mbox{\normalfont\AA}}

\def\nn{\nonumber\\}

\def\angstrom{\mbox{\normalfont\AA}}

\def\q{\mathbf{q}}

\def\nn{\nonumber\\}

\begin{document}

\title{Lattice distortions, moir\'e phonons, and relaxed electronic band structures in magic-angle twisted bilayer graphene}

\author{Bo Xie}
\affiliation{School of Physical Science and Technology, ShanghaiTech University, Shanghai 201210, China}%
 
\author{Jianpeng Liu}
\email[]{liujp@shanghaitech.edu.cn}
\affiliation{School of Physical Science and Technology, ShanghaiTech University, Shanghai 201210, China}%
\affiliation{ShanghaiTech laboratory for topological physics, ShanghaiTech University, Shanghai 201210, China}
\affiliation{Liaoning Academy of Materials, Shenyang 110167, China}
\begin{abstract}
In this work, we present a theoretical research on the lattice relaxations, phonon properties, and relaxed electronic structures in magic-angle  twisted bilayer graphene (TBG). We construct a continuum elastic model in order to study the lattice dynamics of magic-angle TBG, where both in-plane and out-of-plane lattice displacements are take into account.  The fully relaxed lattice structure calculated using such a model is in quantitative agreement with experimental measurements. Furthermore, we investigate the phonon properties in magic-angle TBG using the continuum elastic model, where both the in-plane and out-of-plane phonon modes are included and treated on equal footing. We identify different types of moir\'e phonons including  in-plane sliding modes, soft out-of-plane flexural modes, as well as out-of-plane breathing modes. The latter two types of phonon modes exhibit interesting monopolar, dipolar, quadrupolar, and octupolar-type out-of-plane vibration patterns. Additionally, we explore the impact of the relaxed moir\'e superlattice structure on the electronic band structures of magic-angle TBG using an effective continuum model, which shows nearly exact agreement with those calculated using a  microscopic atomistic tight-binding approach. Our work lays foundation for further studies on the electron-phonon coupling effects and their interplay with $e$-$e$ interactions in magic-angle TBG.
\end{abstract}
             
\maketitle


\section{Introduction}
Twisted bilayer graphene (TBG) is consisted of two layers of graphene which are twisted with respect to each other by a small angle $\theta$. Recently, a number of remarkable phenomena have been observed in TBG around the magic angle of approximately $1.05^{\circ}$, including the correlated insulator states \cite{cao-nature18-mott,efetov-nature19,tbg-stm-pasupathy19,tbg-stm-andrei19,tbg-stm-yazdani19, tbg-stm-caltech19, young-tbg-science19, efetov-nature20, young-tbg-np20, li-tbg-science21}, quantum anomalous Hall effect \cite{young-tbg-science19, sharpe-science-19, efetov-arxiv20,yazdani-tbg-chern-arxiv20,andrei-tbg-chern-arxiv20,efetov-tbg-chern-arxiv20,pablo-tbg-chern-arxiv21,yang-tbg-cpl21}, unconventional superconductivity \cite{cao-nature18-supercond,dean-tbg-science19,marc-tbg-19, efetov-nature19, efetov-nature20, young-tbg-np20, li-tbg-science21, cao-tbg-nematic-science21} and so on. These intriguing experimental observations of magic-angle TBG have stimulated intensive  theoretical research. Around magic angle, there are two topologically nontrivial flat bands for each valley and spin with narrow bandwidth \cite{macdonald-pnas11,song-tbg-prl19, yang-tbg-prx19, po-tbg-prb19, origin-magic-angle-prl19, jpliu-prb19}. As a result, electron-electron ($e$-$e$) Coulomb interactions play crucial roles for the flat-band electrons around the magic angle.
A lot of the intriguing phenomena observed in magic-angle TBG, such as correlated insulators and quantum anomalous Hall effects, can be attributed to the interplay between the nontrivial band topology and the strong $e$-$e$ Coulomb interactions in the flat bands \cite{balents-review-tbg,andrei-review-tbg,jpliu-nrp21,kang-tbg-prl19,Uchoa-ferroMott-prl,xie-tbg-2018, wu-chiral-tbg-prb19,  zaletel-tbg-2019, wu-tbg-collective-prl20, zaletel-tbg-hf-prx20,jpliu-tbghf-prb21,zhang-tbghf-arxiv20,hejazi-tbg-hf,kang-tbg-dmrg-prb20,kang-tbg-topomott,yang-tbg-arxiv20,meng-tbg-arxiv20,Bernevig-tbg3-arxiv20,Lian-tbg4-arxiv20,regnault-tbg-ed,zaletel-dmrg-prb20,macdonald-tbg-ed-arxiv21,meng-tbg-qmc-cpl21,lee-tbg-qmc-arxiv21,bultinck-tbg-strain-prl21,tbg-ome-nc20,zhu-tbg-orb-prl20,Macdonald-ome-prl21,balents-cur-prb21}.

However, some other phenomena reported in magic-angle TBG, such as linear in temperature resistivity \cite{young-linear-np19,cao-strange-prl20} and the competition between  correlated insulator  and  superconductivity \cite{young-tbg-np20,li-tbg-science21}, are relatively less understood. One of the perspectives is that only considering $e$-$e$ interactions and band topology is not enough to explain these puzzling experiments. Electron-phonon couplings may play important roles \cite{wu-prl18, lian-tbg-prl19, sharma-tbg-phonon-nc21}. Especially, recent angle-resolved photoemission experiments suggest the presence of phonon replicas of the moir\'e flat bands in magic-angle TBG \cite{chen-replica-arxiv23}, which provides direct experimental evidence of strong electron-phonon coupling effects in this system.  
 However, from the theoretical side, in spite of several pioneering works \cite{tbg-phonon-nature21,balandin-prb13,choi-prb18,wu-prl18,lian-tbg-prl19,tbg-raman-nc18,wu-linear-prb19,angeli-tbg-prx19,koshino-phonon-prb19,koshino-ep-prb20,sharma-tbg-phonon-nc21,choi-tbg-prl21,eliel2018intralayer,lamparski2020soliton,liu-phonon-nano22,kaxiras-phonon-prb22,DaiXi-phonon-2023prb},  understanding of the phononic properties and electron-phonon couplings  remains incomplete. For example, despite a few researches based on microscopic atomistic dynamical matrices \cite{angeli-prx19,liu-phonon-nano22,DaiXi-phonon-2023prb}, most of the previous studies based on continuum elastic model only consider the in-plane lattice displacements \cite{koshino-tbg-prb17,koshino-phonon-prb19,koshino-ep-prb20,carr-relaxation-18prb,lamparski2020soliton}.  It has been shown that the out-of-plane lattice vibrational degrees of freedom would give rise to extremely soft flexural phonon modes \cite{liu-phonon-nano22}, which could be the driving force for the peculiar charge order observed in this system \cite{liu-phonon-nano22, tbg-stm-andrei19}.  The out-of-plane interlayer ``breathing modes" \cite{he-breathing-nano13} are directly coupled with electronic interlayer hopping events, which may have strong effects on the flat-band electrons in magic-angle TBG. Up to date, to the best of knowledge, a continuum elastic model describing the lattice dynamics of TBG including both the in-plane and out-of-plane vibrational degrees of freedom, is still lacking.  Compared to calculations based on  microscopic atomistic force constants,  the construction of a reliable continuum elastic model would allow for much more efficient calculations of structural relaxations, phonons, and effective electronic band structures (with relaxed lattice structures). It would also pave the way for further comprehensive studies on electron-phonon couplings and their interplay with $e$-$e$ interactions.

In this paper, we report such a continuum elastic model for TBG including both in-plane and out-of-plane lattice degrees of freedom. Specially, we treat TBG at small twist angles (with large moir\'e superlattice constants) as  continuum elastic medium. Its low-energy dynamics is fully captured by a number of elastic parameters such as stiffness constants and interlayer binding energy parameters, which are extracted either from first principles density functional theory (DFT) calculations or from experiments. With such a continuum elastic model, we first perform structural relaxation calculations for TBG at small twist angles $0.8^{\circ}\leq\theta\leq 1.5^{\circ}$. The fully relaxed moir\'e superlattice structure calculated using such a model is in \textit{quantitative agreement} with experimental measurements. For example, within the range of twist angles $0.8^{\circ}\leq\theta\leq 1.5^{\circ}$, the deviations between the calculated area ratio of the $AA$ region and the reconstructed rotational angle at the edge of $AA$ region and the corresponding experimental measured values  \cite{kazmierczak2021strain} are about 20\%. The small discrepancy between theory and experiment may be induced by heterostrains and/or twist-angle disorder in the experimental system which are not considered in the calculations. With the fully relaxed superlattice structure, we continue to study the phonon properties of TBG at the magic angle $\theta=1.05^{\circ}$, where both in-plane and out-of-plane vibrational modes are taken into account and are treated on equal footing. We find different types of moir\'e phonon modes which may have important effects on the electronic properties. This includes gapless in-plane sliding modes, gapped but extremely soft  out-of-plane flexural modes, as well as out-of-plane interlayer breathing modes. The in-plane interlayer sliding modes are gapless in the long-wavelength limit (wavevector $\q\to 0$), as they are considered as a kind of  Goldstone modes within the continuum elastic model framework. These sliding modes are gapped if a commensurate microscopic atomic superlattice is considered, with the gap $\sim 2.4\,$meV at the magic angle according to deep-potential molecular dynamics calculations \cite{liu-phonon-nano22}. The  flexural modes  are out-of-plane ``center-of-mass" modes, while the
out-of-plane breathing modes are anti-phase modes, both of which exhibit interesting monopolar, dipolar, quadrupolar, and octupolar-type out-of-plane vibration patterns in real space. Additionally ,we explore the impact of the relaxed moir\'e superlattice structure on the electronic band structures of magic-angle TBG using an effective continuum model, which shows nearly exact agreement with those calculated using a microscopic atomistic tight-binding approach.

This paper is organized as follows. In Sec.~\II, we introduce the continuum elastic model and demonstrate how the parameters of the model are determined. In Sec.~\III, we present the workflow for the lattice relaxation calculations including both in-plane and out-of-plane lattice distortions and perform the structural relaxation calculations within a range of twist angles $0.8^{\circ}\leq\theta\leq 1.5^{\circ}$. In Sec.~\IV, we sketch the formalism for the phonon calculations using the elastic model and present the phonon spectrum and vibrational modes of different types of moir\'e phonons in magic-angle TBG. In Sec.~\V, we study the influence of the lattice distortions on the electronic band structures for magic-angle TBG. A summary and an outlook are given in Sec.~\VI.

\section{Continuum Elastic Model}
\label{sec:continu_elas_model}

We first introduce the lattice geometry of TBG. 
Two rigid graphene layers are considered to be stacked together and are twisted with respect to each other by a small twist angle $\theta$. The lattice vectors of the single layer graphene are $\textbf{a}_{1}\!=\!a(1,0,0)\,$ and $\textbf{a}_{2}\!=\!a(1/2,\sqrt{3}/2,0)\,$, with $a\!=\!2.46\,\angstrom$. The corresponding primitive reciprocal lattice vectors are $\textbf{a}^{*}_{1}=(2\pi/a)(1,-1/\sqrt{3},0)$ and $\textbf{a}^{*}_{2}=(2\pi/a)(0,2/\sqrt{3},0)$. Each graphene layer consists of two sublattices, the positions of which are denoted by $\bm{\tau}_{A}\!=\!(0,0,0)\,$ and $\bm{\tau}_{B}\!=\!(a/\sqrt{3})(0,-1,0)\,$. The atomic lattice vectors of the two layers are defined by $\textbf{a}^{(1)}_{i}=R(-\theta/2)\textbf{a}_{i}\,$ and $\textbf{a}^{(2)}_{i}\!=\!R(\theta/2)\textbf{a}_{i}\,$ with $i\!=\!1,2\,$, and the corresponding primitive reciprocal lattice vectors are $\textbf{a}^{*,(1)}_{i}\!=\!R(-\theta/2)\textbf{a}^{*}_{i}\,$ and $\textbf{a}^{*,(2)}_{i}\!=\!R(\theta/2)\textbf{a}^{*}_{i}\,$, with $i\!=\!1,2\,$. Here $R(\pm\theta/2)$ denotes rotation operation counter-clockwise/clockwise by $\theta/2$.

\begin{figure}[b]
\begin{center}
    \includegraphics[width=9cm]{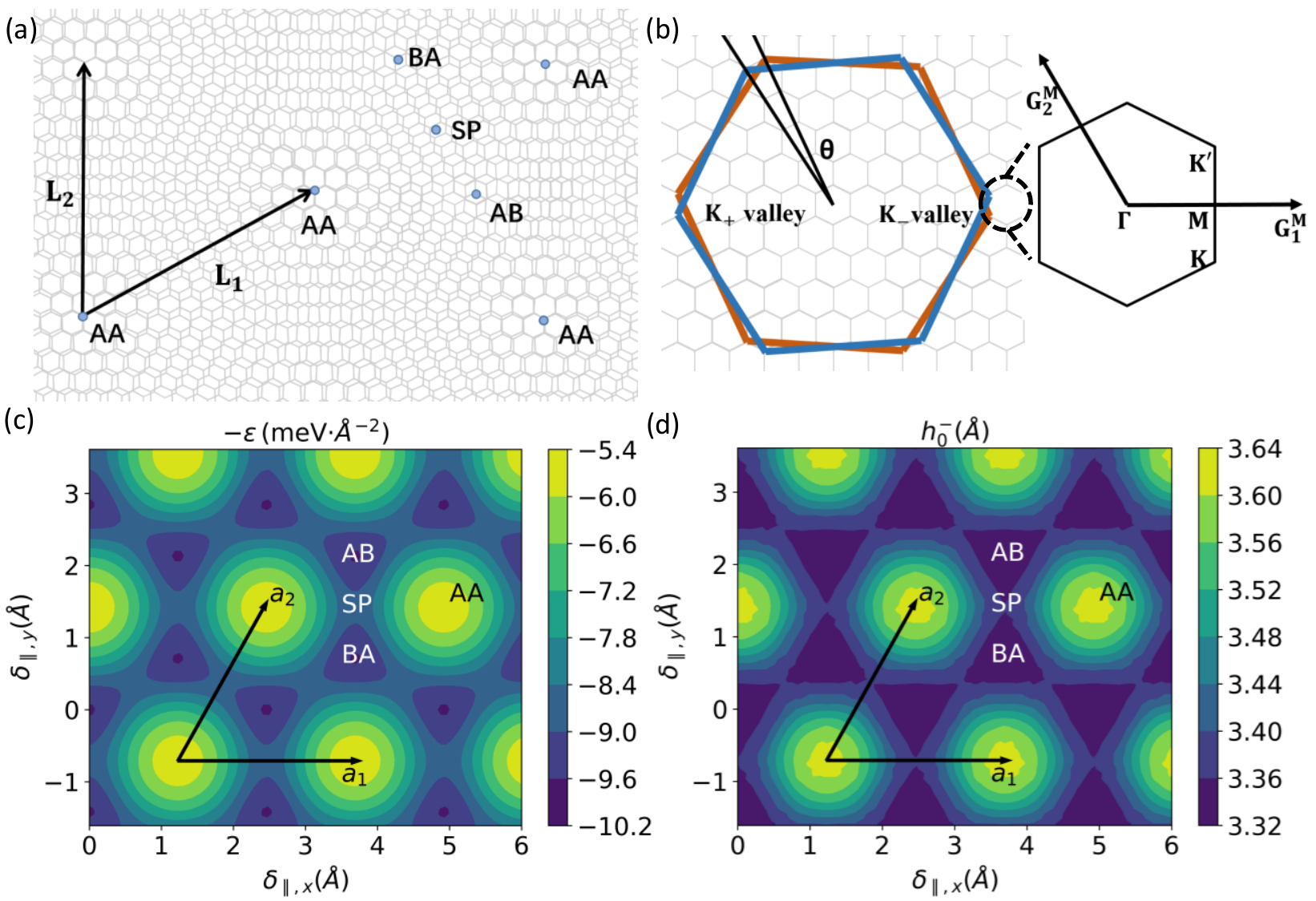}
\caption{
(a) Schematic illustration of  moir\'e superlattice of twisted bilayer graphene in real space. $\textbf{L}^{M}_{1}$ and $\textbf{L}^{M}_{2}$ are the primitive superlattice vectors. The $AA$ point, $AB/BA$ points and the saddle point (SP) are highlighted in the moir\'e supercell. (b) The moir\'e Brillouin zone of TBG with twist angle $\theta$. The two large hexagons with red and blue colors represent  atomic Brillouin zones of the two twisted graphene monolayers, and the small gray hexagon denotes the Brillouin zone of the moir\'e supercell. $\textbf{G}^{M}_{1}$ and $\textbf{G}^{M}_{2}$ are the primitive reciprocal vectors of the moir\'e supercell. (c) $-\epsilon$ as a function of in-plane displacement vector $\bm{\delta}_{\parallel}$ (see text). (d) $h^{-}_{0}$ as a function of in-plane displacement vector $\bm{\delta}_{\parallel}$ (see text).}
\label{fig1}
\end{center}
\end{figure}

In the presence of a small twist, a moir\'e pattern appears in real space, as depicted in Fig.~\ref{fig1}(a). The primitive lattice vectors of the moir\/e supercell are $\textbf{L}^{M}_{1}\!=\!(\sqrt{3}/2,1/2,0)L_{s}\,$ and $\textbf{L}^{M}_{2}\!=\!(0,1,0)L_{s}\,$, with the lattice constant $L_{s}\!=\!a/(2\sin{\theta/2})\,$. We highlight  several points in the moir\'e supercell, i.e., the $AA$, $AB/BA$, and saddle point (SP), as schematically shown in Fig.~\ref{fig1}(a). The local stackings around the $AA$ and $AB/BA$ points resemble those of $AA$ and $AB/BA$ stacked bilayer graphene, respectively. The $AB$ and $BA$ point is connected by the saddle point, as shown in Fig.~\ref{fig1}(a).
In Fig.~\ref{fig1}(b) we present a schematic diagram of the moir\'e Brillouin zone in TBG, where the high-symmetry points are marked. The reciprocal lattice vectors are given by: $\textbf{G}^{M}_{1}\!=\!(\,4\pi/(\sqrt{3}L_{s}),0,0\,)$ and $\textbf{G}^{M}_{2}\!=\!(\,-2\pi/(\sqrt{3}L_{s}), 2\pi/L_{s},0\,)$. 
The single layer graphene's Dirac points are located at $\textbf{K}^{\mu}\!=\!\mu(\,-4\pi/(3a),0,0\,)$, where $\mu\!=\!\pm1$ represents the $\pm$ valley. Under small twist angle, the Dirac points in the $l$th layer are rotated by $\mp\theta/2$. 

In order to analyze the long wavelength lattice distortions in TBG, we adopt a continuum elastic approach, which treats TBG as a continuum medium instead of an atomic superlattice. This method was first introduced by Nam and Koshino \cite{koshino-tbg-prb17}, which only includes in-plane lattice displacements.  In order to perform full lattice relaxation calculations  and to calculate the complete phonon spectrum, here we propose a generalized continuous elastic model including both in-plane and out-of-plane displacements.  
Before discussing the detailed formalism, we first introduce notations for the moir\'e superlattice distortions. 
At position $\textbf{r}$ in the moir\'e supercell, the local atomic shift $\bm{\delta}(\textbf{r})$ is defined as the relative displacement vector  from position $\textbf{r}$ in the first layer to its counterpart in the second layer. With an ideal moir\'e superlattice formed by twisting graphene layers, the displacement vector $\bm{\delta}^{0}(\textbf{r})=R(\theta/2)\,\bm{r}-R(-\theta/2)\,\bm{r}$. Then, $AA$ point can be defined as the position at which the displacement vector satisfies $\bm{\delta}^{0}(\mathbf{r})=\mathbf{0}$; $AB$ points are identified as the positions at which  $\bm{\delta}^{0}(\textbf{r})\!=\!R^{n}(2\pi/3)(\textbf{a}_{1}+\textbf{a}_{2})/3$, with $n=0,1,2$; $BA$ points are identified as the positions at which $\bm{\delta}^{0}(\textbf{r})\!=\!R^{n}(2\pi/3)(2\textbf{a}_{2}-\textbf{a}_{1})/3$, with $n=0,1,2$; and the saddle points (SPs) can be defined as the locations at which $\bm{\delta}^{0}(\textbf{r})\!=\!R^{n}(\pi/3)\textbf{a}_{1}/2$, with $n=0,1,\dots,5$.
In realistic TBG system, the atomic positions may deviate from the ideal displacement vectors $\{\bm{\delta}^{0}(\mathbf{r})\}$. Specifically, 
we introduce lattice distortions  both in the in-plane directions, denoted as $\textbf{u}^{(l)}_{X}(\textbf{r})$, and in the out-of-plane directions, denoted as $h^{(l)}_{X}(\textbf{r})$, where $l=1,2$ and $X\!=\!A,B$ are the layer and sublattice indices, respectively.  Then the local atomic displacement vector under lattice distortion is given by 
\begin{equation}
\bm{\delta}(\textbf{r})\!=\!\bm{\delta}_{0}(\textbf{r})+\textbf{u}^{(2)}_{X}(\textbf{r})-\textbf{u}^{(1)}_{X}(\textbf{r})+(h^{(2)}_{X}(\textbf{r})-h^{(1)}_{X}(\textbf{r}))\mathbf{\hat{z}}\,.
\end{equation}
Since the lattice distortion of interest is a smooth function on the moir\'e length scale, which is much greater than atomic length scale, it is legitimate to omit the sublattice indices in the lattice distortion, i.e.,
$\textbf{u}^{(l)}_{X}(\textbf{r})\!=\!\textbf{u}^{(l)}(\textbf{r})$ and $h^{(l)}_{X}(\textbf{r})\!=\!h^{(l)}(\textbf{r})$.

In our model, the impact of stretches and curvatures are taken into account in the elastic energy $U_{E}\!=\!\int \mathrm{d}^{2}\textbf{r}\,V_{E}$, where
\begin{equation}
\begin{split}
V_{E}=&\sum^{2}_{l=1} \frac{1}{2} \left\{ \left( \lambda+\mu \right)\left( u^{(l)}_{xx}+u^{(l)}_{yy} \right)^{2}\right.\\
&\left. + \mu\left[ \left( u^{(l)}_{xx}-u^{(l)}_{yy} \right)^{2} + 4(u^{(l)}_{xy})^{2} \right] +\right.\\
&\left.\kappa\left[ (\frac{\partial^{2}}{\partial^{2}x}+\frac{\partial^{2}}{\partial^{2}y}) h^{(l)} \right]^{2} \right\}.\\
\end{split}
\end{equation}
Here, $\lambda\approx 3.25\,$eV/$\angstrom^{2}$ and $\mu\approx 9.57\,$eV/$\angstrom^{2}$ are the Lam\'e factors, $\kappa\!=\!1.6\,$eV is the curvature modulus \cite{jung2015origin}. $u^{(l)}_{\alpha\beta}$ ($\alpha,\beta=x,y$) is the strain tensor defined as \cite{landau1986theory}:
\begin{equation}
\begin{split}
u^{(l)}_{xx}=&\frac{\partial u^{(l)}_{x}}{\partial x}+\frac{1}{2}\left(\frac{\partial h^{(l)}}{\partial x}\right)^{2}, \\
u^{(l)}_{yy}=&\frac{\partial u^{(l)}_{y}}{\partial y}+\frac{1}{2}\left(\frac{\partial h^{(l)}}{\partial y}\right)^{2}, \\
u^{(l)}_{xy}=&\frac{1}{2}\left(\frac{\partial u^{(l)}_{x}}{\partial y} + \frac{\partial u^{(l)}_{y}}{\partial x} \right)+\frac{1}{2}\frac{\partial h^{(l)}}{\partial x}\frac{\partial h^{(l)}}{\partial y}.
\label{eq:strain}
\end{split}
\end{equation}

We continue to discuss the binding energy.
Typically the binding energy is only dependent on the  relative distortions between the two layers. We can  make a linear combination of the lattice distortions: $h^{\pm}=h^{(2)}\pm h^{(1)}$, $\textbf{u}^{\pm}=\textbf{u}^{(2)}\pm \textbf{u}^{(1)}$, here the ``$-$" sign represents the relative distortions and the ``$+$" sign represents the center-of-mass distortions for the two layers. Then the binding energy is given by $U_{B}\!=\!\int\mathrm{d}^{2}\textbf{r} V_{B}[\bm{u}^{-}(\mathbf{r}),h^{-}(\mathbf{r})]$, where
\begin{equation}\label{eq:bindener}
\begin{split}
&V_{B}[\bm{u}^{-}(\mathbf{r}),h^{-}(\mathbf{r})]\\
=&\,\epsilon[\bm{u}^{-}(\mathbf{r})]\,\left[ \left( \frac{h^{-}_{0}[\bm{u}^{-}(\textbf{r})]}{h^{-}(\textbf{r})} \right)^{12} - 2\left( \frac{h^{-}_{0}[\bm{u}^{-}(\textbf{r})]}{h^{-}(\textbf{r})} \right)^{6} \right]  \\
\approx&\,\epsilon[\bm{u}^{-}(\textbf{r})]\,\left[ -1+36\left(\frac{h^{-}(\textbf{r})-h^{-}_{0}[\bm{u}^{-}(\textbf{r})]}{h^{-}_{0}[\bm{u}^{-}(\textbf{r})]}\right)^{2} \right]. \\
\end{split}
\end{equation}
In general, the binding energy is a functional of both in-plane distortion $\mathbf{u}^{-}(\mathbf{r})$ and out-of-plane distortion $h^{-}(\mathbf{r})$. In Eq.~(\ref{eq:bindener}), we assume that the binding energy can be written in a separable-variable-like form, i.e., 
\begin{equation}
V_{B}[\bm{u}^{-}(\mathbf{r}),h^{-}(\mathbf{r})]=\epsilon[\bm{u}^{-}(\textbf{r})]\,H[h^{-}(\mathbf{r})],  
\label{eq:bind}
\end{equation}
where $\epsilon[\bm{u}^{-}(\textbf{r})]$ only depends on in-plane relative distortion $\bm{u}^{-}(\textbf{r})$, and $H[h^{-}(\mathbf{r})]$ is only \textit{explicitly} dependent on the out-of-plane relative distortion $h^{-}(\mathbf{r})$. We further assume that $H[h^{-}(\mathbf{r})]$ takes the form of Lennard-Jones potential as shown in the second line of Eq.~(\ref{eq:bindener}), and is expanded to the second order of $(h^{-}(\mathbf{r})-h^{-}_0(\mathbf{r}))$, which leads to the expression in the last line of  Eq.~(\ref{eq:bindener}). Here $h^{-}_{0}(\bm{u}^{-}(\textbf{r}))$ is the reference equilibrium interlayer distance, which also depends on the in-plane local atomic distortion $\bm{\delta}^{0}(\mathbf{r})+\mathbf{u}^{-}(\mathbf{r})$. Therefore, the Lennard-Jones type potential $H[h^{-}(\mathbf{r})]$ is  implicitly dependent on in-plane lattice distortions through $h^{-}_{0}[\bm{u}^{-}(\textbf{r})]$.

We can evaluate both $\epsilon[\bm{u}^{-}(\textbf{r})]$ and $h^{-}_{0}[\bm{u}^{-}(\textbf{r})]$ from first principles density functional theory (DFT) calculations. To be specific, at position $\mathbf{r}$ in the moir\'e supercell, the local lattice structure can be viewed as bilayer graphene with a relative shift vector with respect to $AA$ stacking, and in the meanwhile the local frames of the two layers are also rotated with respect to each other  by an angle $\theta$. When the twist angle $\theta$ is small, we can neglect the small rotation of the local frame and depict the local lattice structure at  position $\textbf{r}$ by a relative shift $\bm{\delta}(\textbf{r})$. Then, we start with two rigid  graphene layers and apply a small in-plane shift to one  layer. With a fixed in-plane shift vector for the untwisted bilayer graphene, we utilize Vienna Ab initio Simulation Package (VASP) \cite{VASP1,VASP2,VASP3,VASP4} with PBE functional \cite{PBE} to relax the interlayer distance. 
With the relaxed interlayer distance, we further calculate the total energy of the untwisted bilayer graphene system. The binding energy is obtained by subtracting the total energy of two isolated  graphene monolayers (with infinite interlayer distance) from  that of the coupled bilayer graphene with relaxed interlayer distance and fixed in-plane shift  $\bm{\delta}_{\parallel}$. In Fig.~\ref{fig1}(c) and Fig.~\ref{fig1}(d)  we present $-\epsilon(\bm{\delta}_{\parallel})$ and $h^{-}_{0}(\bm{\delta}_{\parallel})$ as a function of in-plane  displacement vector $\bm{\delta}_{\parallel}$, which are subtracted from DFT calculations.  

Both $\epsilon[\bm{\delta}_{\parallel}(\textbf{r})]$  and $h^{-}_{0}[\bm{\delta}_{\parallel}(\textbf{r})]$ can be expanded in terms of Fourier series
\begin{align}
&\epsilon[\bm{\delta}_{\parallel}(\textbf{r})]=\sum_{m}\,\epsilon_{m}e^{i\textbf{a}^{*}_{m}\cdot\bm{\delta}_{\parallel}}\\
&h^{-}_{0}(\bm{\delta}_{\parallel}(\textbf{r}))=\sum_{m}\,h^{-}_{0,m}e^{i\textbf{a}^{*}_{m}\cdot\bm{\delta}_{\parallel}},
\end{align} 
where $\textbf{a}^{*}_{m}\!=\!m_{1}\textbf{a}^{*}_{1}+m_{2}\textbf{a}^{*}_{2}$ denotes the reciprocal lattice vectors of monolayer graphene. We note that 
\begin{equation}
\textbf{a}^{*}_{m}\cdot\bm{\delta}_{\parallel}(\textbf{r})\!=\!\textbf{a}^{*}_{m}\cdot(\bm{\delta}^{0}(\textbf{r})+\bm{u}^{-}(\textbf{r}))\!=\!\textbf{G}_{m}\cdot\textbf{r}+\textbf{a}^{*}_{m}\cdot\textbf{u}^{-}(\textbf{r})\;,
\label{eq:au}
\end{equation}
where $\textbf{G}_{m}\!=\!m_{1}\textbf{G}^{M}_{1}+m_{2}\textbf{G}^{M}_{2}$ is the moir\'e reciprocal lattice vector. With Eq.~(\ref{eq:au}) we can map $\epsilon[\bm{\delta}_{\parallel}(\textbf{r})]$ and $h^{-}_{0}[\bm{\delta}_{\parallel}(\textbf{r})]$ to a point $\mathbf{r}$ in the moir\'e supercell. The AA region of the moir\'e supercell has weaker binding energy density and a larger interlayer distance, and the AB/BA region has stronger binding energy density and a smaller interlayer distance.  In Table.~\ref{tab:bindingfourier}, we list several leading-order Fourier coefficients $-\epsilon_{m}$ and $h^{-}_{0,m}$. In the following calculations, $\epsilon[\bm{\delta}_{\parallel}(\textbf{r})]$ and $h^{-}_{0}[\bm{\delta}_{\parallel}(\textbf{r})]$ will be Fourier expanded up to the cutoff vector ($\bm{G}_{\textrm{max}}$) of the reciprocal space, with $\vert\bm{G}_{\textrm{max}}\vert= m_1^{\textrm{max}}\mathbf{G}_{1}^{M}+m_2^{\textrm{max}}\mathbf{G}_{2}^{M}$, and $\vert m_{1,2}^{\textrm{max}}\vert=6$.
\begin{table}[]
    \centering
    \caption{Fourier component of $-\epsilon_{m}$ and $h^{-}_{0,m}$. ($m_{1}$, $m_{2}$) represents the Fourier component with the wavevector at $\textbf{a}^{*}_{m}\!=\!m_{1}\textbf{a}^{*}_{1}+m_{2}\textbf{a}^{*}_{2}$. }
    \begin{tabular}{l|c|c|c|c}
    \hline\hline
        & (0,0) & (1,0)  & (2,1) & (2,0) \\
      \hline
       $-\epsilon_{m}\,$(meV$\cdot\angstrom^{-2}$)  & -7.924 & 0.4635 & -0.0595 &-0.0182 \\
       $h^{-}_{0,m}\,$($\angstrom$)   & 3.433 & 0.0343 & -0.0010 &-0.0014 \\
       \hline\hline
    \end{tabular}
    \label{tab:bindingfourier}
\end{table}

 It is worthwhile to note again that the separable-variable-like form of the binding energy Eq.~(\ref{eq:bindener}) is an approximation. We have explicitly checked the validity of such an approximation by  calculating the second-order expansion coefficients of $(h^{-}(\mathbf{r})-h^{-}_0(\mathbf{r}))$ as a function of in-plane shift $\bm{\delta}_{\parallel}$, which is presented in Appendix \I. 
It turns out that Eq.~(\ref{eq:bindener}) is a fairly good approximation to the interlayer binding energy.

\section{Lattice relaxations}

\subsection{Model and Formalism}
\label{sec:lattice}

In our model, the total energy $U\!=\!U_{E}+U_{B}$ is a functional of the lattice distortions. We can minimize the total energy with respect to the lattice distortions by solving the Euler-Lagrange equation:
\begin{equation}
\begin{split}
&\frac{\delta U}{\delta f}-\sum_{\alpha}\frac{\partial}{\partial \alpha}\frac{\delta U}{\delta f_{\alpha}}+\sum_{\alpha,\beta}\frac{\partial^{2}}{\partial \alpha\partial \beta}\frac{\delta U}{\delta f_{\alpha,\beta}}=0,
\end{split}
\end{equation}
with 
\begin{align}
&f_{\alpha}=\frac{\partial f}{\partial \alpha}\;,\\ 
&f_{\alpha,\beta}=\frac{\partial^{2} f}{\partial \alpha\partial\beta}\;,
\end{align}
where $\alpha,\beta\!=\!x,y$ and $f\!=\!u^{\pm}_{x},u^{\pm}_{y},h^{\pm}$. We define the Fourier transformation of lattice distortions:
\begin{equation}\label{eq:fts}
\begin{split}
\textbf{u}^{\pm}(\textbf{r})=\sum_{\textbf{G}_{m}}\textbf{u}^{\pm}_{\textbf{G}_{m}}e^{i\textbf{G}_{m}\cdot\textbf{r}},\\
h^{\pm}(\textbf{r})=\sum_{\textbf{G}_{m}}h^{\pm}_{\textbf{G}_{m}}e^{i\textbf{G}_{m}\cdot\textbf{r}},
\end{split}
\end{equation} 
where $\textbf{G}_{m}\!=\!m_{1}\textbf{G}^{M}_{1}+m_{2}\textbf{G}^{M}_{2}$ is the moir\'e reciprocal lattice vector. Then we can solve the Euler-Lagrange equations in reciprocal space and obtain the relaxed moir\'e superlattice structure with long-wavelength lattice distortions. Furthermore, 
we assume that the center-of-mass component of out-of-plane distortion vanishes in the relaxed structure, i.e. $h^{+}(\textbf{r})=0$. This is an excellent approximation since nonzero $h^{+}(\textbf{r})$ means that there are center-of-mass ripples in the TBG system, which typically occurs as  thermal excitation effects and/or strain effects. At zero temperature and in the absence of strain, it is legitimate to set $h^{+}(\textbf{r})=0$.
Then, the Euler-Lagrange equations in the reciprocal space are given by: 
\begin{widetext}
\begin{align}
\begin{split}
&\left[\begin{array}{cc}
				(\lambda+2\mu)G^{2}_{m,x}+\mu G^{2}_{m,y} & (\lambda+\mu)G_{m,x}G_{m,y} \\
				(\lambda+\mu)G_{m,x}G_{m,y} & (\lambda+2\mu)G^{2}_{m,y}+\mu G^{2}_{m,x}
\end{array}\right]\left[\begin{array}{c}
                     u^{-}_{\textbf{G}_{m},x}\\
                     u^{-}_{\textbf{G}_{m},y}
\end{array}\right]=-2\left[\begin{array}{c}
                     F_{\textbf{G}_{m},x}\\
                     F_{\textbf{G}_{m},y}
\end{array}\right]\\
&\left[\begin{array}{cc}
				(\lambda+2\mu)G^{2}_{m,x}+\mu G^{2}_{m,y} & (\lambda+\mu)G_{m,x}G_{m,y} \\
				(\lambda+\mu)G_{m,x}G_{m,y} & (\lambda+2\mu)G^{2}_{m,y}+\mu G^{2}_{m,x}
\end{array}\right]\left[\begin{array}{c}
                     u^{+}_{\textbf{G}_{m},x}\\
                     u^{+}_{\textbf{G}_{m},y}
\end{array}\right]=\left[\begin{array}{c}
                     M_{x}\\
                     M_{y}
\end{array}\right],\\
&\frac{\partial V}{\partial h^{-}}-\frac{\partial}{\partial x}\frac{\partial V}{\partial \frac{\partial h^{-}}{\partial x}}-\frac{\partial}{\partial y}\frac{\partial V}{\partial \frac{\partial h^{-}}{\partial y}}+\frac{\partial^{2}}{\partial^{2} x}\frac{\partial V}{\partial \frac{\partial^{2} h^{-}}{\partial^{2} x}}+\frac{\partial^{2}}{\partial^{2} y}\frac{\partial V}{\partial \frac{\partial^{2} h^{-}}{\partial^{2} y}}=0.
\end{split}
\end{align}
\end{widetext}
where $V$ is the energy density, the integral of which is the total elastic and binding energy, i.e., $U=U_E+U_B=\int^2d\mathbf{r}\,V[\mathbf{r}]$. 
$F_{\textbf{G}_m,\alpha}$ (with $\alpha\!=\!x,y$) is the Fourier coefficient of $\partial V/\partial u_{\alpha}^{-}=\sum_{\textbf{G}_m} F_{\textbf{G}_m,\alpha}\,e^{i\textbf{G}_m\cdot\mathbf{r}}$. $M_{x,y}$ are to the third order terms of moire reciprocal vectors, which originates from the $h^{(l)}$ dependence of the strain tensor as expressed in Eq.~(\ref{eq:strain}). The detailed expressions of  $F_{\textbf{G}_{m},\alpha}$ and $M_{\alpha}$, as well as the detailed formalism of lattice relaxations are presented in Appendix \II.

It is  important to note that $F_{\textbf{G}_m,\alpha}$ is a function of both $\textbf{u}^{-}$ and $h^{-}$. For  fixed $\{h^{-}(\mathbf{r})\}$, $\textbf{u}^{-}(\mathbf{r})$ can be solved iteratively. Likewise, for  fixed $\{\textbf{u}^{-}(\mathbf{r})\}$, $\textbf{u}^{+}(\mathbf{r})$ and $h^{-}(\mathbf{r})$ can be solved iteratively as well. Thus, we can divide the full Euler-Lagrange equations into two subsets of equations. In one subset, $u^{+}_{\textbf{G}_{m},\alpha}$ and $h^{-}_{\textbf{G}_{m}}$ are kept fixed, and  $\{u^{-}_{\textbf{G}_{m},\alpha}\}$ are treated as variables. 
In the other subset, $\{u^{-}_{\textbf{G}_{m},\alpha}\}$ are kept fixed, whereas $u^{+}_{\textbf{G}_{m},\alpha}$ and $h^{-}_{\textbf{G}_{m}}$ are treated as variables and solved iteratively.
Thus, we outline the work flow for solving the coupled Euler-Lagrange equations as follows: 
\begin{itemize}
\item (\i) Setting initial $h^{-}(\mathbf{r})\!=\!h^{-}_0(\mathbf{r})$ and initial $\mathbf{u}^{+}(\mathbf{r})=0$, we first iteratively solve the first subset of Lagrange equations of $\textbf{u}^{-}$, until a converged solution is obtained.
\item (\ii) Treating the converged solution of $\{\textbf{u}^{-}(\mathbf{r})\}$ from the previous step as fixed parameters, we solve the other subset of Lagrange equations for $u^{+}_{\textbf{G}_{m},\alpha}$ and $h^{-}_{\textbf{G}_{m}}$ iteratively, until convergence is reached.
\item(\iii) We repeat step (\i) with fixed  $\textbf{u}^{+}(\mathbf{r})$ and $h^{-}(\mathbf{r})$ obtained from step (\ii), and continue to solve these equations until all components of the lattice distortions are converged.
\end{itemize}

\subsection{Results}
\label{sec:lattice-results}

\begin{figure}[b]
\begin{center}
    \includegraphics[width=9cm]{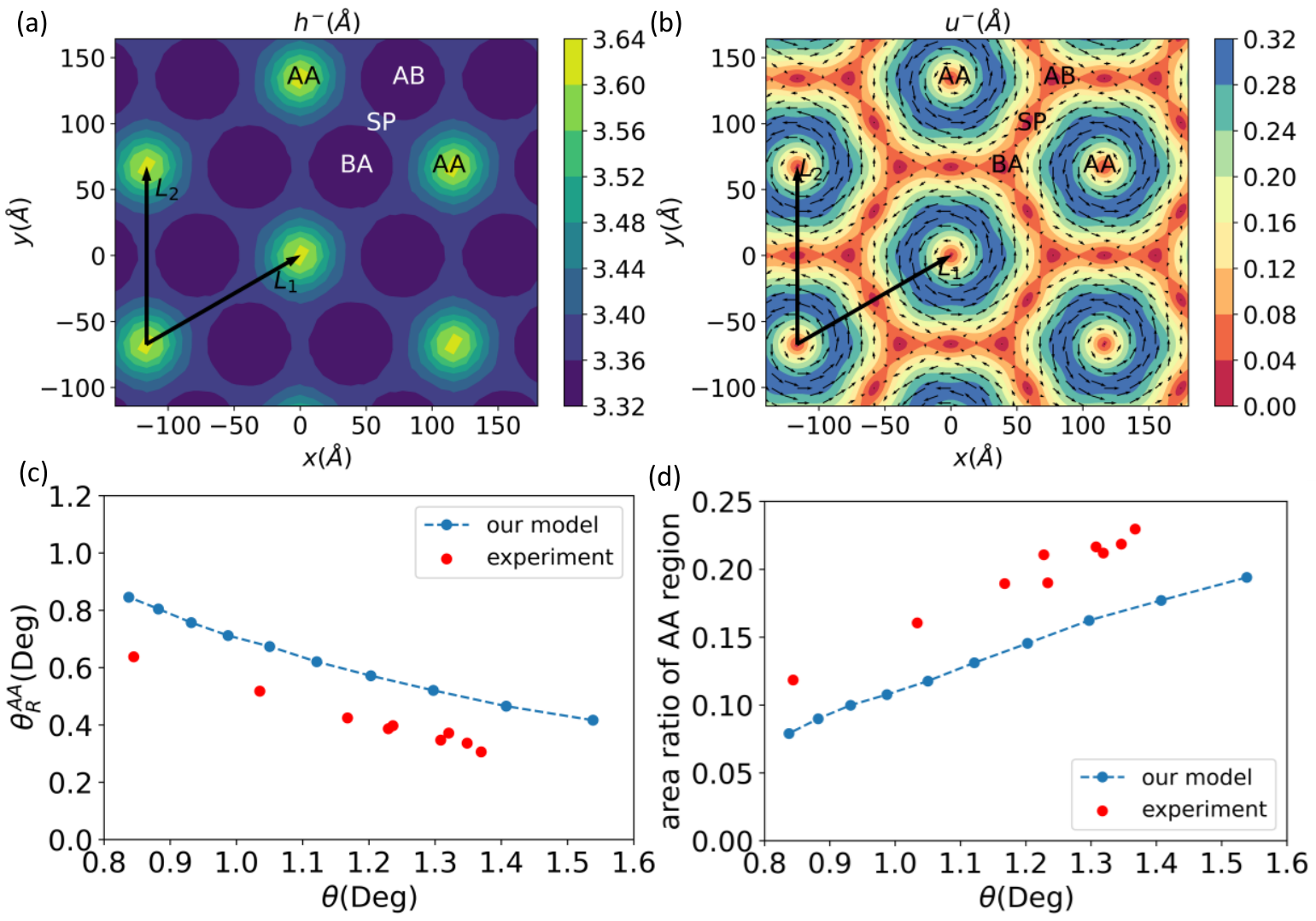}
\caption{
(a) The real-space distribution of out-of-plane relative distortion $h^{-}(\textbf{r})$ of magic-angle TBG. The colorbar represents the value of $h^{-}(\textbf{r})$. (b) The real-space distribution of the in-plane relative distortion $\textbf{u}^{-}(\textbf{r})$ of magic-angle TBG. The arrows denote the directions of $\textbf{u}^{-}(\textbf{r})$,  and the colorbar represents the amplitudes of $\textbf{u}^{-}(\textbf{r})$.  (c) The reconstruction rotation angle at the edge of the $AA$ region $\theta^{AA}_{R}$ as a function of twist angle $\theta$. (d) The area ratio of the $AA$ region as a function of twist angle $\theta$. The red circles represent data from Ref.\cite{kazmierczak2021strain,  van_winkle_madeline_2021_4459671, van_winkle_madeline_2021_4459674, van_winkle_madeline_2021_4459676, van_winkle_madeline_2021_4459678, van_winkle_madeline_2021_4459680, van_winkle_madeline_2021_4459682, van_winkle_madeline_2021_4459687, van_winkle_madeline_2021_4459685, van_winkle_madeline_2021_4460605, van_winkle_madeline_2021_4460881, van_winkle_madeline_2021_4460609, van_winkle_madeline_2021_4460611, van_winkle_madeline_2021_4460615, van_winkle_madeline_2021_4460885, van_winkle_madeline_2021_4460883, van_winkle_madeline_2021_4460887, van_winkle_madeline_2021_4460879, van_winkle_madeline_2021_4464219, van_winkle_madeline_2021_4463612}}
\label{fig2}
\end{center}
\end{figure}
The Euler-Lagrange equations are solved numerically in reciprocal space with a $13\times13$ mesh. In Fig.~\ref{fig2}(a), we display the real-space distribution of interlayer distance of magic-angle TBG (with $\theta\!=\!1.05^{\circ}$). The interlayer distance exhibits a maximum value of $3.617\,\angstrom$ at the $AA$ point and a minimum value of $3.335\,\angstrom$  at the $AB/BA$ point. The domain wall between the $AB$ and $BA$ regions is also clearly seen.  
Prior to presenting further results of lattice relaxations, the $AA$ region should be defined more explicitly and precisely: here we define $AA$ region  as  the region within which the amplitude of the in-plane local shift is smaller than $\sqrt{3}a/6\!\approx\!0.71\,\angstrom$. In Fig.~\ref{fig2}(b), we present the in-plane relative distortions of magic-angle TBG. The color coding represent the amplitudes of the local in-plane distortion fields, while the directions of the  distortion fields are depicted by the black arrows. 
A rotational distortion field circling around the $AA$ region is obtained. As a result, the area of $AA$ region is decreased while that of the $AB/BA$ region is increased, minimizing the total energy of TBG. The maximal amplitude of the in-plane relative displacement is about $0.32\,\angstrom$. At $AB/BA$ point and SP, the relative in-plane distortions vanish and the lattice remains undistorted. Besides, the in-plane center-of-mass distortion of magic-angle TBG is $10^{3}$ times smaller than the in-plane relative distortion (see Fig.~\ref{sif2} in Appendix \III), which can be neglected.
Moreover, in order to investigate the variation of lattice distortions at different twist angles, we perform the lattice relaxation calculations with the twist angle ranging from $\theta\!=\!0.76^{\circ}$ to $\theta\!=\!1.54^{\circ}$.  The rotational distortion field at some given position $\mathbf{r}$ in the relaxed superlattice structure can be characterized  by a reconstruction rotation angle $\theta_R$.  At the edge of $AA$ region, the reconstruction rotation angle is denoted as $\theta^{AA}_R$.
In Fig.~\ref{fig2}(c),  we plot $\theta^{AA}_{R}$  as a function of twist angle  $\theta$. 
We see that $\theta^{AA}_R$ increases approximately linearly with the decrease of $\theta$.  We also evaluate the area ratio of $AA$ region in the fully relaxed moir\'e supercell and plot it as a function of twist angle $\theta$ in Fig.~\ref{fig2}(d). 
Clearly the area ratio of $AA$ region is also increased with the decrease of $\theta$. 
Both of the two results indicate that the lattice relaxation effects become more pronounced when the twist angle decreases. Our results show decent consistency with the corresponding experimental data \cite{kazmierczak2021strain}, marked as red dots in Fig.~\ref{fig2}(c) and Fig.~\ref{fig2}(d). We see that the discrepancy between the calculated values and the experimental measured ones  is on of the order of $20\%$. However, we also note that heterostrain and twist-angle disorder are present in the realistic device of TBG reported in Ref.~\onlinecite{kazmierczak2021strain}, which leads to an additional rotational angle $\sim 0.15^{\circ}$ at $AB/BA$ and SP positions \cite{kazmierczak2021strain}. While our structural relaxation calculations start from an ideal unstrained moir\'e superlattice, which naturally results in vanishing $\theta_R$ at $AB/BA$ and SP positions.
Therefore, the 20\% discrepancy between theory and experiment  may originate from the heterostrain in the device reported in Ref.~\onlinecite{kazmierczak2021strain}. Moreover, necessary smoothing of the experimental data may also lead  to  discrepancy between theory and experiment \cite{kazmierczak2021strain}.

\section{Phonons in TBG}

\subsection{Model and Formalism}
\label{sec:phonon}
With the converged lattice distortions characterized by $\{\bm{u}^{\pm}(\mathbf{r}), h^{\pm}(\mathbf{r})\}$ as discussed in the previous subsection
, we proceed to investigate the phonon properties in TBG using the continuum elastic model introduced in Sec.~\ref{sec:continu_elas_model}.  
To start with, we introduce a time-dependent displacement field near the equilibrium position:
\begin{equation}
\begin{split}
\textbf{u}^{\pm}(\textbf{r},t)=\textbf{u}^{\pm}_{c}(\textbf{r})+\delta\textbf{u}^{\pm}(\textbf{r},t),\\
h^{\pm}(\textbf{r},t)=h^{\pm}_{c}(\textbf{r})+\delta h^{\pm}(\textbf{r},t),
\end{split}
\end{equation}
where $\{\textbf{u}^{\pm}_{c}(\mathbf{r})\}$ and $\{h^{\pm}_{c}(\mathbf{r})\}$ are the converged lattice displacement fields obtained from the structural relaxation calculations. We define the Fourier transformations of the displacement fields as follows:
\begin{equation}
\begin{split}
\delta\textbf{u}^{\pm}(\textbf{r},t)=e^{-i\omega t}\sum_{\textbf{q}}\delta\textbf{u}^{\pm}_{\textbf{q}}e^{i\textbf{q}\cdot\textbf{r}},\\
\delta h^{\pm}(\textbf{r},t)=e^{-i\omega t}\sum_{\textbf{q}}\delta h^{\pm}_{\textbf{q}}e^{i\textbf{q}\cdot\textbf{r}},
\end{split}
\label{eq:phonon-fourier}
\end{equation}
where $\omega$ is the frequency of displacement fields. Then we expand the time dependent elastic energy and binding energy to the second order of $\delta\textbf{u}^{\pm}(\textbf{r},t)$ and $\delta h^{\pm}(\textbf{r},t)$, from which the dynamical matrix can be constructed. The derivations of the dynamical matrix are straightforward but tedious. The details of the derivations are presented in Appendix \IV. 

\begin{widetext}
After carefully evaluating the second-order functional derivatives of the total energy with respect to the displacement fields, we obtain the  equations of motions for the displacement fields in  Fourier space as follows:
\begin{align}
&\frac{\rho\omega^{2}}{4}\delta\tilde{\mathbf{u}}_{\textbf{G}+\textbf{q}}=\tilde{\textbf{U}}^{1,(2)}_{E,\textbf{G},\textbf{q}}\delta\tilde{\mathbf{u}}_{\textbf{G}+\textbf{q}}+\sum_{\textbf{G}'}\tilde{\textbf{U}}^{2,(2)}_{E,\textbf{G},\textbf{G}',\textbf{q}}\delta\tilde{\mathbf{u}}_{\textbf{G}'+\textbf{q}}+\sum_{\textbf{G}'}\tilde{\textbf{U}}^{(2)}_{B,\textbf{G},\textbf{G}'}\delta\tilde{\mathbf{u}}_{\textbf{G}'+\textbf{q}}
\label{eq:equation-of-motion}
\end{align}
where $\rho\!=\!7.61\times10^{-7}\,$kg/$\textrm{m}^{2}$ is the mass density of monolayer graphene.  Note that the wavevector $\mathbf{q}$ in Eq.~(\ref{eq:phonon-fourier}) has been written as the sum of  the wavevector within moir\'e Brillouin zone and a moir\'e reciprocal vector $\mathbf{G}$, i.e., $\mathbf{q}\to\mathbf{q}+\mathbf{G}$, so that $\textbf{q}$ in Eq.~(\ref{eq:equation-of-motion}) denotes the wavevector within moir\'e Brillouin zone.
$\textbf{G}\,(\textbf{G}')\,\!=\!m_{1}\, (m_1')\, \textbf{G}^{M}_{1}+m_{2}\, (m_2')\, \textbf{G}^{M}_{2}$  represents the reciprocal moir\'e lattice vector. The generalized displacement vector $\delta \tilde{\mathbf{u}}_{\mathbf{G}+\mathbf{q}}$ in Eq.~(\ref{eq:equation-of-motion}) is defined as
\begin{align}
&\delta\tilde{\textbf{u}}_{\textbf{G}+\textbf{q}}=[\delta u^{+}_{\textbf{G}+\textbf{q},x}\ \delta u^{+}_{\textbf{G}+\textbf{q},y}\ \delta h^{+}_{\textbf{G}+\textbf{q}}\ \delta u^{-}_{\textbf{G}+\textbf{q},x}\ \delta u^{-}_{\textbf{G}+\textbf{q},y}\ \delta h^{-}_{\textbf{G}+\textbf{q}} ]^{\mathrm{T}}\;,
\label{eq:u-field}
\end{align}
including both the relative and the center-of-mass vibrational modes in all three spatial directions. $\tilde{\textbf{U}}^{1,(2)}_{E,\textbf{G},\textbf{q}}$ represents the force constant contributed by the leading (second-order) terms of  the elastic energy with respect to the moir\'e reciprocal vectors (see Appendix \IV), which can be written as a block diagonal matrix. 
\begin{align}
&\tilde{\textbf{U}}^{1,(2)}_{E,\textbf{G},\textbf{q}}\;\nn
=&\frac{1}{4}\left[\begin{array}{cccccc}
				(\lambda+2\mu)\tilde{G}^{2}_{x}+\mu \tilde{G}^{2}_{y} & (\lambda+\mu)\tilde{G}_{x}\tilde{G}_{y}&0&0&0&0 \\
				(\lambda+\mu)\tilde{G}_{x}\tilde{G}_{y} & (\lambda+2\mu)\tilde{G}^{2}_{y}+\mu \tilde{G}^{2}_{x}&0&0&0&0 \\
				0&0&\kappa(\tilde{G}^{2}_{x}+\tilde{G}^{2}_{y})^{2}&0&0&0\\
				0&0&0&(\lambda+2\mu)\tilde{G}^{2}_{x}+\mu \tilde{G}^{2}_{y} & (\lambda+\mu)\tilde{G}_{x}\tilde{G}_{y}&0\\
				0&0&0&(\lambda+\mu)\tilde{G}_{x}\tilde{G}_{y} & (\lambda+2\mu)\tilde{G}^{2}_{y}+\mu \tilde{G}^{2}_{x}&0\\
				0&0&0&0&0&\kappa(\tilde{G}^{2}_{x}+\tilde{G}^{2}_{y})^{2}
\end{array}\right],
\label{eq:UE1}
\end{align}
where $\tilde{\textbf{G}}\!=\!\textbf{G}+\textbf{q}$. $\tilde{\textbf{U}}^{2,(2)}_{E,\textbf{G},\textbf{G}',\textbf{q}}$ is the force constant  contributed by the third-order terms (with respect to moir\'e reciprocal vectors)  of the elastic energy. Despite its higher-order nature, the $\tilde{\textbf{U}}^{2,(2)}_{E,\textbf{G},\textbf{G}',\textbf{q}}$ term couples all  components of the displacement fields together, giving rise to  fruitful phononic properties in TBG, which will be discussed in detail in Sec.~\ref{sec:phonon-results}. $\tilde{\textbf{U}}^{(2)}_{B,\textbf{G},\textbf{G}'}$ is the force constant contributed by the binding energy. The binding energy is a functional of the relative displacements, which only couples the $\delta h^{-}$ and $\delta\mathbf{u}^{-}$ displacement fields, and can be  expressed in the following matrix form 
\begin{align}
&\tilde{\textbf{U}}^{(2)}_{B,\textbf{G},\textbf{G}'}=\frac{1}{2}\left[\begin{array}{cccccc}
				0 &0&0&0&0&0 \\
				0 & 0&0&0&0&0 \\
				0&0&0&0&0&0\\
				0&0&0&V^{(2)}_{B,uu,\textbf{G}-\textbf{G}',xx} & V^{(2)}_{B,uu,\textbf{G}-\textbf{G}',xy}&V^{(2)}_{B,uh,\textbf{G}-\textbf{G}',x}\\
				0&0&0&V^{(2)}_{B,uu,\textbf{G}-\textbf{G}',xy} & V^{(2)}_{B,uu,\textbf{G}-\textbf{G}',yy}&V^{(2)}_{B,uh,\textbf{G}-\textbf{G}',y}\\
				0&0&0&V^{(2)}_{B,uh,\textbf{G}-\textbf{G}',x} & V^{(2)}_{B,uh,\textbf{G}-\textbf{G}',y}&V^{(2)}_{B,hh,\textbf{G}-\textbf{G}'}
\end{array}\right],
\end{align}
\end{widetext}
The explicit expressions of all the force-constant matrix elements  in Eq.~(\ref{eq:equation-of-motion}) can be found in Appendix \IV. We can solve the eigenvalues and eigenfunctions of the dynamical matrix to obtain the phonon frequencies and the vibrational modes at certain moir\'e wavevector $\textbf{q}$. We note that In Ref.~\cite{koshino-phonon-prb19}, part of the phonon modes of TBG have been calculated based on an elastic model which only includes the in-plane displacement fields. Here we present an elastic model in which both in-plane and out-of-plane displacements have been taken into account and are treated on equal footing. With such a model, we can capture all the essential features of the low-frequency phonon modes with a much lower computational cost compared to direct molecular dynamics simulations \cite{angeli-tbg-prx19,liu-phonon-nano22}, which will be discussed in detail in Sec.~\ref{sec:phonon-results}.

\subsection{Results}
\label{sec:phonon-results}

We numerically solve the equation of motion for the displacement fields to obtain the phonon properties of magic-angle TBG with $\theta=1.05^{\circ}$. 
For each moir\'e phonon wavevector, the displacement field is expanded by a plane-wave basis set on a $13\times 13$ mesh in reciprocal space, so that the total number of basis functions is $13\times 13\times3\times 2=1014$, which is more than 30 times smaller than that of atomistic molecular dynamics simulations.
Nevertheless, we can still capture all the key properties of low-frequency phonons accurately. 

\begin{figure*}
\begin{center}
    \includegraphics[width=16cm]{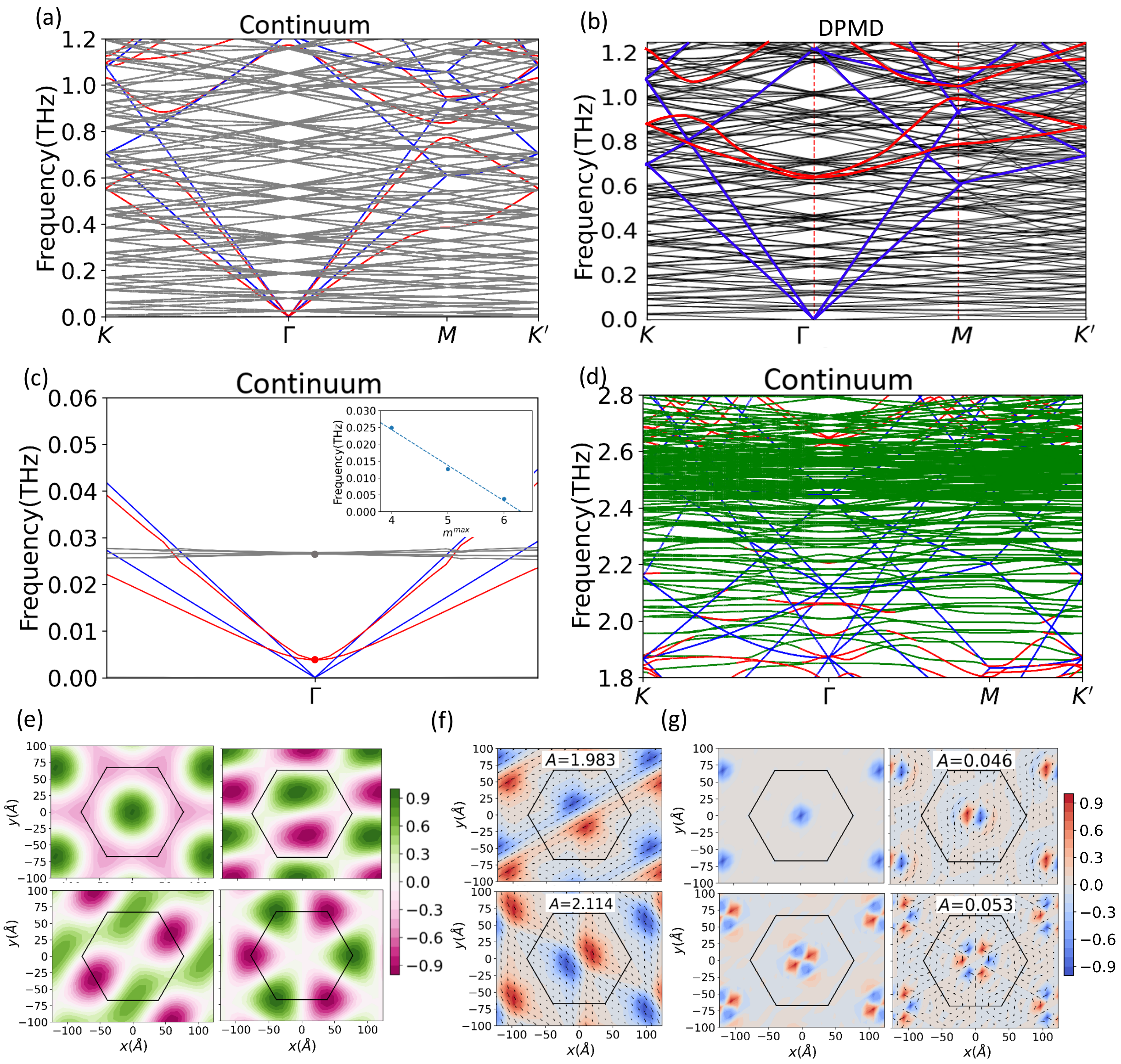}
\caption{
The phonon band structures of magic-angle TBG ($\theta\!=\!1.05^{\circ}$) calculated with the continuum elastic model (a) in the low-frequency regime from 0 to 1.2\,THz, (c) a zoom-in view near the $\Gamma$ point, and (d) in the (relatively) high-frequency regime from 1.8 to 2.8\,THz. (b) The phonon band structure of magic-angle TBG calculated with the DPMD method \cite{liu-phonon-nano22}. (e) The center-of-mass vibrational patterns of four out-of-plane flexural ($h^{+}$) phonon modes  at $\Gamma$ point with frequencies $\sim 0.027\,$THz, highlighted as a gray dot in (c). The colorbar represents the renormalized amplitudes of the out-of-plane center-of-mass vibrations. (f) The real-space vibrational patterns of the two gapless sliding ($u^{-}$) modes at $\Gamma$ point, highlighted as a red dot in (c). (g) The real-space vibrational patterns of four interlayer breathing ($h^{-}$) phonon modes at $\Gamma$ point with the frequencies $\sim 2\,$THz. The colorbar in (f) and (g) represents the renormalized amplitudes of the relative out-of-plane displacements. The arrows in (f) and (g) represents the directions of the relative in-plane vibrations. The ratios between the average amplitudes of the relative in-plane vibrations and the maximal amplitudes of the out-of-plane vibrations are marked as $A\!=\vert u^{-}\vert/\!\vert h^{-}\vert_{max}$ in the insets of (f) and (g). }
\label{fig3}
\end{center}
\end{figure*}

In Fig.~\ref{fig3}(a), (c) and (d), we present the projected phonon band structure calculated with the continuum elastic model. The green lines denote phonon modes with dominant $h^{-}$  components, and the red lines represent phonon modes with dominant $\textbf{u}^{-}$  components. Both of them are relative vibration modes between the two layers. The gray lines show the phonon band structures with dominant $h^{+}$ modes, and the blue lines show the phonon band structure with dominant $\textbf{u}^{+}$ modes, representing the center-of-mass vibrations. As a comparison, we also present the phonon band structures calculated with the deep potential molecular dynamics (DPMD) method \cite{liu-phonon-nano22} in Fig.~\ref{fig3}(b). We see that, except for two in-plane relative sliding modes, the results from continuum elastic model are in very well agreement with the DPMD results. More specifically, in Fig.~\ref{fig3}(c), we obtain three acoustic phonon modes (two $\mathbf{u}^{+}$ modes marked by blue lines  and one $h^{+}$ mode marked by gray lines), two gapless interlayer sliding modes ($\mathbf{u}^{-}$ modes marked by red lines), and a plethora of $h^{+}$, $\mathbf{u}^{-}$, and $h^{-}$ optical phonon modes. We note that although the sliding modes have a tiny gap of about $0.003\,$THz at $\Gamma$ point from our elastic model calculations (see Fig.~\ref{fig3}(c)), they are expected to be gapless since in the continuum model treatment, the ground state energy is invariant under arbitrary continuous interlayer shift, thus these sliding modes are gapless Goldstone modes.  In the inset of Fig.~\ref{fig3}(c), we present the gap of the sliding modes at $\Gamma$ point as a function of $m^{max}$, which indeed extrapolates to zero for $m^{max}\gtrapprox 7$. 
In realistic situation, TBG is an superlattice consisted of carbon atoms located at discrete positions. If the relative in-plane shift vector is incommensurate with graphene's lattice vector, the ground state energy of the system would be changed. This would  lead to gapped interlayer sliding modes with the gap about $0.65\,$THz at $\Gamma$ point, as calculated from DPMD approach \cite{liu-phonon-nano22}. In Fig.~\ref{fig3}(d), we present the phonon band structures in the (relatively) high-frequency regime, which are mostly consisted of $h^{-}$ modes. The $h^{-}$ optical modes are nearly dispersionless and have a large band gap of about $2\,$THz. As $\partial^2 V/\partial^2 h^{-}$ ($V$ denotes interlayer binding energy) 
is much larger than the other components of force constants, it costs much more energy to change the interlayer distance. This results in  the ``hard" interlayer breathing ($h^{-}$) modes with a large gap. 

In Fig,~\ref{fig3}(e), we present the real-space vibrational patterns of several optical $h^{+}$ modes at $\Gamma$ point with frequency of about $0.027\,$THz (gray point in Fig.~\ref{fig3}(c)). The colorbar denotes the amplitude of normalized center-of-mass out-of-plane vibration ($h^{+}$), where the opposite signs represent opposite vibrational directions. We obtain several ``flexural" moir\'e phonon modes exhibiting monopolar, dipolar, quadrupolar, and octupolar vibrational patterns, which are consistent with those reported in Ref.~\onlinecite{liu-phonon-nano22}. The center-of-mass in-plane vibrations ($\mathbf{u}^{+}$) lead to two gapless acoustic phonons, but have negligible contributions to optical phonon modes, which do not deserve further discussions. 
In Fig.~\ref{fig3}(f), we plot the real-space vibrational patterns of the two interlayer sliding modes at $\Gamma$ point (red point in Fig.~\ref{fig3}(c)). Since the out-of-plane and in-plane displacements are coupled together in our elastic model,  the sliding modes involve both in-plane and out-of-plane vibrational components.
We denote the maximal value of the out-of-plane vibration component of the sliding mode by $\vert h^{-}\vert_{max}$. The arrows in Fig.~\ref{fig3}(f) denote the vectors of relative in-plane vibrations, with the average amplitude $\vert \mathbf{u}^{-}\vert$. 
 In the inset of Fig.~\ref{fig3}(f) we list the ratio between the average in-plane   amplitude and the maximal out-of-plane amplitude of the sliding modes, 
denoted by $A\!=\!\vert \mathbf{u}^{-}\vert/\vert h^{-}\vert_{max}$. We see that $A\approx 2$ for both gapless sliding modes. The colorbar in Fig.~\ref{fig3}(f) represents the  relative out-of-plane vibration ($h^{-}$) component, with the maximal amplitude normalized to $1$. 
In Fig.~\ref{fig3}(g), we present the vibrational patterns of interlayer breathing modes ($h^{-}$ modes) at $\Gamma$ point, which also exhibit the monopolar, dipolar, the quadrupolar, and octupolar vibrational patterns.  Again, since the in-plane and out-of-plane vibrational components are coupled together in our elastic model, the interlayer breathing modes are not 100\% contributed by  the relative out-of-plane components, there are also small amount of mixtures of the relative in-plane components. 
The colorbar in Fig.~\ref{fig3}(g) represents the relative out-of-plane vibration with the maximal amplitude normalized to 1, i.e., $\vert h^{-}\vert_{max}=1$. The arrows in Fig.~\ref{fig3}(g) represent the directions of the relative in-plane components,  with their  average amplitudes  denoted by $\vert \mathbf{u}^{-}\vert$.  The in-plane amplitudes turn out to be much smaller than the out-of-plane ones for these breathing modes, with the average ratio $A=\vert \mathbf{u}^{-}\vert/\vert h^{-}\vert_{max}$ listed in the insets of Fig.~\ref{fig3}(g), with $A\sim 5\%$ for the dipolar and octupolar-type breathing modes. The in-plane components for the monopolar and quadrupolar breathing modes are negligible ($<\!1\%$).

Our results are in good agreement with those calculated by the DPMD method \cite{liu-phonon-nano22}. 
The main difference between our results and the ones from the DPMD calculations are the gaps of the sliding modes (see Fig.~\ref{fig3}(a) and (b)), which has been explained above. The monopolar-type, dipolar-type, quadrupolar-type, and octupolar-type out-of-plane vibration patterns for the out-of-plane flexural modes are identified by both methods with similar frequencies $\sim0.03\textrm{-}0.1\,$THz. Besides, we present more results about the interlayer sliding modes and out-of-plane breathing modes with various vibrational patterns.
Our model provides an efficient approach to capture the essential features of the moir\'e phonons in TBG. In particular, the low-frequency flexural, sliding, and breathing modes discussed above may play important roles in the electron-phonon coupling effects in magic-angle TBG.

\section{Electronic band structure}

\subsection{Model and Formalism}
\label{sec:bands}
In this section, we study the influence of the lattice distortions on the electronic band structures. 
Inspired by the treatment of in-plane lattice distortion effects on electronic structures in TBG reported in Ref.~\onlinecite{koshino-ep-prb20}, we start our discussions with the Slater-Koster tight binding model for graphene introduced in Ref.~\onlinecite{moon-tbg-prb13}. The hopping amplitude between two $p_{z}$ orbitals at different sites  is expressed as \cite{moon-tbg-prb13}
\begin{equation}
-T(\textbf{d})=V_{\sigma}\left(\frac{\textbf{d}\cdot\hat{\textbf{z}}}{d}\right)+V_{\pi}\left[1-\left(\frac{\textbf{d}\cdot\hat{\textbf{z}}}{d}\right)^{2}\right]\;,
\label{eq:sk-hopping}
\end{equation}
where $V_{\sigma}\!=\!V^{0}_{\sigma}e^{-(\left|\textbf{d}\right|-d_{c})/\delta_{0}}$ and $V_{\pi}\!=\!V^{0}_{\pi}e^{-(\left|\textbf{d}\right|-a_{0})/\delta_{0}}$. $\textbf{d}\!=\!(d_{x},d_{y},d_{z})$ is the displacement vector between the two sites. $d_{c}\!=\!3.35\,\angstrom$ is the interlayer distance of Bernal bilayer graphene. $a_{0}\!=\!a/\sqrt{3}\!=\!1.42\,\angstrom$ is the distance between two nearest neighbor carbon atoms in monolayer graphene. $\delta_{0}\!=\!0.184a$. $V^{0}_{\sigma}\!=\!0.48$\,eV and $V^{0}_{\pi}=-2.7\,$eV. We consider the Wannier function localized at position $\textbf{R}^{(l)}_{X}+\textbf{u}^{(l)}_{X}(\textbf{R})+h^{(l)}_{X}(\textbf{R})\,\hat{\mathbf{z}}$, where $\textbf{R}^{(l)}_X=\mathbf{R}_X + \tau_z^{(l)}\hat{\mathbf{z}}$ is the position of a given carbon atom in the undistorted $l$th  graphene monolayer ($l=1, 2$). $\textbf{R}_X$ and $\tau_z^{(l)}$ denote the in-plane and out-of-plane components of $\textbf{R}^{(l)}_X$,  with $\mathbf{R}_X=\mathbf{R}+\bm{\tau}_X$. 
Here $\bm{\tau}_X$ denotes the in-plane  position within graphene's primitive cell of sublattice $X$ ($X\!=\!A, B$), and $\mathbf{R}$ denotes the lattice vector of monolayer graphene. $\textbf{u}^{(l)}_{X}(\textbf{R}_X)$ and $h^{(l)}_{X}(\textbf{R}_X)$ characterize the in-plane and out-of-plane lattice distortions of a carbon atom in the $l$th layer  belonging to $X$ sublattice. The Fourier transform of the real-space hopping amplitude is defined as 
\begin{equation}
t(\textbf{q})=\frac{1}{S_{0}d_{0}}\int\mathrm{d}^{3}r\,T(\textbf{r})e^{-i\textbf{q}\cdot\textbf{r}}\;.
\label{eq:tq}
\end{equation}

The Bloch state of layer $l$, sublattice $X$, is defined as:
\begin{equation}
\ket{\textbf{k},X,l}=\frac{1}{\sqrt{N}}\sum_{\textbf{R}\in\textbf{R}^{(l)}_{X}}e^{i\textbf{k}\cdot(\textbf{R}+\mathbf{\tau}_{X})}\ket{\textbf{R}^{(l)}_X+\textbf{u}^{(l)}_{X}+h^{(l)}_{X}\hat{\mathbf{z}}},
\end{equation} 
where $\textbf{k}$ is the two-dimensional Bloch wave vector and $N$ is the number of single-layer graphene primitive cells in the system.

\subsubsection{Interlayer Hamiltonian}
The interlayer hopping from site $\textbf{R}^{(2)}_{X'}$ to site $\textbf{R}^{(1)}_{X}$ is $\bra{\textbf{R}^{(2)}_{X'}} U\ket{\textbf{R}^{(1)}_X}\!=\!-T(\textbf{R}^{(1)}_X-\textbf{R}^{(2)}_{X'})$, where the expression of $T(\textbf{R})$ is given in Eq.~(\ref{eq:sk-hopping}). Then the interlayer hopping matrix element in the Bloch basis is expressed as:
\begin{widetext}
\begin{equation}
\begin{split}
&\bra{\textbf{k}',X',2}U\ket{\textbf{k},X,1}\;\\
=&-\frac{1}{N}\sum_{\textbf{R}\in\textbf{R}^{(1)}_{X}}\sum_{\textbf{R}'\in\textbf{R}^{(2)}_{X'}}\frac{S_{0}d_{0}}{(2\pi)^{3}}\int\mathrm{d}^{3}p\ t(\textbf{p})\,e^{i\,(\,(\textbf{k}-\textbf{p}_{\parallel})\cdot(\textbf{R}+\bm{\tau}_{X})\,-\,\textbf{p}_{\parallel}\cdot\textbf{u}^{(1)}_{X}\,)}\,e^{-i\, p_{z}\,(\tau^{(1)}_z+h^{(1)}_{X})}\\
&\times e^{-i\,(\,(\textbf{k}'-\textbf{p}_{\parallel})\cdot(\textbf{R}^{'}+\bm{\tau}_{X'})-\textbf{p}_{\parallel}\cdot\textbf{u}^{(2)}_{X'}\,)}\,e^{i\, p_{z}\,(\tau^{(2)}_z+h^{(2)}_{X'})},
\end{split}
\label{eq:hopping-bloch}
\end{equation}
\end{widetext}
where $S_{0}$ is the area of the  graphene's primitive cell, and $d_{0}$ is the average interlayer distance of the relaxed lattice structure of TBG, which can be extracted from structural relaxation calculations as explained in Sec.~\ref{sec:lattice}. $\textbf{p}_{\parallel}$ and $p_z$ denote the in-plane and out-of-plane  components of a three dimensional wavevector $\textbf{p}$, respectively. $t(\textbf{p})$ is the Fourier transformation of the Slater-Koster hopping amplitude, which is expressed in Eq.~(\ref{eq:tq}). 

We further perform the Fourier transformation to both  in-plane and out-of-plane lattice distortions of layer $l$ and sublattice $X$:
\begin{equation}
\begin{split}
\textbf{u}^{(l)}_{X}(\textbf{R}_X)=\sum_{\textbf{G}_{m}}e^{i\textbf{G}_{m}\cdot(\textbf{R}+\bm{\tau}_{X})}\textbf{u}^{(l)}_{\textbf{G}_{m},X},\\
h^{(l)}_{X}(\textbf{R}_X)=\sum_{\textbf{G}_{m}}e^{i\textbf{G}_{m}\cdot(\textbf{R}+\bm{\tau}_{X})}h^{(l)}_{\textbf{G}_{m},X}\;,
\end{split}
\end{equation}
where $\bm{\tau}_{X}$ denotes the in-plane atomic position of sublattice $X$ within graphene's primitive cell.
Then, the exponentials of the lattice displacement fields can be expanded into Taylor series:
\begin{equation}
\begin{split}
&e^{i\textbf{p}_{\parallel}\cdot\textbf{u}^{(l)}_{X}(\textbf{R}_X)}
=\prod_{\textbf{G}_{m}}\sum_{n}\frac{1}{n!}(i\textbf{q}_{\parallel}\cdot\textbf{u}^{(l)}_{X,\textbf{G}_{m}})^{n}\,e^{i\,n\,\textbf{G}_{m}\cdot(\textbf{R}+\bm{\tau}_{X})},\\
&e^{i\,p_{z}h^{(l)}_{X}(\textbf{R}_X)}
=\prod_{\textbf{G}_{m}}\sum_{n_{h}}\frac{1}{n_{h}!}(i\,p_{z}h^{(l)}_{X,\mathbf{G}_m})^{n_{h}}\,e^{i\,n_{h}\,\mathbf{G}_m\cdot(\textbf{R}+\bm{\tau}_{X})}\;.
\end{split}
\label{eq:uh-taylor}
\end{equation}
The summation over lattice vectors $\mathbf{R}$, $\mathbf{R}'$ in Eq.~(\ref{eq:hopping-bloch}) can be eliminated using the identity: $\sum_{\textbf{R}}e^{i\textbf{p}\cdot\textbf{R}^{(l)}_X}\!=\!N\sum_{\textbf{g}}e^{i\textbf{g}^{(l)}\cdot\bm{\tau}^{(l)}_{X}}e^{i\,p_{z}\tau^{(l)}_{z}}\delta_{\textbf{p}_{\parallel},\textbf{g}^{(l)}}$, where $\textbf{g}^{(l)}\!=\!m_{1}\textbf{a}^{*,(l)}_{1}+m_{2}\textbf{a}^{*,(l)}_{2}$ is the atomic reciprocal lattice vector in the $l$th layer. As a result, we obtain the following interlayer hopping matrix element:
\begin{widetext}
\begin{equation}
\begin{split}
&\bra{\textbf{k}',X',2}U\ket{\textbf{k},X,1}\\
=&\sum_{\textbf{g}^{(1)},\textbf{g}^{(2)}}\sum_{n_{1},n_1' \dots}\sum_{n_{h,1},n_{h,1}' \dots}\gamma(\textbf{Q})\,e^{-i(\textbf{g}^{(1)}\cdot\bm{\tau}^{(1)}_{X}-\textbf{g}^{(2)}\cdot\bm{\tau}^{(2)}_{X'})}\delta_{\textbf{k}+\textbf{g}^{(1)}+n_{1}\textbf{G}_{1}+n_{h,1}\textbf{G}^{h}_{1}+\dots\, \mathbf{,}\,\textbf{k}'+\textbf{g}^{(2)}-n'_{1}\textbf{G}_{1}-n'_{h,1}\textbf{G}^{h}_{1}+\dots},
\end{split}
\label{eq:hopping-bloch2}
\end{equation}
\end{widetext}
where $\textbf{Q}\!=\!\textbf{Q}_{\parallel}+p_{z}\hat{\mathbf{z}}$, and 
\begin{equation}
\begin{split}
&\textbf{Q}_{\parallel}\\
=&\textbf{k}+\textbf{g}^{(1)}+n_{1}\textbf{G}_{1}+n_{h,1}\textbf{G}^{h}_{1}+n_{2}\textbf{G}_{2}+n_{h,2}\textbf{G}^{h}_{2}+\dots\\
=&\textbf{k}'+\textbf{g}^{(2)}-n'_{1}\textbf{G}_{1}-n'_{h,1}\textbf{G}^{h}_{1}-n'_{2}\textbf{G}_{2}-n'_{h,2}\textbf{G}^{h}_{2}+\dots\;,
\end{split}
\end{equation}
where $\{\mathbf{G}_1,\mathbf{G}_2,\mathbf{G}^{h}_1,\mathbf{G}^{h}_2,...\}$ are moir\'e reciprocal vectors.
The effective interlayer hopping amplitude $\gamma(\textbf{Q})$ is given by:
\begin{widetext}
\begin{equation}
\label{eq:gammaexp}
\begin{split}
&\gamma(\textbf{Q}_{\parallel}+p_{z}\textbf{e}_{z})\\
\approx&-\frac{d_{0}}{2\pi}\int\mathrm{d}p_{z}\,t(\textbf{Q})\,e^{i\,p_{z}d_{0}}\,\frac{\left[i\textbf{Q}_{\parallel}\cdot\textbf{u}^{-}_{\textbf{G}_{1}}/2\right]^{n_{1}+n'_{1}}}{n_{1}!n'_{1}!}\frac{\left[i\textbf{Q}_{\parallel}\cdot\textbf{u}^{-}_{\textbf{G}_{2}}/2\right]^{n_{2}+n'_{2}}}{n_{2}!n'_{2}!}\dots\times\frac{\left[i\,p_{z}h^{-}_{\textbf{G}^{h}_{1}}/2\right]^{n_{h,1}+n'_{h,1}}}{n_{h,1}!n'_{h,1}!}\frac{\left[i\,p_{z}h^{-}_{\textbf{G}^{h}_{2}}/2\right]^{n_{h,2}+n'_{h,2}}}{n_{h,2}!n'_{h,2}!}\dots\\
& \times e^{-i(\textbf{G}_{1}\cdot\bm{\tau}_{X})n_{1}}e^{-i(\textbf{G}_{1}\cdot\bm{\tau}_{X'})n'_{1}}\dots\times e^{-i(\textbf{G}^{h}_{1}\cdot\bm{\tau}_{X})n_{h,1}}e^{-i(\textbf{G}^{h}_{1}\cdot\bm{\tau}_{X'})n'_{h,1}}\dots
\end{split}
\end{equation}
\end{widetext}
When going from Eq.~(\ref{eq:hopping-bloch}) to Eq.~(\ref{eq:hopping-bloch2}) and (\ref{eq:gammaexp}), we have dropped the effects of the in-plane center-of-mass displacements $\mathbf{u}^{+}$, which turns out to be three orders of magnitudes smaller than the relative displacement fields $\mathbf{u}^{-}$ according to our structural relaxation calculations (see Sec.~\ref{sec:lattice-results} for details).
We note that the phase factor in Eq.~(\ref{eq:gammaexp}) $\exp(-i\textbf{G}\cdot\bm{\tau}_X)\sim\exp(-i\theta)\approx 1$ for small twist angle $\theta$, which can be dropped if we are interested in small angle TBG, e.g. when $\theta$ is around the magic angle.

The interlayer hopping amplitude $t(\mathbf{Q})$ varies on the length scale of $1/a$ in reciprocal space, while the  Fourier transformed displacement field decays quickly on the scale of $1/L_s$. Thus, to the leading-order approximation, it is legitimate to assume: $\mathbf{Q}_{\parallel}\approx \mathbf{k}+\mathbf{g}^{(1)}=\mathbf{k'}+\mathbf{g}^{(2)}$. Since we are interested in the low-energy Dirac electrons around $\mathbf{K}$ and $\mathbf{K'}$ points, one can further set $\mathbf{k}$ and $\mathbf{k}'$ to the Dirac points in atomic Brillouin zone. This would give rise to three pairs of reciprocal vectors $\{(\mathbf{g}^{(2)}_j,\mathbf{g}^{(1)}_j)\}$ ($j=1, 2, 3$) satisfying the condition $\mathbf{k}+\mathbf{g}^{(1)}=\mathbf{k}'+\mathbf{g}^{(2)}$ \cite{macdonald-pnas11,castro-neto-prb12,koshino-ep-prb20}.
It is convenient to define $\textbf{Q}_{1}^{\mu}\!=\!\textbf{K}^{\mu}$, $\textbf{Q}_{2}^{\mu}\!=\!\textbf{K}^{\mu}+\mu\textbf{a}^{*}_{1}$, and $\textbf{Q}_{3}^{\mu}\!=\!\textbf{K}^{\mu}+\mu(\textbf{a}^{*}_{1}+\textbf{a}^{*}_{2})$, where $\mu=\pm$ is the valley index, and $\mathbf{K}^{-} (\mathbf{K}^{+})=\mathbf{K} (\mathbf{K}')$. Then the  interlayer hopping matrix element can be approximated as:
\begin{equation}
\begin{split}
&\bra{\textbf{k}',X',2}U\ket{\textbf{k},X,1}\\
=&\sum_{j=1}^{3}\sum_{n_{1},n_{h,1},\dots}\sum_{n_{1}',n'_{h,1}\dots}\gamma(\textbf{Q}_{j}+p_{z}\textbf{e}_{z})M^{j}_{X'X}\,\\
&\times \delta_{\textbf{k}',\textbf{k}+\mathbf{G}_j+(n_{1}+n'_{1})\textbf{G}_{1}+(n_{h,1}+n'_{h,1})\textbf{G}^{h}_{1}+\dots},
\end{split}
\label{eq:hopping-simple}
\end{equation}
where $\mathbf{G}_{1}\!=\!\mathbf{0}$, $\mathbf{G}_{2}\!=\!\mu\textbf{G}^{M}_{1}$, $\textbf{G}_{3}\!=\!\mu(\textbf{G}^{M}_{1}+\textbf{G}^{M}_{2})$. $M^{j}_{X'X}$ is given by:
\begin{align}
\begin{split}
M^{1}=\left[\begin{array}{cc}
				1 & 1 \\
				1 & 1 
\end{array}\right],
M^{2}=\left[\begin{array}{cc}
				1 & \omega^{-\mu} \\
				\omega^{\mu} & 1 
\end{array}\right],
M^{3}=\left[\begin{array}{cc}
				1 & \omega^{\mu} \\
				\omega^{-\mu} & 1 
\end{array}\right]\;,
\end{split}
\end{align}
where $\omega\!=\!e^{i2\pi/3}$. 

Taking use of the identity $\delta_{\mathbf{k}',\mathbf{k}+\mathbf{q}}=(1/S)\int d^2r \, e^{i(\mathbf{k}+\mathbf{q}-\mathbf{k}')\cdot\mathbf{r})}$,  the interlayer hopping matrix element can be  expressed in real-space representation as:
\begin{equation}
\begin{split}
\bra{\textbf{k}',X',l'}U\ket{\textbf{k},X,l}=&\frac{1}{S}\int\mathrm{d}^{2}\textbf{r}\,e^{i(\textbf{k}-\textbf{k}')\cdot\textbf{r}}\,U_{X'X}(\textbf{r})
\end{split}
\end{equation}
where the ``moir\'e potential" with relaxed lattice structure is given by \cite{koshino-ep-prb20}
\begin{equation}
\begin{split}
&U_{X'X}(\textbf{r})\\
=&-\sum_{j=1}^{3}\frac{d_{0}}{2\pi}\int^{\infty}_{-\infty}\mathrm{d}p_{z}\,M^{j}_{X'X}e^{i(\textbf{Q}_{j}\cdot\textbf{u}^{-}(\textbf{r})+\mathbf{G}_{j}\cdot\textbf{r}+\,p_{z}(d_0+h^{-}(\textbf{r}))}\,\\
&\times t(\textbf{Q}_{j}+p_{z}\textbf{e}_{z})
\end{split}
\end{equation}

The integration over $p_z$ in Eq.~(\ref{eq:gammaexp}) can be carried out analytically using the trick of integration by parts. We first consider the terms with $n_{h,1}+n'_{h,1}=0, n_{h,2}+n'_{h,2}=0, \dots$, i.e. the effects from the out-of-plane distortions are completely neglected. Taking the approximation $\mathbf{Q}_{\parallel}\approx\mathbf{Q}_j$,  the integration over $p_z$ leads to:
\begin{equation}
\begin{split}
-\frac{d_{0}}{2\pi}\int\mathrm{d}p_{z}\,t(\textbf{Q}_j+p_{z}\textbf{e}_{z})e^{i\,p_{z}d_{0}}
\approx 0.101\,\textrm{eV},
\end{split}
\label{eq:t0}
\end{equation}
where $d_0=3.3869\,$\angstrom\ is the average interlayer distance of the fully relaxed lattice structure.
Then, we consider the first order effects from out-of-plane distortions.  Specifically, for certain moir\'e reciprocal lattice vector $\textbf{G}^{h}_{m_{1}}$, we have $n_{h,m_{1}}+n'_{h,m_{1}}\!=\!1$ and $n_{h,m}+n'_{h,m}\!=\!0$ with $m\!\neq\!m_{1}$. The integration over $p_z$ for such first-order term can also be done analytically:
\begin{equation}
\begin{split}
-&\frac{d_{0}}{2\pi}\int\mathrm{d}p_{z}\,t(\textbf{Q}_j+p_{z}\textbf{e}_{z})e^{i\,p_{z}d_{0}}\times\left[i\,p_{z}h^{-}_{\textbf{G}_{m_{1}}^{h}}\right]\\
\approx & -0.248h^{-}_{\textbf{G}_{m_{1}}^{h}}\,\textrm{eV}
\end{split}
\label{eq:t1}
\end{equation}
We also consider the second-order effects to the interlayer hopping from out-of-plane lattice distortions. For some moir\'e reciprocal lattice vectors $\textbf{G}^{h}_{m_{1}}$ and $\textbf{G}^{h}_{m_{2}}$, we have $n_{h,m_{1}}+n'_{h,m_{1}}=1, n_{h,m_{2}}+n'_{h,m_{2}}\!=\!1$ and $n_{h,m}+n'_{h,m}\!=\!0$ with $m\!\neq\!m_{1}, m_{2}$ and $m_{1}\!\neq\!m_{2}$, which leads to
\begin{equation}
\begin{split}
-&\frac{d_{0}}{2\pi}\int\mathrm{d}p_{z}\,t(\textbf{Q}_j+p_{z}\textbf{e}_{z})e^{i\,p_{z}d_{0}}\times (ip_z h^{-}_{\mathbf{G}^{h}_{m_{1}}})\times(ip_z h^{-}_{\mathbf{G}^{h}_{m_{2}}})\\
\approx &0.508 \,h^{-}_{\mathbf{G}^{h}_{m_{1}}}\,h^{-}_{\mathbf{G}^{h}_{m_{2}}}\,\textrm{eV}\;.
\end{split}
\label{eq:t2}
\end{equation}
The detailed derivations of Eqs.~(\ref{eq:t0})-(\ref{eq:t2}) are presented in Appendix \V.
Finally  we obtain the following expression for the effective interlayer hopping amplitude:
\begin{widetext}
\allowdisplaybreaks
\begin{align}\label{eq:intert0t1t2}
&\gamma(\textbf{Q}_{\parallel}+p_{z}\textbf{e}_{z})\nonumber\\
\approx&\,+\, t_{0}(\mathbf{Q}_j)\times\frac{\left[i\textbf{Q}_j\cdot\textbf{u}^{-}_{\textbf{G}_{1}}/2\right]^{n_{1}+n'_{1}}}{n_{1}!n'_{1}!}\times\frac{\left[i\textbf{Q}_j\cdot\textbf{u}^{-}_{\textbf{G}_{2}}/2\right]^{n_{2}+n'_{2}}}{n_{2}!n'_{2}!}\times \dots \,\nn
&-\,t_{1}(\mathbf{Q}_j)h_{\textbf{G}_{1}^{h}}^{-}\times\frac{\left[i\textbf{Q}_j\cdot\textbf{u}^{-}_{\textbf{G}_{1}}/2\right]^{n_{1}+n'_{1}}}{n_{1}!n'_{1}!}\times\frac{\left[i\textbf{Q}_j\cdot\textbf{u}^{-}_{\textbf{G}_{2}}/2\right]^{n_{2}+n'_{2}}}{n_{2}!n'_{2}!}\times\dots\nonumber\\
&\,+\,t_{2}(\mathbf{Q}_j)h_{\textbf{G}_{1}^{h}}^{-}h_{\textbf{G}_{2}^{h}}^{-}\times\frac{\left[i\textbf{Q}_j\cdot\textbf{u}^{-}_{\textbf{G}_{1}}/2\right]^{n_{1}+n'_{1}}}{n_{1}!n'_{1}!}\times\frac{\left[i\textbf{Q}_j\cdot\textbf{u}^{-}_{\textbf{G}_{2}}/2\right]^{n_{2}+n'_{2}}}{n_{2}!n'_{2}!}\times\dots +\dots\;\nn
\end{align}
\end{widetext}
where $t_{0}(\mathbf{Q}_j)\!\approx\!0.101\,$eV, $t_{1}(\mathbf{Q}_j)\!\approx\!0.248\,$eV/\angstrom\ and $t_{2}(\mathbf{Q}_j)\!\approx\!0.508\,$eV/\angstrom$^2$.

\begin{figure}[bth!]
\begin{center}
    \includegraphics[width=3.2in]{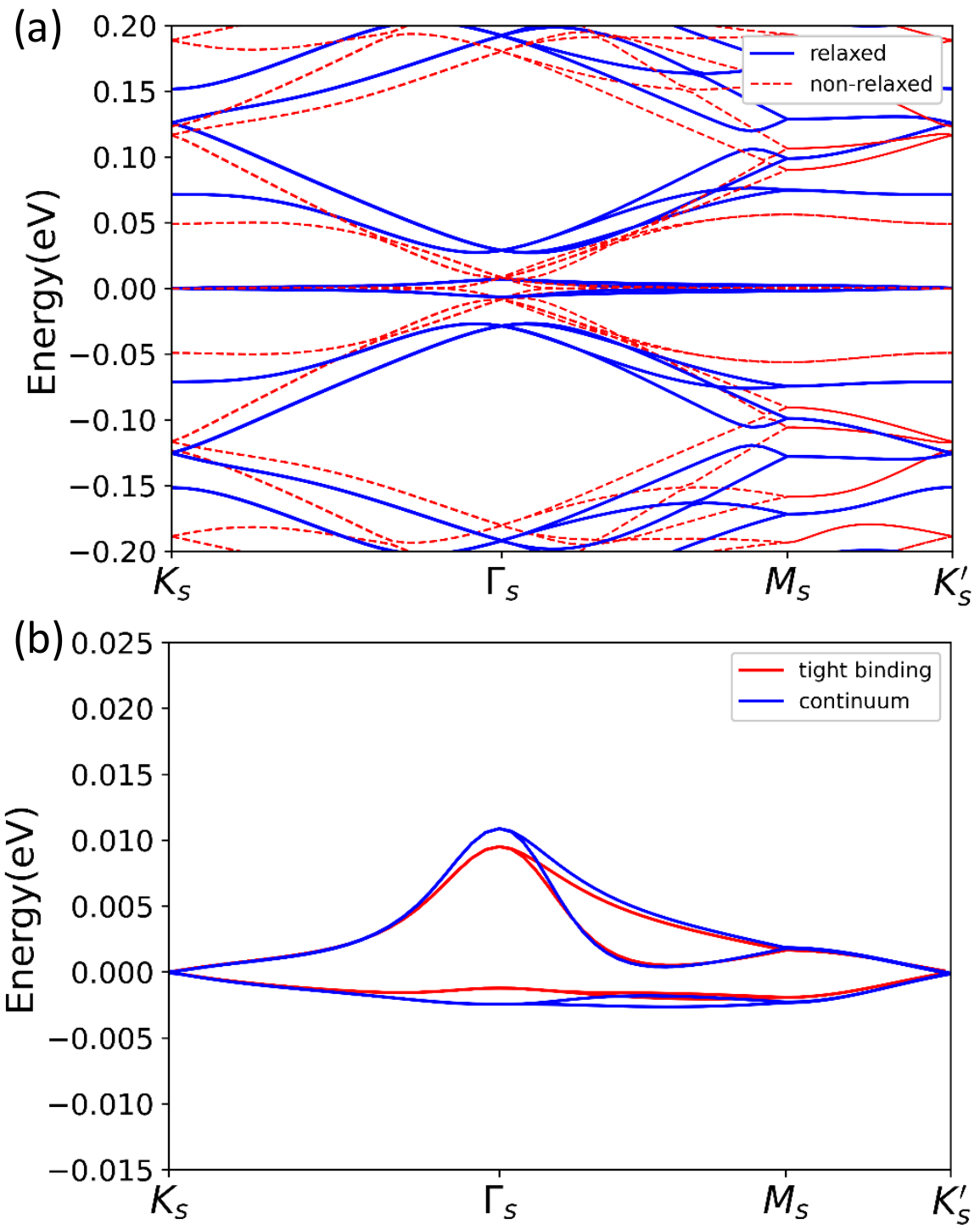}
\caption{
The electronic band structures of TBG magic-angle twist angle ($\theta\!=\!1.05^{\circ}$). (a) The band structures calculated with the effective continuum model, where the red dashed lines denote the energy bands with an ideal moir\'e superlattice, and the blue solid lines represent the energy bands with fully relaxed lattice structure with $\mathbf{k}$-independent interlayer coupling. (b) The band structure with fully relaxed lattice structure, where the red lines are the energy bands calculated with the atomistic Slater-Koster tight binding model, and the blue lines are the energy bands calculated with the effective continuum model taking into account the $\textbf{k}$-dependent interlayer coupling and the $\textbf{k}^{2}$ terms in the intralayer Hamiltonian).}
\label{fig4}
\end{center}
\end{figure}

In Eq.~(\ref{eq:hopping-simple}), the interlayer hopping matrix element has been simplified in such a way that the wavevector dependence of the effective hopping amplitude has been omitted, i.e., $\gamma(\mathbf{Q}_{\parallel}+p_z\hat{\mathbf{z}})\approx\gamma(\mathbf{Q}_j+p_z\hat{\mathbf{z}})$ ($j=1, 2, 3$). As will be shown in Sec.~\ref{sec:bands-results}, such a simplified treatment to the effective interlayer hopping can already capture the essential features of the low-energy band structures of magic-angle TBG, such as the significant reduction of the flat bandwidth and the dramatically enhanced gaps between flat bands and the remote bands (see Fig.~\ref{fig4}(a)). However, there is still non-negligible discrepancy compared to the band structures directly calculated from atomistic Slater-Koster tight-binding model with fully relaxed lattice structure (see Sec.~\ref{sec:bands-results}), which comes from the wavevector dependence of interlayer hopping $\gamma(\mathbf{Q})$ as well as the $\mathcal{O}(k^2)$ terms of the intralayer Hamiltonian. Both of the high order terms would lead to particle-hole asymmetry in the electronic band structures, which will be discussed as follows.

Following the hierarchy shown in Eq.~(\ref{eq:intert0t1t2}),  
we first consider the $\textbf{k}$ dependence of the bare interlayer hopping amplitude $t_0$ without any lattice relaxation effects, i.e. $\textbf{Q}_{\parallel}\!=\!\textbf{k}+\textbf{g}^{(1)}=\textbf{k}'+\textbf{g}^{(2)}$. Specifically, an analytic expression $t_{0}(\textbf{q})\!=\!t^{0}_{0}\,\exp(-\alpha(\left|\textbf{q}\right|d_{\perp})^{\gamma})$ is  adopted from Refs.~\onlinecite{macdonald-prb10,SongZhida-tbg2-prb21} , where 
$d_{\perp}\!=\!3.3869\,\angstrom$. $t^{0}_{0}\!=\!1\,$eV, $\alpha\!=\!0.1565$ and $\gamma\!=\!1.5313$. Then, $t_{0}(\mathbf{Q}_j)$ is substituted by $t_{0}(\textbf{Q}_{j}+\bar{\textbf{k}})$ in Eq.~(\ref{eq:intert0t1t2}), where $\bar{\mathbf{k}}$ measures the deviation of the electron's wavevector with respect to the Dirac point. Second, we consider the $\textbf{k}$ dependence for the $t_0$ term in Eq.~(\ref{eq:intert0t1t2}) including the first-order in-plane relaxation effects. Specifically, for certain moir\'e reciprocal lattice vector $\textbf{G}_{m_{1}}$, we have $n_{m_{1}}+n'_{m_{1}}\!=\!1$ and $n_{m}+n'_{m}\!=\!0$ with $m\!\neq\!m_{1}$, i.e. $\textbf{Q}_{\parallel}\!=\!\textbf{k}+\textbf{g}^{(1)}+n_{m_{1}}\textbf{G}_{m_{1}}=\textbf{k}'+\textbf{g}^{(2)}-n'_{m_{1}}\textbf{G}_{m_{1}}$. This term is thus expressed as $t_{0}(\textbf{Q}_{j}+\bar{\textbf{k}}+n_{m_{1}}\textbf{G}_{m_{1}})i(\textbf{Q}_{j}+\bar{\textbf{k}}+n_{m_{1}}\textbf{G}_{m_{1}})\cdot\textbf{u}^{-}_{G_{1}}$ , which have $\textbf{k}$ dependence in both $t_{0}$ and $\textbf{Q}_{\parallel}$. 
Third, we consider the $\textbf{k}$ dependence for the $t_{1}$ in Eq.~(\ref{eq:intert0t1t2}) including the leading order out-of-plane corrugation effects, with  $n_{h,m_{1}}+n'_{h,m_{1}}\!=\!1$ and $n_{h,m}+n'_{h,m}\!=\!0$, $m\!\neq\!m_{1}$. We take a linear function $t_{1}(\textbf{q})\!=\!t^{0}_{1}+t^{1}_{1}\left|\textbf{q}\right|$ to fit the wavevector dependence of $t_{1}(\mathbf{q})$ near $\textbf{K}^{\pm}$ point, where $t^{0}_{1}=-1.0235\,$eV  and $t^{1}_{1}\!=\!0.4531\,$eV$\cdot\angstrom$.  As a result, we substitute constant $t_{1}(\mathbf{Q}_j)$ by $t_{1}(\textbf{Q}_{j}+\overline{\textbf{k}}+n_{h_{m_{1}}}\textbf{G}^{h}_{m_{1}})$.

Besides, the high order $\textbf{k}$ dependence in the intralayer hopping amplitude would break the particle-hole symmetry as well. We start with the tight binding Hamiltonian of the single layer graphene introduced above. We can expand the Hamiltonian near $\textbf{K}^{\mu}$ ($\mu=\pm$) point:
\begin{equation}
\begin{split}
&H^{(l)}(\textbf{k})\\
=&-\hbar\,v_{F}\,[\,\bar{k}_{x}\,\mu\sigma_{x}+\bar{k}_{y}\sigma_{y}+m_{\alpha}(\bar{k}^{2}_{x}-\bar{k}^{2}_{y})\sigma_{x}\;\\
&-2m_{\alpha}\bar{k}_{x}\bar{k}_{y}\,\mu\sigma_{y}+m_{\beta}(\bar{k}_{x}^{2}+\bar{k}_{y}^{2})\sigma_{0}\,]\,,
\end{split}
\end{equation}
where $\bar{\textbf{k}}=\textbf{k}-\textbf{K}^{\mu}$, $m_{\alpha}\!=\!0.4563\,\angstrom$ and $m_{\beta}\!=\!0.2345\,\angstrom$.

\subsubsection{Intralayer Hamiltonian}
The lattice distortions also have significant influences on the intralayer hopping terms of TBG. Microscopically, the  strain field would change the position of each carbon atom and affect the in-plane hopping amplitudes.
In order to treat both in-plane and out-of-plane lattice distortions on equal footing, here we  introduce a three dimensional strain tensor $\hat{u}$, which is expressed as
\begin{align}\label{strain}
\begin{split}
\hat{u}=\left[\begin{array}{ccc}
				u_{xx} & u_{xy} & u_{hx} \\
				u_{xy} & u_{yy} & u_{hy} \\
				u_{hx} & u_{hy} & u_{hh}
\end{array}\right],\\
\end{split}
\end{align}
 where $u_{ij}\!=\!(\partial u_{i}/\partial r_j + \partial u_{j}/\partial r_i)/2$, $i,j\!=\!x, y$, and $u_{hi}\!=\!\partial h/\partial r_i$, $i\!=\!x, y$. We consider the hopping events between $B$ and $A$ sublattices which are connected to each other by the three first neighbor vectors  $\textbf{r}^{0}_{1}\!=\!(0,a_{0},0)$, $\textbf{r}^{0}_{2}\!=\!(\sqrt{3}/2,-1/2,0)a_{0}$ and $\textbf{r}^{0}_{3}\!=\!(-\sqrt{3}/2,-1/2,0)a_{0}$ with $a_{0}\!=\!a/\sqrt{3}$, with the hopping amplitude given by Eq.~(\ref{eq:sk-hopping}).
The three vectors pointing from B sublattice to A sublattice undergoes small shifts from $\textbf{r}^{0}_{i}$ to $\textbf{r}_{i}\!=\!(1+\hat{u})\textbf{r}^{0}_{i}$ ($i=1, 2, 3$). As a result, the hopping amplitudes change from $t(\textbf{r}^{0}_{i})$ to $t(\textbf{r}_{i})$ for $i=1, 2, 3$. We have carefully derived the differences for the three nearest neighbor hopping amplitudes $\delta t_i=t(\textbf{r}_{i})-t(\textbf{r}_{i}^{0})$, and the detailed expressions are given in Appendix \V. Then we consider the monolayer graphene Hamiltonian in reciprocal space:
\begin{align}\label{strain}
\begin{split}
H_0(\mathbf{k})=\left[\begin{array}{cc}
				0 & g(\textbf{k},\hat{u})  \\
				g^{*}(\textbf{k},\hat{u}) & 0
\end{array}\right],\\
\end{split}
\end{align}
where $g(\textbf{k},\hat{u})\!=\!\sum^{3}_{i=1}t(\textbf{r}_{i})e^{i\textbf{k}\cdot\textbf{r}_{i}}$. We can expand $g(\textbf{k})$ near the $K^{\mu}$ point, with $K_{\mu}\!=\!\mu(-4\pi/3a,0,0)$ with $\mu=\pm$. Then we have $g(\bar{\textbf{k}},\hat{u})\!=\!-\hbar v_{F}[\mu(\bar{k}_{x}+\mu A_{x})-i(\bar{k}_{y}+\mu A_{y})]$, with $\overline{\textbf{k}}\!=\!\textbf{k}-\textbf{K}^{\mu}$. In other words, the strain field acts as a pseudo vector potential coupled with low-energy Dirac electrons \cite{castro-neto-strain-prl09}.
Including effects of both in-plane  and out-of-plane lattice distortions, the pseudo vector potential induced by strain in the $l$th layer graphene $\mathbf{A}^{(l)}=[A_x^{(l)}, A_y^{(l)}]$ is expressed by
\begin{widetext}
\allowdisplaybreaks
\begin{align}
A_{x}^{(l)}=&\frac{\beta\gamma_{0}}{ev}\left[\frac{3}{4}(\frac{\partial u^{(l)}_{x}}{\partial x}-\frac{\partial u^{(l)}_{y}}{\partial y})+\frac{3}{32}\left((1-3\beta)\left(\frac{\partial u^{(l)}_{x}}{\partial x}\right)^{2}-(4+4\beta)\left(\frac{\partial u^{(l)}_{x}}{\partial y}+\frac{\partial u^{(l)}_{y}}{\partial x}\right)^{2}\right.\right.\nonumber\\
&\left.\left.-(2+2\beta)\frac{\partial u^{(l)}_{x}}{\partial x}\frac{\partial u^{(l)}_{y}}{\partial y}+(1+5\beta)\left(\frac{\partial u^{(l)}_{y}}{\partial y}\right)^{2}\right)\right]\nonumber+\left(\mu\frac{\beta\gamma_{0}}{ev}\frac{3}{8}+\mu\frac{M}{ev}\frac{3}{4}\right)\left[\left(\frac{\partial h^{(l)}}{\partial x}\right)^{2}-\left(\frac{\partial h^{(l)}}{\partial y}\right)^{2}\right],\nonumber\\
A_{y}^{(l)}=&\frac{\beta\gamma_{0}}{ev}\left[-\frac{3}{4}\left(\frac{\partial u^{(l)}_{x}}{\partial y}+\frac{\partial u^{(l)}_{y}}{\partial x}\right)+\frac{3}{16}\left((1+3\beta)\frac{\partial u^{(l)}_{x}}{\partial x}\left(\frac{\partial u^{(l)}_{x}}{\partial y}+\frac{\partial u^{(l)}_{y}}{\partial x}\right)-(1-\beta)\left(\frac{\partial u^{(l)}_{x}}{\partial y}+\frac{\partial u^{(l)}_{y}}{\partial x}\right)\frac{\partial u^{(l)}_{y}}{\partial y}\right)\right]\nonumber\\
&-\left(\mu\frac{\beta\gamma_{0}}{ev}\frac{3}{4}+\mu\frac{M}{ev}\frac{3}{2}\right)\frac{\partial h^{(l)}}{\partial x}\frac{\partial h^{(l)}}{\partial y},
\label{eq:Avec}
\end{align}
\end{widetext}
where $M\!=\!-V^{0}_{\pi}+V^{0}_{\sigma}e^{-(a_{0}-d_{0})/r_{0}}\!\approx\!36\,$eV, $\beta\!=\!a_0/r_0\!\approx\!3.14$ and $\gamma_{0}=\left|V^{0}_{\pi}\right|\!=\!2.7\,$eV. $v$ in the above equation is just the Fermi velocity $v_F$.
 Thus, the intralayer Hamiltonian of valley $\mu$ for the $l$th layer is given by
\begin{equation}
H^{\mu,(l)}(\textbf{k})=-\hbar v_{F}\left[\left(\textbf{k}+\mu\frac{e}{\hbar}\textbf{A}^{(l)}-\textbf{K}^{\mu,(l)}\right)\right]\cdot(\mu\sigma_{x},\sigma_{y}),
\end{equation}
where $\sigma_{x}$ and $\sigma_{y}$ are Pauli matrices defined in the sublattice space and $\textbf{K}^{\mu,(l)}$ is the Dirac point in the $l$th layer from valley $\mu$. The total Hamiltonian of TBG for $\mu$ valley is given by
\begin{align}\label{eq:conham}
\begin{split}
H=\left[\begin{array}{cc}
				H^{\mu,(1)} & U^{\dagger}  \\
				U & H^{\mu,(2)} 
\end{array}\right]\;,\\
\end{split}
\end{align}
where the interlayer coupling $U$ has been discussed in the previous subsection.

\subsection{Results}
\label{sec:bands-results}

In this subsection, we present the effects of the lattice distortions on the electronic band structure of magic-angle TBG using both the effective continuum model and the atomistic tight binding model. To be specific, the effective continuum model Eq.~(\ref{eq:conham}) is expanded in reciprocal space with a $7\times7$ mesh. As discussed in Sec.~\ref{sec:bands}, the  interlayer Hamiltonian matrix element is expanded as a power series of lattice distortion $\textbf{u}^{\pm}_{\textbf{G}_{m}}$ and $h^{\pm}_{\textbf{G}_{m}}$. 
The cutoff of the expansion is set at the third order of the product of $\textbf{u}^{\pm}_{\textbf{G}_{m}}$ and $h^{\pm}_{\textbf{G}_{m}}$. For the intralayer Hamiltonian, strain couples to Dirac electrons as a pseudo vector potential $\mathbf{A}$ as expressed in Eqs.~(\ref{eq:Avec}).

In Fig.~\ref{fig4}(a), we present the band structure calculated by the effective continuum model with $\mathbf{k}$ independent interlayer coupling as given by Eq.~(\ref{eq:hopping-simple}), and the $\textbf{k}^{2}$ terms in the intralayer Hamiltonian are also neglected. 
The red dashed lines in Fig.~(\ref{fig4})(a) denote the energy bands of magic-angle TBG ($\theta=1.05^{\circ}$) without lattice relaxation effects. Although two flat bands (per spin per valley) with bandwidth $\sim 15\,$meV are obtained, we note that there is no band gap between the flat bands and the remote bands for magic-angle TBG with an ideal moir\'e superlattice. Including the lattice relaxation effects would open up a gap $\sim$20\,meV between the flat bands and the remote energy bands, as shown by the blue lines in Fig.~(\ref{fig4})(a).

In Fig.~\ref{fig4}(b), we compare the band structures calculated using both the effective continuum model and the atomistic tight binding model. The red lines represent the band structure calculated by the tight binding model (Eq.~(\ref{eq:sk-hopping})) with the fully relaxed lattice structure; while the blue lines denote the energy band structures calculated by the effective continuum model including both $\textbf{k}$-dependent interlayer hopping terms and the $\textbf{k}^{2}$ terms in the intralayer hopping terms. These two terms would strongly break particle-hole symmetry in the electronic band structures, consistent with the results reported in literatures \cite{koshino-ep-prb20,fang-tbg-arxiv19,SongZhida-tbg2-prb21,kv-prb23}. The band structure calculated by the continuum model is in perfect agreement with that calculated by the tight binding approach as shown in Fig.~\ref{fig4}(b).

\section{Summary and Outlook}

To summarize, we have employed a continuum elastic model to describe the lattice dynamics of twisted bilayer graphene. Based on this model, we have  calculated the lattice distortions,  phonon properties, and  effective electronic band structures with the fully relaxed lattice structure for TBG. 
We have introduced a  work flow for the structural relaxation calculations of TBG, in which both the in-plane and out-of-plane lattice distortions are taken into account and are treated on equal footing. The calculated lattice distortions show pleasant agreement with the experimental findings. Moreover, We have studied the phonon properties in magic-angle TBG based on the elastic model. Our results are in good agreement with those obtained from the DPMD method, despite  a much lower computational cost. 
These results indicate that our continuum elastic model is reliable, accurate, and computational efficient.  
Finally, we evaluate the influence of the lattice distortions on the electronic band structures for magic-angle TBG. We find that the lattice distortions would  open a gap between the flat bands and remote bands, and strongly break particle-hole symmetry of the flat band dispersions. Moreover, the electronic band structure calculated using the continuum model is in perfect agreement with that calculated using the tight binding approach.

Previous studies already reveal that  
the coupling strengths between the flat band electrons and moir\'e phonons are considerable in magic-angle TBG,  which may give rise to fruitful physics including charge order \cite{angeli-tbg-prx19,liu-phonon-nano22}, linear in temperature resistivity \cite{wu-linear-prb19,sharma-tbg-phonon-nc21}, and superconductivity \cite{lian-prl18,wu-prl18,choi-tbg-prl21,liu-bernevig-Kphonon-arxiv23}. In our framework, it is straightforward to evaluate the electron-phonon coupling effects by introducing the phonon excitations into the effective continuum model. More importantly, in our framework both the in-plane and out-of-plane moir\'e phonon modes are taken into account in an unbiased manner. 
On the one hand, It is expected that the relative in-plane and out-of-plane phonon modes would be strongly coupled to the flat-band electrons in magic-angle TBG via the interlayer hopping terms.
On the other hand, although all the moir\'e phonon modes reported in this paper  would be coupled with electrons through the intralayer  Hamiltonian,  they are of higher order compared to the intervalley, intralayer electron-phonon couplings mediated by the optical phonons at atomic $\mathbf{K}$/$\mathbf{K'}$ points \cite{liu-bernevig-Kphonon-arxiv23}.
However, given that there are plenty of soft optical moir\'e phonons with frequencies $\sim 0.01\textrm{-}0.1\,$THz, the relatively weak intravalley, intralayer electron-phonon couplings may still have significant contributions to the phonon self energies. 
Our work paves the way for further comprehensive studies of electron-phonon coupling effects and their interplay with $e$-$e$ interactions in magic-angle TBG.

\acknowledgements

We thank Yanran Shi, Zhida Song,  Jian Kang, and D. Kwabena Bediako for valuable discussions. This work is supported by the National Natural Science Foundation of China (grant No. 12174257), the National Key R \& D program of China (grant No. 2020YFA0309601),  and the start-up grant of ShanghaiTech University. 

\begin{widetext}
\clearpage

\begin{center}
\textbf{\large Appendix}
\end{center}

\vspace{12pt}
\begin{center}
\textbf{\large \I. Separation of variables in the binding energy}
\end{center}

In order to validate the separation-of-variables treatment to the binding energy as expressed in Eq.~(\ref{eq:bind}), we explicitly calculate the second order expansion coefficients of the interlayer binding energy  with respect to the interlayer distance $(h^{-}(\textbf{r})-h^{-}{0}(\textbf{r}))$ (see Eq.~(\ref{eq:bindener})) based on first principles density functional theory. In our model, this expansion coefficient is taken to be 36, adopted from Lennard-Jones potential. We would like to inspect how this coefficient varies with respect to in-plane interlayer shift vector $\bm{\delta}_{\parallel}$.
More specifically, with  fixed $\bm{\delta}_{\parallel}$ between two graphene monolayers(without twist angle), we vary the interlayer distance near the equilibrium value. Then, we fit the binding energy to a polynomial function of the interlayer distance for each $\bm{\delta}_{\parallel}$, and determine the expansion coefficients of the second order term. Fig.~\ref{sif1} presents the second-order expansion coefficient as a function of the in-plane shift vector, which mimics the distribution of the second-order expansion coefficients at different locations within the moir\'e supercell. The expansion coefficient has a maximal value $\sim 50$ near the $AA$ point, while remains as a constant $\sim$32 in the other regions. The average value of the second order coefficient is estimated to be about 34, which is very close to 36 used in our model. Certainly, it would be more accurate if one could also take into account the real-space variation of the second-order coefficients, but this would lead to an unmanageable number of terms when taking derivatives of the elastic energy in the Euler-Lagrange equations. Hence, we choose to use a constant value of 36 (adopted from Lennard-Jones potential) for the second-order expansion coefficient in our model.

\begin{figure}[htb!]
\begin{center}
    \includegraphics[width=8cm]{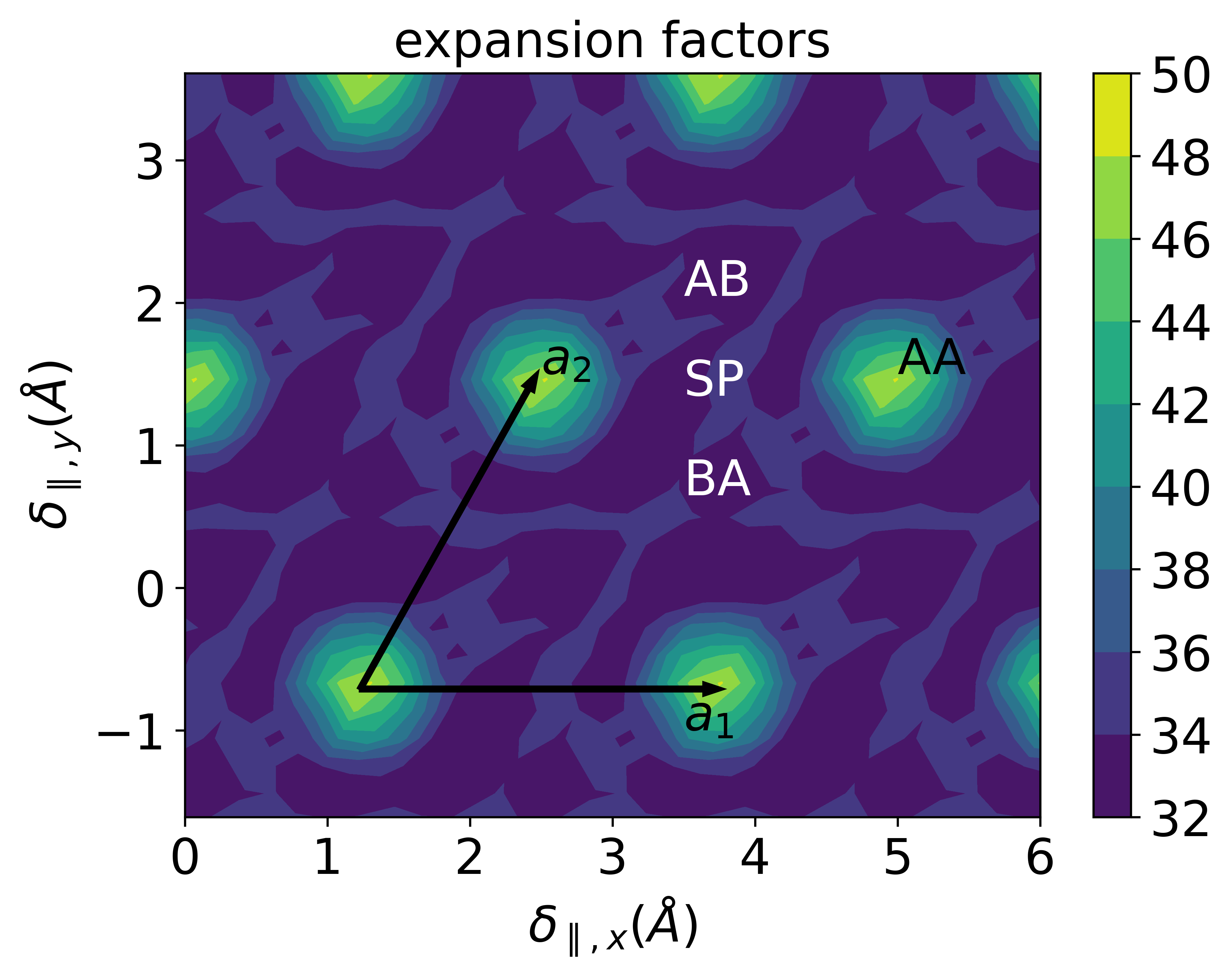}
\caption{The second order expansion coefficients of $(h^{-}(\textbf{r})-h^{-}_{0}(\textbf{r}))$ in Eq.~(\ref{eq:bindener}) as a function of the in-plane displacement vector $\bm{\delta}_{\parallel}$.}
\label{sif1}
\end{center}
\end{figure}

\vspace{12pt}
\begin{center}
\textbf{\large \II. Details in lattice relaxation calculations}
\end{center}

In the present model, the total energy $U\!=\!U_{E}+U_{B}$ is a functional of lattice distortions. We can minimize the total energy with respect to the lattice distortions by solving the Euler-Lagrange equation:
\begin{equation}
\begin{split}\label{eq:eulerlagrange}
&\frac{\delta U}{\delta f}-\sum_{\alpha}\frac{\partial}{\partial \alpha}\frac{\delta U}{\delta f_{\alpha}}+\sum_{\alpha,\beta}\frac{\partial^{2}}{\partial \alpha\partial \beta}\frac{\delta U}{\delta f_{\alpha,\beta}}=0,
\end{split}
\end{equation}
with 
\begin{align}\label{eq:partial}
&f_{\alpha}=\frac{\partial f}{\partial \alpha}\;,\nonumber\\ 
&f_{\alpha,\beta}=\frac{\partial^{2} f}{\partial \alpha\partial\beta}\;,
\end{align}
where $\alpha,\beta\!=\!x,y$ and $f\!=\!u^{\pm}_{x},u^{\pm}_{y},h^{\pm}$. The interlayer binding energy is given by Eq.~\ref{eq:bindener}. We consider the impact of stretches and curvatures into account in our model. Then, the elastic energy is given by:
\allowdisplaybreaks
\begin{align}
U_{E}=&\sum^{2}_{l=1} \frac{1}{2} \left\{ \left( \lambda+\mu \right)\left( u^{(l)}_{xx}+u^{(l)}_{yy} \right)^{2} + \mu\left[ \left( u^{(l)}_{xx}-u^{(l)}_{yy} \right)^{2} + 4(u^{(l)}_{xy})^{2} \right] +\kappa\left[ (\frac{\partial^{2}}{\partial^{2}x}+\frac{\partial^{2}}{\partial^{2}y}) h^{(l)} \right]^{2} \right\}.\nonumber\\
=&\sum_{l}\int d^{2}\textbf{r}\frac{1}{2}\left\{\kappa\left[\left(\frac{\partial^{2}}{\partial^{2}x}+\frac{\partial^{2}}{\partial^{2}y} \right)h^{(l)}\right]^{2} \right. + \nonumber\\
&(\lambda+\mu)\left[ \left(\frac{\partial u^{(l)}_{x}}{\partial x}\right)^{2}+\left(\frac{\partial u^{(l)}_{y}}{\partial y}\right)^{2}+2\frac{\partial u^{(l)}_{x}}{\partial x}\frac{\partial u^{(l)}_{y}}{\partial y}+\frac{\partial u^{(l)}_{x}}{\partial x}\left(\frac{\partial h^{(l)}}{\partial x}\right)^{2}+\frac{\partial u^{(l)}_{y}}{\partial y}\left(\frac{\partial h^{(l)}}{\partial y}\right)^{2} \right. \nonumber\\
&\left.+\frac{\partial u^{(l)}_{y}}{\partial y}\left(\frac{\partial h^{(l)}}{\partial x}\right)^{2}+\frac{\partial u^{(l)}_{x}}{\partial x}\left(\frac{\partial h^{(l)}}{\partial y}\right)^{2}+\frac{1}{4}\left(\frac{\partial h^{(l)}}{\partial x}\right)^{4}+\frac{1}{4}\left(\frac{\partial h^{(l)}}{\partial y}\right)^{4}+\frac{1}{2}\left(\frac{\partial h^{(l)}}{\partial x}\right)^{2}\left(\frac{\partial h^{(l)}}{\partial y}\right)^{2}\right]\nonumber\\
&+(\mu)\left[ \left(\frac{\partial u^{(l)}_{x}}{\partial x}\right)^{2}+\left(\frac{\partial u^{(l)}_{y}}{\partial y}\right)^{2}-2\frac{\partial u^{(l)}_{x}}{\partial x}\frac{\partial u^{(l)}_{y}}{\partial y}+\frac{\partial u^{(l)}_{x}}{\partial x}\left(\frac{\partial h^{(l)}}{\partial x}\right)^{2}+\frac{\partial u^{(l)}_{y}}{\partial y}\left(\frac{\partial h^{(l)}}{\partial y}\right)^{2} \right.\nonumber\\
&-\frac{\partial u^{(l)}_{y}}{\partial y}\left(\frac{\partial h^{(l)}}{\partial x}\right)^{2}-\frac{\partial u^{(l)}_{x}}{\partial x}\left(\frac{\partial h^{(l)}}{\partial y}\right)^{2}+\frac{1}{4}\left(\frac{\partial h^{(l)}}{\partial x}\right)^{4}+\frac{1}{4}\left(\frac{\partial h^{(l)}}{\partial y}\right)^{4}-\frac{1}{2}\left(\frac{\partial h^{(l)}}{\partial x}\right)^{2}\left(\frac{\partial h^{(l)}}{\partial y}\right)^{2}\nonumber\\
&\left.\left. +\left(\frac{\partial u^{(l)}_{x}}{\partial y}+\frac{\partial u^{(l)}_{y}}{\partial x}\right)^{2}+2\left(\frac{\partial u^{(l)}_{x}}{\partial y}+\frac{\partial u^{(l)}_{y}}{\partial x}\right)\frac{\partial h^{(l)}}{\partial x}\frac{\partial h^{(l)}}{\partial y}+\left(\frac{\partial h^{(l)}}{\partial x}\right)^{2}\left(\frac{\partial h^{(l)}}{\partial y}\right)^{2} \right]\right\},
\end{align}
where, $\lambda\approx 3.25\,$eV/$\angstrom^{2}$ and $\mu\approx 9.57\,$eV/$\angstrom^{2}$ are the Lam\'e factors, $\kappa\!=\!1.6\,$eV is the curvature modulus \cite{jung2015origin}.

Before we construct the Euler-Lagrange equation, we evaluate the partial derivatives given in Eq.~(\ref{eq:partial}).
\allowdisplaybreaks
\begin{align}\label{eq:partialde}
\frac{\partial V}{\partial \frac{\partial u^{-}_{x}}{\partial x}}=&\sum_{l}\frac{\partial V}{\partial \frac{\partial u^{(l)}_{x}}{\partial x}}\frac{\partial \frac{\partial u^{(l)}_{x}}{\partial x}}{\partial \frac{\partial u^{-}_{x}}{\partial x}}\nonumber\\
=&\frac{1}{2}\left\{ (\lambda+\mu)\left[2\frac{\partial u^{(2)}_{x}}{\partial x}+2\frac{\partial u^{(2)}_{y}}{\partial y}+\left(\frac{h^{(2)}}{\partial x}\right)^{2}+\left(\frac{h^{(2)}}{\partial y}\right)^{2}\right] \right.\nonumber\\
&\left.+\mu\left[2\frac{\partial u^{(2)}_{x}}{\partial x}-2\frac{\partial u^{(2)}_{y}}{\partial y}+\left(\frac{h^{(2)}}{\partial x}\right)^{2}-\left(\frac{h^{(2)}}{\partial y}\right)^{2}\right]\right\}\times\frac{1}{2}\nonumber\\
&+ \frac{1}{2} \left\{ \dots (2)\leftrightarrow(1)\dots \right\}\times\frac{-1}{2}\nonumber\\
&=\frac{1}{2}\left\{(\lambda+\mu)\left[ \frac{\partial u^{-}_{x}}{\partial x}+\frac{\partial u^{-}_{y}}{\partial y}+\frac{1}{2}\frac{\partial h^{+}}{\partial x}\frac{\partial h^{-}}{\partial x}+\frac{1}{2}\frac{\partial h^{+}}{\partial y}\frac{\partial h^{-}}{\partial y} \right]\right.\nonumber\\
&\left.+\mu\left[ \frac{\partial u^{-}_{x}}{\partial x}-\frac{\partial u^{-}_{y}}{\partial y}+\frac{1}{2}\frac{\partial h^{+}}{\partial x}\frac{\partial h^{-}}{\partial x}-\frac{1}{2}\frac{\partial h^{+}}{\partial y}\frac{\partial h^{-}}{\partial y} \right]\right\},\nonumber\\
\frac{\partial V}{\partial \frac{\partial u^{-}_{y}}{\partial y}}=&\frac{1}{2}\left\{(\lambda+\mu)\left[ \frac{\partial u^{-}_{x}}{\partial x}+\frac{\partial u^{-}_{y}}{\partial y}+\frac{1}{2}\frac{\partial h^{+}}{\partial x}\frac{\partial h^{-}}{\partial x}+\frac{1}{2}\frac{\partial h^{+}}{\partial y}\frac{\partial h^{-}}{\partial y} \right]\right.\nonumber\\
&\left.+\mu\left[- \frac{\partial u^{-}_{x}}{\partial x}+\frac{\partial u^{-}_{y}}{\partial y}-\frac{1}{2}\frac{\partial h^{+}}{\partial x}\frac{\partial h^{-}}{\partial x}+\frac{1}{2}\frac{\partial h^{+}}{\partial y}\frac{\partial h^{-}}{\partial y} \right]\right\},\nonumber\\
\frac{\partial V}{\partial \frac{\partial u^{-}_{x}}{\partial y}}=&\frac{\partial V}{\partial \frac{\partial u^{-}_{y}}{\partial x}}=\frac{1}{2}\left\{ \mu\left[ \frac{\partial u^{-}_{x}}{\partial y}+\frac{\partial u^{-}_{y}}{\partial x}+\frac{1}{2}\frac{\partial h^{+}}{\partial x}\frac{\partial h^{-}}{\partial y}+\frac{1}{2}\frac{\partial h^{-}}{\partial x}\frac{\partial h^{+}}{\partial y} \right] \right\}\nonumber\\
\frac{\partial V}{\partial \frac{\partial u^{+}_{x}}{\partial x}}=&\frac{1}{2}\left\{(\lambda+\mu)\left[ \frac{\partial u^{+}_{x}}{\partial x}+\frac{\partial u^{+}_{y}}{\partial y}+\frac{1}{4}\left(\frac{\partial h^{+}}{\partial x}\right)^{2}+\frac{1}{4}\left(\frac{\partial h^{-}}{\partial x}\right)^{2}+\frac{1}{4}\left(\frac{\partial h^{+}}{\partial y}\right)^{2}+\frac{1}{4}\left(\frac{\partial h^{-}}{\partial y}\right)^{2}\right]\right.\nonumber\\
&\left.+\mu\left[\frac{\partial u^{+}_{x}}{\partial x}-\frac{\partial u^{+}_{y}}{\partial y}+\frac{1}{4}\left(\frac{\partial h^{+}}{\partial x}\right)^{2}+\frac{1}{4}\left(\frac{\partial h^{-}}{\partial x}\right)^{2}-\frac{1}{4}\left(\frac{\partial h^{+}}{\partial y}\right)^{2}-\frac{1}{4}\left(\frac{\partial h^{-}}{\partial y}\right)^{2}\right]\right\},\nonumber\\
\frac{\partial V}{\partial \frac{\partial u^{+}_{y}}{\partial y}}=&\frac{1}{2}\left\{(\lambda+\mu)\left[ \frac{\partial u^{+}_{x}}{\partial x}+\frac{\partial u^{+}_{y}}{\partial y}+\frac{1}{4}\left(\frac{\partial h^{+}}{\partial x}\right)^{2}+\frac{1}{4}\left(\frac{\partial h^{-}}{\partial x}\right)^{2}+\frac{1}{4}\left(\frac{\partial h^{+}}{\partial y}\right)^{2}+\frac{1}{4}\left(\frac{\partial h^{-}}{\partial y}\right)^{2}\right]\right.\nonumber\\
&\left.+\mu\left[-\frac{\partial u^{+}_{x}}{\partial x}+\frac{\partial u^{+}_{y}}{\partial y}-\frac{1}{4}\left(\frac{\partial h^{+}}{\partial x}\right)^{2}-\frac{1}{4}\left(\frac{\partial h^{-}}{\partial x}\right)^{2}+\frac{1}{4}\left(\frac{\partial h^{+}}{\partial y}\right)^{2}+\frac{1}{4}\left(\frac{\partial h^{-}}{\partial y}\right)^{2}\right]\right\},\nonumber\\
\frac{\partial V}{\partial \frac{\partial u^{+}_{x}}{\partial y}}=&\frac{\partial V}{\partial \frac{\partial u^{+}_{y}}{\partial x}}=\frac{1}{2}\left\{\mu\left[ \frac{\partial u^{+}_{x}}{\partial y}+\frac{\partial u^{+}_{y}}{\partial x} + \frac{1}{2}\frac{\partial h^{+}}{\partial x}\frac{\partial h^{+}}{\partial y}+\frac{1}{2}\frac{\partial h^{-}}{\partial x}\frac{\partial h^{-}}{\partial y}\right]\right\}\nonumber\\
\frac{\partial V}{\partial \frac{\partial h^{-}}{\partial x}}=&\frac{1}{2}\left\{ (\lambda+\mu)\left[ \frac{1}{2}\left( \frac{\partial h^{+}}{\partial x}\frac{\partial u^{-}_{x}}{\partial x}+\frac{\partial h^{-}}{\partial x}\frac{\partial u^{+}_{x}}{\partial x} \right) +\frac{1}{2}\left( \frac{\partial h^{+}}{\partial x}\frac{\partial u^{-}_{y}}{\partial y}+\frac{\partial h^{-}}{\partial x}\frac{\partial u^{+}_{y}}{\partial y} \right)  \right. \right.\nonumber\\
&\left.+\frac{1}{8}\left(\left( \frac{\partial h^{-}}{\partial x} \right)^{3}+3\left(\frac{\partial h^{+}}{\partial x}\right)^{2}\frac{\partial h^{-}}{\partial x}\right)+\frac{1}{8}\left( \frac{\partial h^{-}}{\partial x}\left(\frac{\partial h^{-}}{\partial y}\right)^{2}+\frac{\partial h^{-}}{\partial x}\left(\frac{\partial h^{+}}{\partial y}\right)^{2}+2\frac{\partial h^{+}}{\partial x}\frac{\partial h^{-}}{\partial y}\frac{\partial h^{+}}{\partial y} \right)\right]\nonumber\\
&+ \mu\left[ \frac{1}{2}\left( \frac{\partial h^{+}}{\partial x}\frac{\partial u^{-}_{x}}{\partial x}+\frac{\partial h^{-}}{\partial x}\frac{\partial u^{+}_{x}}{\partial x} \right) -\frac{1}{2}\left( \frac{\partial h^{+}}{\partial x}\frac{\partial u^{-}_{y}}{\partial y}+\frac{\partial h^{-}}{\partial x}\frac{\partial u^{+}_{y}}{\partial y} \right)  \right.\nonumber\\
&+\frac{1}{8}\left(\left( \frac{\partial h^{-}}{\partial x} \right)^{3}+3\left(\frac{\partial h^{+}}{\partial x}\right)^{2}\frac{\partial h^{-}}{\partial x}\right)-\frac{1}{8}\left( \frac{\partial h^{-}}{\partial x}\left(\frac{\partial h^{-}}{\partial y}\right)^{2}+\frac{\partial h^{-}}{\partial x}\left(\frac{\partial h^{+}}{\partial y}\right)^{2}+2\frac{\partial h^{+}}{\partial x}\frac{\partial h^{-}}{\partial y}\frac{\partial h^{+}}{\partial y} \right)\nonumber\\
&+\frac{1}{2}\left( \frac{\partial u^{+}_{x}}{\partial y}\frac{\partial h^{-}}{\partial y}+\frac{\partial u^{-}_{x}}{\partial y}\frac{\partial h^{+}}{\partial y} \right)+\frac{1}{2}\left( \frac{\partial u^{+}_{y}}{\partial x}\frac{\partial h^{-}}{\partial y}+\frac{\partial u^{-}_{y}}{\partial x}\frac{\partial h^{+}}{\partial y} \right)\nonumber\\
&\left.\left.+\frac{1}{4}\left(\frac{\partial h^{-}}{\partial x}\left(\frac{\partial h^{-}}{\partial y}\right)^{2}+\frac{\partial h^{-}}{\partial x}\left(\frac{\partial h^{+}}{\partial y}\right)^{2}+2\frac{\partial h^{+}}{\partial x}\frac{\partial h^{+}}{\partial y}\frac{\partial h^{-}}{\partial y}\right)\right]\right\},\nonumber\\
\frac{\partial V}{\partial \frac{\partial h^{+}}{\partial x}}=&\frac{1}{2}\left\{ (\lambda+\mu)\left[ \frac{1}{2}\left( \frac{\partial h^{+}}{\partial x}\frac{\partial u^{+}_{x}}{\partial x}+\frac{\partial h^{-}}{\partial x}\frac{\partial u^{-}_{x}}{\partial x} \right) +\frac{1}{2}\left( \frac{\partial h^{+}}{\partial x}\frac{\partial u^{+}_{y}}{\partial y}+\frac{\partial h^{-}}{\partial x}\frac{\partial u^{-}_{y}}{\partial y} \right)  \right. \right.\nonumber\\
&\left.+\frac{1}{8}\left(\left( \frac{\partial h^{+}}{\partial x} \right)^{3}+3\left(\frac{\partial h^{-}}{\partial x}\right)^{2}\frac{\partial h^{+}}{\partial x}\right)+\frac{1}{8}\left( \frac{\partial h^{+}}{\partial x}\left(\frac{\partial h^{+}}{\partial y}\right)^{2}+\frac{\partial h^{+}}{\partial x}\left(\frac{\partial h^{-}}{\partial y}\right)^{2}+2\frac{\partial h^{-}}{\partial x}\frac{\partial h^{-}}{\partial y}\frac{\partial h^{+}}{\partial y} \right)\right]\nonumber\\
&+ \mu\left[ \frac{1}{2}\left( \frac{\partial h^{+}}{\partial x}\frac{\partial u^{+}_{x}}{\partial x}+\frac{\partial h^{-}}{\partial x}\frac{\partial u^{-}_{x}}{\partial x} \right) -\frac{1}{2}\left( \frac{\partial h^{+}}{\partial x}\frac{\partial u^{+}_{y}}{\partial y}+\frac{\partial h^{-}}{\partial x}\frac{\partial u^{-}_{y}}{\partial y} \right)  \right.\nonumber\\
&+\frac{1}{8}\left(\left( \frac{\partial h^{+}}{\partial x} \right)^{3}+3\left(\frac{\partial h^{-}}{\partial x}\right)^{2}\frac{\partial h^{+}}{\partial x}\right)-\frac{1}{8}\left( \frac{\partial h^{+}}{\partial x}\left(\frac{\partial h^{+}}{\partial y}\right)^{2}+\frac{\partial h^{+}}{\partial x}\left(\frac{\partial h^{-}}{\partial y}\right)^{2}+2\frac{\partial h^{-}}{\partial x}\frac{\partial h^{-}}{\partial y}\frac{\partial h^{+}}{\partial y} \right)\nonumber\\
&+\frac{1}{2}\left( \frac{\partial u^{+}_{x}}{\partial y}\frac{\partial h^{+}}{\partial y}+\frac{\partial u^{-}_{x}}{\partial y}\frac{\partial h^{-}}{\partial y} \right)+\frac{1}{2}\left( \frac{\partial u^{+}_{y}}{\partial x}\frac{\partial h^{+}}{\partial y}+\frac{\partial u^{-}_{y}}{\partial x}\frac{\partial h^{-}}{\partial y} \right)\nonumber\\
&\left.\left.+\frac{1}{4}\left(\frac{\partial h^{+}}{\partial x}\left(\frac{\partial h^{+}}{\partial y}\right)^{2}+\frac{\partial h^{+}}{\partial x}\left(\frac{\partial h^{-}}{\partial y}\right)^{2}+2\frac{\partial h^{-}}{\partial x}\frac{\partial h^{-}}{\partial y}\frac{\partial h^{+}}{\partial y}\right)\right]\right\},\nonumber\\
\frac{\partial V}{\partial \frac{\partial h^{\pm}}{\partial y}}=&(\dots x\leftrightarrow y\dots),\nonumber\\
\frac{\partial V}{\partial \frac{\partial^{2} h^{\pm}}{\partial^{2} x}}=&\frac{\partial V}{\partial \frac{\partial^{2} h^{\pm}}{\partial^{2} y}}=\frac{\kappa}{2}\left(\frac{\partial^{2}}{\partial^{2}x}+\frac{\partial^{2}}{\partial^{2}y}\right)h^{\pm}
\end{align}
We also assume that the center-of-mass component of out-of-plane distortion vanishes in the relaxed structure, i.e. $h^{+}(\textbf{r})=0$. This is an excellent approximation since nonzero $h^{+}(\textbf{r})$ means that there are center-of-mass ripples in the TBG system, which typically occurs as  thermal excitation effects and/or strain effects. At zero temperature and in the absence of external strain, it is legitimate to set $h^{+}(\textbf{r})=0$.

Then, we can construct the Euler-Lagrange Equations. We start with functions of the in-plane relative distortion $\textbf{u}^{-}$:
\begin{align}\label{eq:elum}
&\frac{\partial V}{\partial u^{-}_{x}}-\frac{\partial }{\partial x}\frac{\partial V}{\partial \frac{\partial u^{-}_{x}}{\partial x}}-\frac{\partial }{\partial y}\frac{\partial V}{\partial \frac{\partial u^{-}_{x}}{\partial y}}=0,\nonumber\\
&\frac{\partial V}{\partial u^{-}_{y}}-\frac{\partial }{\partial x}\frac{\partial V}{\partial \frac{\partial u^{-}_{y}}{\partial x}}-\frac{\partial }{\partial y}\frac{\partial V}{\partial \frac{\partial u^{-}_{y}}{\partial y}}=0,
\end{align}
where
\begin{equation}
\begin{split}
&\frac{\partial V}{\partial u^{-}_{x}}=\sum_{m}e^{i(\textbf{G}_{m}\cdot\textbf{r}+\textbf{a}^{*}_{m}\cdot\textbf{u}^{-})}i\textbf{a}^{*}_{m,x}\left\{\epsilon_{\textbf{a}^{*}_{m}}\left[-1+36\left(\frac{h^{-}-h^{-}_{0}}{h^{-}_{0}}\right)^{2}\right]+\epsilon(\textbf{r})72\frac{h^{-}-h^{-}_{0}}{h^{-}_{0}}(-1)\frac{h^{-}}{(h^{-}_{0})^{2}}h^{-}_{0,\textbf{a}^{*}_{m}}\right\}\\
-&\frac{\partial }{\partial x}\frac{\partial V}{\partial \frac{\partial u^{-}_{x}}{\partial x}}=-\frac{1}{2}\left\{ (\lambda+\mu)\left[ \frac{\partial^{2}u^{-}_{x}}{\partial^{2}x}+\frac{\partial^{2}u^{-}_{y}}{\partial x \partial y} \right]+ \mu\left[  \frac{\partial^{2}u^{-}_{x}}{\partial^{2}x}-\frac{\partial^{2}u^{-}_{y}}{\partial x \partial y} \right] \right\},\\
-&\frac{\partial }{\partial y}\frac{\partial V}{\partial \frac{\partial u^{-}_{x}}{\partial y}}=-\frac{1}{2}\left\{\mu\left[\frac{\partial^{2}u^{-}_{x}}{\partial^{2}y}+\frac{\partial^{2}u^{-}_{y}}{\partial x \partial y} \right]\right\},\\
&\frac{\partial V}{\partial u^{-}_{y}}=\sum_{m}e^{i(\textbf{G}_{m}\cdot\textbf{r}+\textbf{a}^{*}_{m}\cdot\textbf{u}^{-})}i\textbf{a}^{*}_{m,y}\left\{\epsilon_{\textbf{a}^{*}_{m}}\left[-1+36\left(\frac{h^{-}-h^{-}_{0}}{h^{-}_{0}}\right)^{2}\right]+\epsilon(\textbf{r})72\frac{h^{-}-h^{-}_{0}}{h^{-}_{0}}(-1)\frac{h^{-}}{(h^{-}_{0})^{2}}h^{-}_{0,\textbf{a}^{*}_{m}}\right\}\\
-&\frac{\partial }{\partial y}\frac{\partial V}{\partial \frac{\partial u^{-}_{y}}{\partial y}}=-\frac{1}{2}\left\{ (\lambda+\mu)\left[ \frac{\partial^{2}u^{-}_{y}}{\partial^{2}y}+\frac{\partial^{2}u^{-}_{x}}{\partial y \partial x} \right]+ \mu\left[  \frac{\partial^{2}u^{-}_{y}}{\partial^{2}y}-\frac{\partial^{2}u^{-}_{x}}{\partial y \partial x} \right] \right\},\\
-&\frac{\partial }{\partial x}\frac{\partial V}{\partial \frac{\partial u^{-}_{y}}{\partial x}}=-\frac{1}{2}\left\{\mu\left[\frac{\partial^{2}u^{-}_{y}}{\partial^{2}x}+\frac{\partial^{2}u^{-}_{x}}{\partial y \partial x} \right]\right\}.
\end{split}
\end{equation}
Then, we introduce the following Fourier transformation:
\begin{equation}
\begin{split}
&\textbf{u}^{-}(\textbf{r})=\sum_{m}\textbf{u}^{-}_{\textbf{G}_{m}}e^{i\textbf{G}_{m}\cdot\textbf{r}}.\\
&\sum_{m'}e^{i(\textbf{G}_{m}\cdot\textbf{r}+\textbf{a}^{*}_{m'}\cdot\textbf{u}^{-})}i\textbf{a}^{*}_{m',\alpha}\left\{\epsilon_{\textbf{a}^{*}_{m'}}\left[-1+36\left(\frac{h^{-}-h^{-}_{0}}{h^{-}_{0}}\right)^{2}\right]+\epsilon(\textbf{r})72\frac{h^{-}-h^{-}_{0}}{h^{-}_{0}}(-1)\frac{h^{-}}{(h^{-}_{0})^{2}}h^{-}_{0,\textbf{a}^{*}_{m'}}\right\}=\sum_{m}F_{\textbf{G}_{m},\alpha}e^{i\textbf{G}_{m}\cdot\textbf{r}}, \alpha=x, y,
\end{split}
\end{equation}
where $\textbf{a}^{*}_{m}=m_{1}\textbf{a}^{*}_{1}+m_{2}\textbf{a}^{*}_{2}$ is the reciprocal lattice vectors of monolayer graphene and $\textbf{G}_{m}=m_{1}\textbf{G}^{M}_{1}+m_{2}\textbf{G}^{M}_{2}$ is the moir\'e reciprocal lattice vector. We can write Eq.~(\ref{eq:elum}) into a matrix form as follows:
\begin{align}
\begin{split}
&\left[\begin{array}{cc}
				(\lambda+2\mu)G^{2}_{m,x}+\mu G^{2}_{m,y} & (\lambda+\mu)G_{m,x}G_{m,y} \\
				(\lambda+\mu)G_{m,x}G_{m,y} & (\lambda+2\mu)G^{2}_{m,y}+\mu G^{2}_{m,x}
\end{array}\right]\left[\begin{array}{c}
                     u^{-}_{\textbf{G}_{m},x}\\
                     u^{-}_{\textbf{G}_{m},y}
\end{array}\right]=-2\left[\begin{array}{c}
                     F_{\textbf{G}_{m},x}\\
                     F_{\textbf{G}_{m},y}
\end{array}\right]
\end{split}
\end{align}
It is important to note that $F_{\textbf{G}_{m},\alpha}$ is a function of both $\textbf{u}^{-}$ and $h^{-}$. For fixed $\{h^{-}(\textbf{r})\}$, $\textbf{u}^{-}$ can be solved iteratively.

The Euler-Lagrange equations of center-of-mass in-plane distortion $\textbf{u}^{+}$ are given by:
\begin{equation}\label{eq:elup}
\begin{split}
&\frac{\partial V}{\partial u^{+}_{x}}-\frac{\partial }{\partial x}\frac{\partial V}{\partial \frac{\partial u^{+}_{x}}{\partial x}}-\frac{\partial }{\partial y}\frac{\partial V}{\partial \frac{\partial u^{+}_{x}}{\partial y}}=0,\\
&\frac{\partial V}{\partial u^{+}_{y}}-\frac{\partial }{\partial x}\frac{\partial V}{\partial \frac{\partial u^{+}_{y}}{\partial x}}-\frac{\partial }{\partial y}\frac{\partial V}{\partial \frac{\partial u^{+}_{y}}{\partial y}}=0,
\end{split}
\end{equation}
where
\begin{equation}
\begin{split}
&\frac{\partial V}{\partial u^{+}_{x}}=0,\\
-&\frac{\partial }{\partial x}\frac{\partial V}{\partial \frac{\partial u^{+}_{x}}{\partial x}}=-\frac{1}{2}\left\{ (\lambda+\mu)\left[ \frac{\partial^{2}u^{+}_{x}}{\partial^{2}x}+ \frac{\partial^{2}u^{+}_{y}}{\partial x\partial y}+\frac{1}{2}\frac{\partial h^{-}}{\partial x}\frac{\partial^{2}h^{-}}{\partial^{2}x}+\frac{1}{2}\frac{\partial h^{-}}{\partial y}\frac{\partial^{2}h^{-}}{\partial x\partial y} \right] \right.\\
&\ \ \ \ \ \ \ \ \ \ \ \ \ \ \ \ \ \ \ \ \ \ \left.+\mu\left[\frac{\partial^{2}u^{+}_{x}}{\partial^{2}x}- \frac{\partial^{2}u^{+}_{y}}{\partial x\partial y}+\frac{1}{2}\frac{\partial h^{-}}{\partial x}\frac{\partial^{2}h^{-}}{\partial^{2}x}-\frac{1}{2}\frac{\partial h^{-}}{\partial y}\frac{\partial^{2}h^{-}}{\partial x\partial y}\right]\right\},\\
-&\frac{\partial }{\partial y}\frac{\partial V}{\partial \frac{\partial u^{+}_{x}}{\partial y}}=-\frac{1}{2}\left\{\mu\left[\frac{\partial^{2}u^{-}_{x}}{\partial^{2}y}+\frac{\partial^{2}u^{-}_{y}}{\partial x \partial y}+\frac{1}{2}\frac{\partial^{2}h^{-}}{\partial x\partial y}\frac{\partial h^{-}}{\partial y}+\frac{1}{2}\frac{\partial h^{-}}{\partial x}\frac{\partial^{2}h^{-}}{\partial^{2}y} \right]\right\}.\\
&\frac{\partial V}{\partial u^{+}_{y}}=0,\\
-&\frac{\partial }{\partial y}\frac{\partial V}{\partial \frac{\partial u^{+}_{y}}{\partial y}}=-\frac{1}{2}\left\{ (\lambda+\mu)\left[ \frac{\partial^{2}u^{+}_{y}}{\partial^{2}y}+ \frac{\partial^{2}u^{+}_{x}}{\partial y\partial x}+\frac{1}{2}\frac{\partial h^{-}}{\partial y}\frac{\partial^{2}h^{-}}{\partial^{2}y}+\frac{1}{2}\frac{\partial h^{-}}{\partial x}\frac{\partial^{2}h^{-}}{\partial y\partial x} \right] \right.\\
&\ \ \ \ \ \ \ \ \ \ \ \ \ \ \ \ \ \ \ \ \ \ \left.+\mu\left[\frac{\partial^{2}u^{+}_{y}}{\partial^{2}y}- \frac{\partial^{2}u^{+}_{x}}{\partial y\partial x}+\frac{1}{2}\frac{\partial h^{-}}{\partial y}\frac{\partial^{2}h^{-}}{\partial^{2}y}-\frac{1}{2}\frac{\partial h^{-}}{\partial x}\frac{\partial^{2}h^{-}}{\partial y\partial x}\right]\right\},\\
-&\frac{\partial }{\partial x}\frac{\partial V}{\partial \frac{\partial u^{+}_{y}}{\partial x}}=-\frac{1}{2}\left\{\mu\left[\frac{\partial^{2}u^{-}_{y}}{\partial^{2}x}+\frac{\partial^{2}u^{-}_{x}}{\partial y \partial x}+\frac{1}{2}\frac{\partial^{2}h^{-}}{\partial y\partial x}\frac{\partial h^{-}}{\partial x}+\frac{1}{2}\frac{\partial h^{-}}{\partial y}\frac{\partial^{2}h^{-}}{\partial^{2}x} \right]\right\}.
\end{split}
\end{equation}
We introduce the Fourier transformation for the lattice distortion:
\begin{equation}
\textbf{u}^{+}(\textbf{r})=\sum_{m}\textbf{u}^{+}_{\textbf{G}_{m}}e^{i\textbf{G}_{m}\cdot\textbf{r}}, h^{-}(\textbf{r})=\sum_{m}h^{-}_{\textbf{G}_{m}}e^{i\textbf{G}_{m}\cdot\textbf{r}}
\end{equation}
Then, the Euler-Lagrange equations of $\textbf{u}^{+}$ in reciprocal space in matrix form is given by: 
\begin{align}\label{eq:a24}
\begin{split}
&\left[\begin{array}{cc}
				(\lambda+2\mu)G^{2}_{m,x}+\mu G^{2}_{m,y} & (\lambda+\mu)G_{m,x}G_{m,y} \\
				(\lambda+\mu)G_{m,x}G_{m,y} & (\lambda+2\mu)G^{2}_{m,y}+\mu G^{2}_{m,x}
\end{array}\right]\left[\begin{array}{c}
                     u^{+}_{\textbf{G}_{m},x}\\
                     u^{+}_{\textbf{G}_{m},y}
\end{array}\right]=\left[\begin{array}{c}
                     M_{x}\\
                     M_{y}
\end{array}\right],
\end{split}
\end{align}
where
\begin{align}
\begin{split}
M_{x}=&\frac{1}{2}\sum_{\textbf{P}_{1},\textbf{P}_{2}}h^{-}_{P_{1}}h^{-}_{P_{2}}\left\{(\lambda+\mu)*i\left[-P_{1,x}P^{2}_{2,x}-P_{1,y}P_{2,x}P_{2,y}\right]\right.+\\
&\left.\mu *i\left[-P_{1,x}P^{2}_{2,x}+P_{1,y}P_{2,x}P_{2,y}-P_{1,x}P_{1,y}P_{2,y}-P_{1,x}P^{2}_{2,y}\right]\right\}\delta_{\textbf{P}_{1}+\textbf{P}_{2},\textbf{G}_{m}},\\
M_{y}=&\frac{1}{2}\sum_{\textbf{P}_{1},\textbf{P}_{2}}h^{-}_{P_{1}}h^{-}_{P_{2}}\left\{(\lambda+\mu)*i\left[-P_{1,y}P^{2}_{2,y}-P_{1,x}P_{2,y}P_{2,x}\right]\right.+\\
&\left.\mu *i\left[-P_{1,y}P^{2}_{2,y}+P_{1,x}P_{2,y}P_{2,x}-P_{1,y}P_{1,x}P_{2,x}-P_{1,y}P^{2}_{2,x}\right]\right\}\delta_{\textbf{P}_{1}+\textbf{P}_{2},\textbf{G}_{m}}.
\end{split}
\end{align}
Here $\textbf{P}_{1}, \textbf{P}_{2}\!=\!m_{1}\textbf{G}^{M}_{1}+m_{2}\textbf{G}^{M}_{2}$ is the moir\'e reciprocal lattice vector. We note that $M_{\alpha}$ is a function of $h^{-}$. As a result, $\textbf{u}^{+}$ can be solved directly for fixed $\{h^{-}(\textbf{r})\}$, 

The Euler-Lagrange equation of the out-of-plane relative distortions is given by: 
\begin{align}
&\frac{\partial V}{\partial h^{-}}-\frac{\partial}{\partial x}\frac{\partial V}{\partial \frac{\partial h^{-}}{\partial x}}-\frac{\partial}{\partial y}\frac{\partial V}{\partial \frac{\partial h^{-}}{\partial y}}+\frac{\partial^{2}}{\partial^{2} x}\frac{\partial V}{\partial \frac{\partial^{2} h^{-}}{\partial^{2} x}}+\frac{\partial^{2}}{\partial^{2} y}\frac{\partial V}{\partial \frac{\partial^{2} h^{-}}{\partial^{2} y}}=0,
\end{align}
where
\allowdisplaybreaks
\begin{align}
&\frac{\partial V}{\partial h^{-}}=\epsilon(\textbf{r})72\frac{h^{-}-h^{-}_{0}}{\left(h^{-}_{0}\right)^{2}},\nonumber\\
-&\frac{\partial}{\partial x}\frac{\partial V}{\partial \frac{\partial h^{-}}{\partial x}}=-\frac{1}{2}\left\{(\lambda+\mu) \left[ \frac{1}{2}\left( \frac{\partial^{2}h^{-}}{\partial^{2}x}\frac{\partial u^{+}_{x}}{\partial x}+\frac{\partial h^{-}}{\partial x}\frac{\partial^{2}u^{+}_{x}}{\partial^{2}x} \right)+\frac{1}{2}\left( \frac{\partial^{2}h^{-}}{\partial^{2}x}\frac{\partial u^{+}_{y}}{\partial y}+\frac{\partial h^{-}}{\partial x}\frac{\partial^{2}u^{+}_{y}}{\partial^{2}y} \right) \right. \right.\nonumber\\
&\ \ \ \ \ \ \ \ \ \ \ \ \ \ \ \ \ \ \ \ \ \ \left.+\frac{1}{8}\left(3\left(\frac{\partial h^{-}}{\partial x}\right)^{2}\frac{\partial^{2}h^{-}}{\partial^{2}x}+\frac{1}{8}\left(\frac{\partial^{2}h^{-}}{\partial^{2}x}\left(\frac{\partial h^{-}}{\partial y}\right)^{2}+2\frac{\partial h^{-}}{\partial x}\frac{\partial h^{-}}{\partial y}\frac{\partial^{2} h^{-}}{\partial x\partial y}\right)\right)\right]\nonumber\\
&\ \ \ \ \ \ \ \ \ \ \ \ \ \ \ \ \ \ \ \ \ +\mu\left[ \frac{1}{2}\left(  \frac{\partial^{2}h^{-}}{\partial^{2}x}\frac{\partial u^{+}_{x}}{\partial x}+\frac{\partial h^{-}}{\partial x}\frac{\partial^{2}u^{+}_{x}}{\partial^{2}x} \right) - \frac{1}{2}\left( \frac{\partial^{2}h^{-}}{\partial^{2}x}\frac{\partial u^{+}_{y}}{\partial y}+\frac{\partial h^{-}}{\partial x}\frac{\partial^{2}u^{+}_{y}}{\partial^{2}y} \right) \right.\nonumber\\
&\ \ \ \ \ \ \ \ \ \ \ \ \ \ \ \ \ \ \ \ \ +\frac{1}{8}\left(3\left(\frac{\partial h^{-}}{\partial x}\right)^{2}\frac{\partial^{2}h^{-}}{\partial^{2}x}-\frac{1}{8}\left(\frac{\partial^{2}h^{-}}{\partial^{2}x}\left(\frac{\partial h^{-}}{\partial y}\right)^{2}+2\frac{\partial h^{-}}{\partial x}\frac{\partial h^{-}}{\partial y}\frac{\partial^{2} h^{-}}{\partial x\partial y}\right)\right)\nonumber\\
&\ \ \ \ \ \ \ \ \ \ \ \ \ \ \ \ \ \ \ \ \ +\frac{1}{2}\left(\frac{\partial^{2}u^{+}_{x}}{\partial x\partial y}\frac{\partial h^{-}}{\partial y}+\frac{\partial u^{+}_{x}}{\partial y}\frac{\partial^{2}h^{-}}{\partial x\partial y}\right)+\frac{1}{2}\left( \frac{\partial^{2}u^{+}_{y}}{\partial^{2}x}\frac{\partial h^{-}}{\partial y}+\frac{\partial u^{+}_{y}}{\partial x}\frac{\partial^{2}h^{-}}{\partial x\partial y } \right)\nonumber\\
&\ \ \ \ \ \ \ \ \ \ \ \ \ \ \ \ \ \ \ \ \ \left.\left.+\frac{1}{4}\left( \frac{\partial^{2}h^{-}}{\partial^{2}x}\left( \frac{\partial h^{-}}{\partial y} \right)^{2}+2\frac{\partial h^{-}}{\partial x}\frac{\partial h^{-}}{\partial y}\frac{\partial^{2} h^{-}}{\partial x\partial y} \right)\right]\right\},\nonumber\\
-&\frac{\partial}{\partial y}\frac{\partial V}{\partial \frac{\partial h^{-}}{\partial y}}=-\frac{1}{2}\left\{(\lambda+\mu) \left[ \frac{1}{2}\left( \frac{\partial^{2}h^{-}}{\partial^{2}y}\frac{\partial u^{+}_{y}}{\partial y}+\frac{\partial h^{-}}{\partial y}\frac{\partial^{2}u^{+}_{y}}{\partial^{2}y} \right)+\frac{1}{2}\left( \frac{\partial^{2}h^{-}}{\partial^{2}y}\frac{\partial u^{+}_{x}}{\partial x}+\frac{\partial h^{-}}{\partial y}\frac{\partial^{2}u^{+}_{x}}{\partial^{2}x} \right) \right. \right.\nonumber\\
&\ \ \ \ \ \ \ \ \ \ \ \ \ \ \ \ \ \ \ \ \ \ \left.+\frac{1}{8}\left(3\left(\frac{\partial h^{-}}{\partial y}\right)^{2}\frac{\partial^{2}h^{-}}{\partial^{2}y}+\frac{1}{8}\left(\frac{\partial^{2}h^{-}}{\partial^{2}y}\left(\frac{\partial h^{-}}{\partial x}\right)^{2}+2\frac{\partial h^{-}}{\partial y}\frac{\partial h^{-}}{\partial x}\frac{\partial^{2} h^{-}}{\partial y\partial x}\right)\right)\right]\nonumber\\
&\ \ \ \ \ \ \ \ \ \ \ \ \ \ \ \ \ \ \ \ \ +\mu\left[ \frac{1}{2}\left(  \frac{\partial^{2}h^{-}}{\partial^{2}y}\frac{\partial u^{+}_{y}}{\partial y}+\frac{\partial h^{-}}{\partial y}\frac{\partial^{2}u^{+}_{y}}{\partial^{2}y} \right) - \frac{1}{2}\left( \frac{\partial^{2}h^{-}}{\partial^{2}y}\frac{\partial u^{+}_{x}}{\partial x}+\frac{\partial h^{-}}{\partial y}\frac{\partial^{2}u^{+}_{x}}{\partial^{2}x} \right) \right.\nonumber\\
&\ \ \ \ \ \ \ \ \ \ \ \ \ \ \ \ \ \ \ \ \ +\frac{1}{8}\left(3\left(\frac{\partial h^{-}}{\partial y}\right)^{2}\frac{\partial^{2}h^{-}}{\partial^{2}y}-\frac{1}{8}\left(\frac{\partial^{2}h^{-}}{\partial^{2}y}\left(\frac{\partial h^{-}}{\partial x}\right)^{2}+2\frac{\partial h^{-}}{\partial y}\frac{\partial h^{-}}{\partial x}\frac{\partial^{2} h^{-}}{\partial y\partial x}\right)\right)\nonumber\\
&\ \ \ \ \ \ \ \ \ \ \ \ \ \ \ \ \ \ \ \ \ +\frac{1}{2}\left(\frac{\partial^{2}u^{+}_{y}}{\partial y\partial x}\frac{\partial h^{-}}{\partial x}+\frac{\partial u^{+}_{y}}{\partial x}\frac{\partial^{2}h^{-}}{\partial y\partial x}\right)+\frac{1}{2}\left( \frac{\partial^{2}u^{+}_{x}}{\partial^{2}y}\frac{\partial h^{-}}{\partial x}+\frac{\partial u^{+}_{x}}{\partial y}\frac{\partial^{2}h^{-}}{\partial y\partial x } \right)\nonumber\\
&\ \ \ \ \ \ \ \ \ \ \ \ \ \ \ \ \ \ \ \ \ \left.\left.+\frac{1}{4}\left( \frac{\partial^{2}h^{-}}{\partial^{2}y}\left( \frac{\partial h^{-}}{\partial x} \right)^{2}+2\frac{\partial h^{-}}{\partial y}\frac{\partial h^{-}}{\partial x}\frac{\partial^{2} h^{-}}{\partial y\partial x} \right)\right]\right\},\nonumber\\
+&\frac{\partial^{2}}{\partial^{2}x}\frac{\partial V}{\partial \frac{\partial^{2}h^{-}}{\partial^{2}x}}=\frac{1}{2}\kappa\left(\frac{\partial^{4}}{\partial^{4}x}+\frac{\partial^{4}}{\partial^{2}x\partial^{2}y}\right)h^{-},\nonumber\\
+&\frac{\partial^{2}}{\partial^{2}y}\frac{\partial V}{\partial \frac{\partial^{2}h^{-}}{\partial^{2}y}}=\frac{1}{2}\kappa\left(\frac{\partial^{4}}{\partial^{2}x\partial^{2}y}+\frac{\partial^{4}}{\partial^{4}y}\right)h^{-}
\end{align}
In order to derive the Euler-Lagrange equations in reciprocal space, we introduce the following Fourier transformations:
\allowdisplaybreaks
\begin{align}
&\textbf{u}^{+}(\textbf{r})=\sum_{m}\textbf{u}^{+}_{\textbf{G}_{m}}e^{i\textbf{G}_{m}\cdot\textbf{r}}, h^{-}(\textbf{r})=\sum_{m}h^{-}_{\textbf{G}_{m}}e^{i\textbf{G}_{m}\cdot\textbf{r}},\nonumber\\
&\frac{\epsilon(\textbf{r})}{(h^{2}_{0})^{2}}=\sum_{m}g_{\textbf{G}_{m}}e^{i\textbf{G}_{m}\cdot\textbf{r}}, \frac{\epsilon(\textbf{r})}{h^{-}_{0}}=\sum_{m}f_{\textbf{G}_{m}}e^{i\textbf{G}_{m}\cdot\textbf{r}}.
\end{align}
Then, we can express the partial derivatives in reciprocal space:
\allowdisplaybreaks
\begin{align}\label{eq5}
\frac{\partial V}{\partial h^{-}}=&72\left[\sum_{m,m'}g_{\textbf{G}_{m'}}h^{-}_{\textbf{G}_{m}}e^{i(\textbf{G}_{m'}+\textbf{G}_{m})\cdot\textbf{r}}-\sum_{m'}f_{\textbf{G}_{m'}}e^{i\textbf{G}_{m'}\cdot\textbf{r}}\right],\nonumber\\
-\frac{\partial}{\partial x}\frac{\partial V}{\partial \frac{\partial h^{-}}{\partial x}}=&-\frac{1}{2}\sum_{m,m'}\left\{ (\lambda+\mu)\frac{i}{2}\left[ h^{-}_{\textbf{G}_{m}}u^{+}_{\textbf{G}_{m'},x}(-G^{2}_{m,x}G_{m',x}-G_{m,x}G^{2}_{m',x})\right.\right. \nonumber\\
&\ \ \ \ \ \ \ \ \ \ \ \ \ \ \ \ \ \ \ \ \ \ +\left.h^{-}_{\textbf{G}_{m}}u^{+}_{\textbf{G}_{m'},y}(-G^{2}_{m,x}G_{m',y}-G_{m,x}G_{m',x}G_{m',y}) \right]\nonumber \\
&\ \ \ \ \ \ \ \ +\mu\frac{i}{2}\left[  h^{-}_{\textbf{G}_{m}}u^{+}_{\textbf{G}_{m'},x}(-G^{2}_{m,x}G_{m',x}-G_{m,x}G^{2}_{m',x})- h^{-}_{\textbf{G}_{m}}u^{+}_{\textbf{G}_{m'},y}(-G^{2}_{m,x}G_{m',y}-G_{m,x}G_{m',x}G_{m',y}) \right.\nonumber\\
&\ \ \ \ \ \ \ \ \ \ \ \ \ \ \ +h^{-}_{\textbf{G}_{m}}u^{+}_{\textbf{G}_{m'},x}(-G_{m,y}G_{m',x}G_{m',y}-G_{m,x}G_{m,y}G_{m',y})\nonumber\\
&\ \ \ \ \ \ \ \ \ \ \ \ \ \ \ \left.\left.+h_{\textbf{G}_{m}}u^{+}_{\textbf{G}_{m'},y}(-G_{m,y}G^{2}_{m',x}-G_{m,x}G_{m,y}G_{m',x})\right]\right\}e^{i(\textbf{G}_{m}+\textbf{G}_{m'})\cdot\textbf{r}},\nonumber\\
-\frac{\partial}{\partial y}\frac{\partial V}{\partial \frac{\partial h^{-}}{\partial y}}=&(\dots x\leftrightarrow y\dots),\nonumber\\
+\frac{\partial^{2}}{\partial^{2}x}\frac{\partial V}{\partial \frac{\partial^{2}h^{-}}{\partial^{2}x}}=&\frac{\kappa}{2}\sum_{m}h^{-}_{\textbf{G}_{m}}(G^{4}_{m,x}+G^{2}_{m,x}G^{2}_{m,y})e^{i\textbf{G}_{m}\cdot\textbf{r}},\nonumber\\
+\frac{\partial^{2}}{\partial^{2}y}\frac{\partial V}{\partial \frac{\partial^{2}h^{-}}{\partial^{2}y}}=&\frac{\kappa}{2}\sum_{m}h^{-}_{\textbf{G}_{m}}(G^{2}_{m,x}G^{2}_{m,y}+G^{4}_{m,y})e^{i\textbf{G}_{m}\cdot\textbf{r}}.
\end{align}
We want to note that  Eq.~(\ref{eq5}) and Eq.~(\ref{eq:elup}) can be solved iteratively for  fixed $\{\textbf{u}^{-}(\textbf{r})\}$. Thus, we can divide the full Euler-Lagrange equations into two sets of equations and solve the coupled equations following the work flow introduced in Sec.~\ref{sec:lattice}.

\vspace{12pt}
\begin{center}
\textbf{\large \III. In-plane center-of-mass distortion of magic-angle TBG}
\end{center}

We present the real-space distribution of the in-plane center-of-mass distortions of magic-angle TBG in Fig.~6. The colorbar represents the amplitudes of local in-plane center-of-mass distortions ($\mathbf{u}^{+}$) of magic-angle TBG. The directions of the distortions fields are depicted by the arrows. The maximal amplitude of the in-plane center-of-mass distortion is about $10^{-4}\angstrom$, which can be neglected.
\begin{figure}[htb!]
\begin{center}
    \includegraphics[width=8cm]{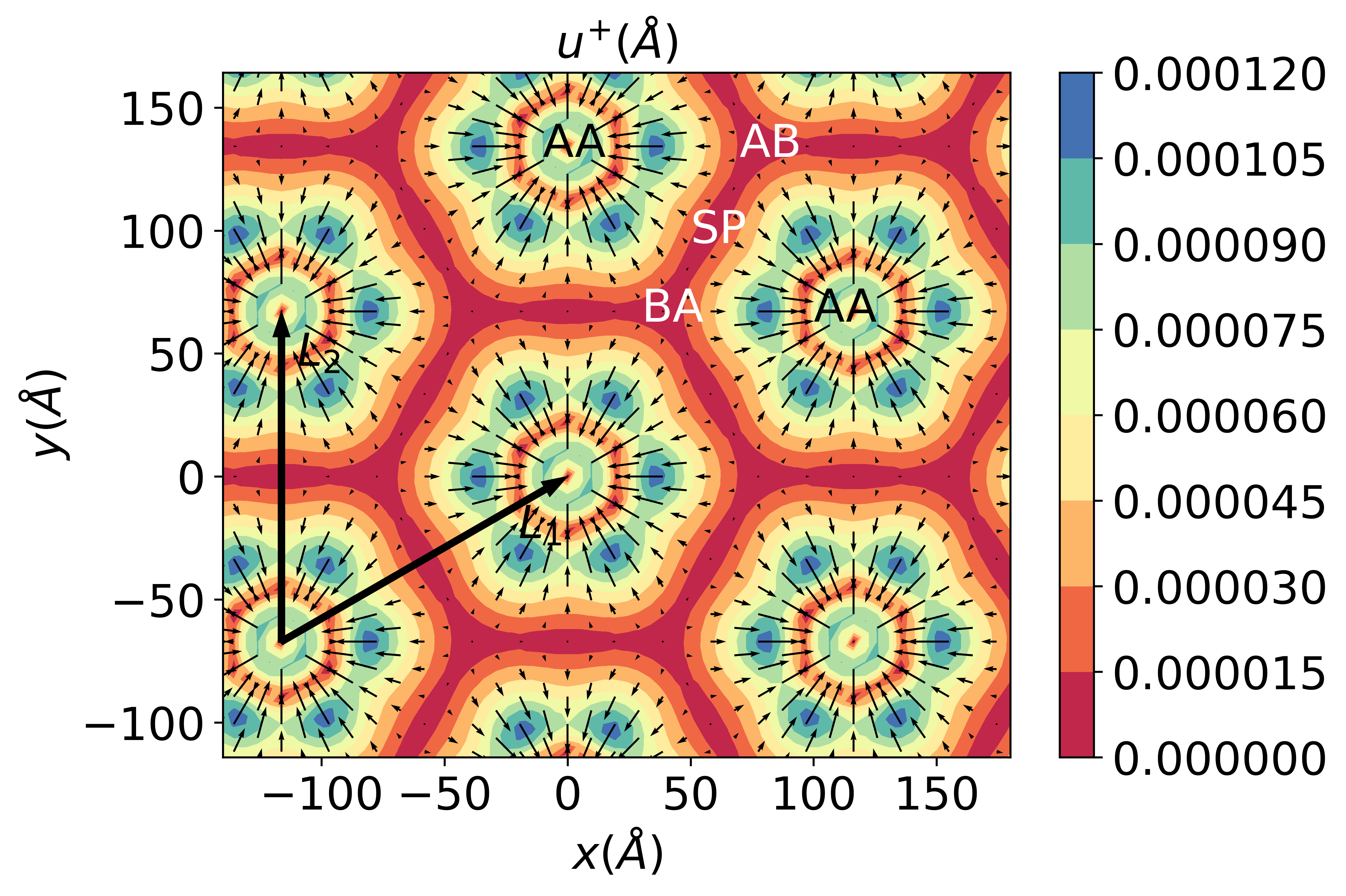}
\caption{The real-space distribution of the in-plane center-of-mass distortion of magic-angle TBG. The colorbar represents the amplitudes of local in-plane center-of-mass distortion. The arrows represent the directions of the distortion fields.}
\label{sif2}
\end{center}
\end{figure}

\vspace{12pt}
\begin{center}
\textbf{\large \IV.  Detailed formalism for  phonon calculations}
\end{center}

After evaluating the converged lattice displacement fields from the structural relaxation calculation, we study the phonon properties in magic-angle TBG using the continuum model. With the converged lattice distortion characterized by $\{\textbf{u}^{\pm}(\textbf{r}), h^{\pm}(\textbf{r})\}$, we introduce the time-dependent displacement field near the equilibrium position:
\begin{equation}
\begin{split}
\textbf{u}^{\pm}(\textbf{r},t)=\textbf{u}^{\pm}_{c}(\textbf{r})+\delta\textbf{u}^{\pm}(\textbf{r},t),\\
h^{\pm}(\textbf{r},t)=h^{\pm}_{c}(\textbf{r})+\delta h^{\pm}(\textbf{r},t),
\end{split}
\end{equation}
where $\textbf{u}^{\pm}_{c}(\textbf{r})$ and h$^{\pm}_{c}(\textbf{r})$ are the converged lattice distortions, $\delta\textbf{u}^{\pm}(\textbf{r},t)$ and $\delta h^{\pm}(\textbf{r},t)$ are the time dependent perturbative excitations near equilibrium positions. We expand the binding energy to the second order of $\delta\textbf{u}^{\pm}(\textbf{r},t)$ and  $\delta h^{\pm}(\textbf{r},t)$.
\begin{equation}
\begin{split}
U_{B}=&U^{(0)}_{B}+\frac{1}{2}\left[\int d^{2}\textbf{r} V^{(2)}_{B,uu}\delta\textbf{u}^{-}(\textbf{r},t)\delta\textbf{u}^{-}(\textbf{r},t) + \int d^{2}\textbf{r}V^{(2)}_{B,hh}\delta h^{-}(\textbf{r},t)\delta h^{-}(\textbf{r},t) \right. \\
&\ \ \ \ \ \ \ \ \ \ \ \ \ \ \left.\int d^{2}\textbf{r}V^{(2)}_{B,uh}\delta\textbf{u}^{-}(\textbf{r},t)\delta h^{-}(\textbf{r},t) +\int d^{2}\textbf{r}V^{(2)}_{B,hu}\delta\textbf{u}^{-}(\textbf{r},t)\delta h^{-}(\textbf{r},t)\right] , 
\end{split}
\end{equation}
where $V^{(2)}_{B,i}$, $i\!=\!uu,hh,uh,hu$, represents the second order functional derivatives of the binding energy density with respect to the different components of displacement fields, which can be expressed as
\begin{equation}
\begin{split}
\left(V^{(2)}_{B,uu}\right)_{\alpha,\beta}=&\frac{\partial^{2} V_{B}}{\partial u_{\alpha}\partial u_{\beta}}=\sum_{m}\epsilon_{\textbf{a}^{*}_{m}}e^{i(\textbf{G}_{m}\cdot\textbf{r}+\textbf{a}^{*}_{m}\cdot\textbf{u}^{-})}i\textbf{a}^{*}_{m,\alpha}i\textbf{a}^{*}_{m,\beta}\left[-1+36\left(\frac{h^{-}-h^{-}_{0}}{h^{-}_{0}}\right)^{2}\right]+\\
&\epsilon(\textbf{r})72\left[(-1)\frac{h^{-}}{(h^{-}_{0})^{2}}\sum_{m'}h_{0,\textbf{a}^{*}_{m'}}e^{i(\textbf{G}_{m'}\cdot\textbf{r}+\textbf{a}^{*}_{m'}\cdot\textbf{u}^{-})}i\textbf{a}^{*}_{m',\beta}\right]\left[(-1)\frac{h^{-}}{(h^{-}_{0})^{2}}\sum_{m}h_{0,\textbf{a}^{*}_{m}}e^{i(\textbf{G}_{m}\cdot\textbf{r}+\textbf{a}^{*}_{m}\cdot\textbf{u}^{-})}i\textbf{a}^{*}_{m,\alpha}\right]+\\
&2\sum_{m}\epsilon_{\textbf{a}^{*}_{m}}e^{i(\textbf{G}_{m}\cdot\textbf{r}+\textbf{a}^{*}_{m}\cdot\textbf{u}^{-})}i\textbf{a}^{*}_{m,\alpha}\left[72\frac{h^{-}-h^{-}_{0}}{h^{-}_{0}}(-1)\frac{h^{-}}{(h^{-}_{0})^{2}}\sum_{m'}h_{0,\textbf{a}^{*}_{m'}}e^{i(\textbf{G}_{m'}\cdot\textbf{r}+\textbf{a}^{*}_{m'}\cdot\textbf{u}^{-})i\textbf{a}^{*}_{m'}}\right]+\\
&\epsilon(\textbf{r})72\frac{h^{-}-h^{-}_{0}}{h^{-}_{0}}(-1)\frac{(-2)h^{-}}{(h^{-}_{0})^{3}}\left[\sum_{m'}h_{0,\textbf{a}^{*}_{m'}}e^{i(\textbf{G}_{m'}\cdot\textbf{r}+\textbf{a}^{*}_{m'}\cdot\textbf{u}^{-})}\right]\left[\sum_{m}h_{0,\textbf{a}^{*}_{m}}e^{i(\textbf{G}_{m}\cdot\textbf{r}+\textbf{a}^{*}_{m}\cdot\textbf{u}^{-})}\right]+\\
&\epsilon(\textbf{r})72\frac{h^{-}-h^{-}_{0}}{h^{-}_{0}}(-1)\frac{h^{-}}{(h^{-}_{0})^{2}}\sum_{m}h_{0,\textbf{a}^{*}_{m}}e^{i(\textbf{G}_{m}\cdot\textbf{r}+\textbf{a}^{*}_{m}\cdot\textbf{u}^{-})}i\textbf{a}^{*}_{m,\alpha}i\textbf{a}^{*}_{m,\beta},\\
V^{(2)}_{B,uh}=&\frac{\partial^{2} V_{B}}{\partial u_{\alpha}\partial h}=\sum_{m}e^{i(\textbf{G}_{m}\cdot\textbf{r}+\textbf{a}^{*}_{m}\cdot\textbf{u}^{-})}i\textbf{a}^{*}_{m,\alpha}\left\{\epsilon_{\textbf{a}^{*}_{m}}72\frac{h^{-}-h^{-}_{0}}{(h^{-}_{0})^{2}}+\epsilon(\textbf{r})72\left[\frac{-h^{-}}{(h^{-}_{0})^{3}}+\frac{-(h^{-}-h^{-}_{0})}{(h^{-}_{0})^{3}}\right]\right\}\\
V^{(2)}_{B,hh}=&\frac{\partial^{2} V_{B}}{\partial^{2}h}=\epsilon(\textbf{r})\frac{72}{(h^{-}_{0})^{2}}, \\
\end{split}
\end{equation}
where $\alpha,\beta\!=\!x,y$. We introduce the following Fourier transformation:
\begin{equation}
\begin{split}
\textbf{u}^{\pm}_{c}(\textbf{r})=&\sum_{m}\textbf{u}^{\pm}_{c,\textbf{G}_{m}}e^{i\textbf{G}_{m}\cdot\textbf{r}}, \delta\textbf{u}^{\pm}(\textbf{r},t)=e^{-i\omega t}\sum_{\textbf{q}}\delta\textbf{u}^{\pm}_{\textbf{q}}e^{i\textbf{q}\cdot\textbf{r}},\\
h^{\pm}_{c}(\textbf{r})=&\sum_{m}h^{\pm}_{c,\textbf{G}_{m}}e^{i\textbf{G}_{m}\cdot\textbf{r}}, \delta h^{\pm}(\textbf{r},t)=e^{-i\omega t}\sum_{\textbf{q}}\delta h^{\pm}_{\textbf{q}}e^{i\textbf{q}\cdot\textbf{r}}\\
V^{(2)}_{B,i}=&\sum_{m}V^{(2)}_{B,i,\textbf{G}_{m}}e^{i\textbf{G}_{m}\cdot\textbf{r}},
\end{split}
\end{equation}
where $i\!=\!uu,hh,uh,hu$, $\textbf{G}_{m}\!=\!m_{1}\textbf{G}^{M}_{1}+m_{2}\textbf{G}^{M}_{2}$ is the reciprocal lattice vectors. Then, we can express the binding energy in reciprocal space, given by $U^{(2)}_{B}=\sum_{\textbf{G},\textbf{G}',\textbf{q}}\delta\tilde{\textbf{u}}^{\dagger}_{{\textbf{G}'}+\textbf{q}}\tilde{\textbf{U}}^{(2)}_{B,\textbf{G},\textbf{G}'}\delta\tilde{\textbf{u}}_{\textbf{G}+\textbf{q}}$, where $\textbf{G}, \textbf{G}'\!=\!m_{1}\textbf{G}^{M}_{1}+m_{2}\textbf{G}^{M}_{2}$ are the moir\'e reciprocal lattice vector and $\textbf{q}$ is the wavevector within moir\'e Brillouin zone. The binding energy is a functional of relative displacements, which couples the $\delta\textbf{u}^{-}$ and $\delta h^{-}$ vibration. The force constant contributed by the binding energy in the matrix form is given by:
%
\allowdisplaybreaks
\begin{align}
\begin{split}
&\tilde{\textbf{U}}^{(2)}_{B,\textbf{G},\textbf{G}'}=\frac{1}{2}\left[\begin{array}{cccccc}
				0 &0&0&0&0&0 \\
				0 & 0&0&0&0&0 \\
				0&0&0&0&0&0\\
				0&0&0&V^{(2)}_{B,uu,\textbf{G}-\textbf{G}',xx} & V^{(2)}_{B,uu,\textbf{G}-\textbf{G}',xy}&V^{(2)}_{B,uh,\textbf{G}-\textbf{G}',x}\\
				0&0&0&V^{(2)}_{B,uu,\textbf{G}-\textbf{G}',xy} & V^{(2)}_{B,uu,\textbf{G}-\textbf{G}',yy}&V^{(2)}_{B,uh,\textbf{G}-\textbf{G}',y}\\
				0&0&0&V^{(2)}_{B,uh,\textbf{G}-\textbf{G}',x} & V^{(2)}_{B,uh,\textbf{G}-\textbf{G}',y}&V^{(2)}_{B,hh,\textbf{G}-\textbf{G}'}
\end{array}\right],\\
\end{split}
\end{align}
with the generalized displacement vector $\delta\tilde{\textbf{u}}_{\textbf{G}+\textbf{q}}$ defined as:
\begin{equation}
\delta\tilde{\textbf{u}}_{\textbf{G}+\textbf{q}}=[\delta u^{+}_{\textbf{G}+\textbf{q},x}\ \delta u^{+}_{\textbf{G}+\textbf{q},y}\ \delta h^{+}_{\textbf{G}+\textbf{q}}\ \delta u^{-}_{\textbf{G}+\textbf{q},x}\ \delta u^{-}_{\textbf{G}+\textbf{q},y}\ \delta h^{-}_{\textbf{G}+\textbf{q}} ]^{\mathrm{T}}.
\end{equation}

Before we derive the force constant contributed by the elastic energy, we divide the elastic energy into three terms according to the energy hierarchy. $U_{E}=U^{1}_{E}+U^{2}_{E}+U^{3}_{E}$, where
\allowdisplaybreaks
\begin{align}
U^{1}_{E}=&\int d^{2}\textbf{r} \frac{\kappa}{4}\left[ \left( \frac{\partial^{2} }{\partial^{2} x}+\frac{\partial^{2} }{\partial^{2} y} \right) h^{+} \right]^{2} + \left[ \left( \frac{\partial^{2} }{\partial^{2} x}+\frac{\partial^{2} }{\partial^{2} y} \right) h^{-} \right]^{2}\nonumber \\
&+\frac{\lambda+\mu}{4}\left[ \left( \frac{\partial u^{+}_{x}}{\partial x} \right)^{2} + \left( \frac{\partial u^{-}_{x}}{\partial x} \right)^{2} + \left( \frac{\partial u^{+}_{y}}{\partial y} \right)^{2} + \left( \frac{\partial u^{-}_{y}}{\partial y} \right)^{2} +2\left( \frac{\partial u^{+}_{x}}{\partial x}\frac{\partial u^{+}_{y}}{\partial y}+\frac{\partial u^{-}_{x}}{\partial x}\frac{\partial u^{-}_{y}}{\partial y} \right) \right]\nonumber\\
&+\frac{\mu}{4}\left[  \left( \frac{\partial u^{+}_{x}}{\partial x} \right)^{2} + \left( \frac{\partial u^{-}_{x}}{\partial x} \right)^{2} + \left( \frac{\partial u^{+}_{y}}{\partial y} \right)^{2} + \left( \frac{\partial u^{-}_{y}}{\partial y} \right)^{2} -2\left( \frac{\partial u^{+}_{x}}{\partial x}\frac{\partial u^{+}_{y}}{\partial y}+\frac{\partial u^{-}_{x}}{\partial x}\frac{\partial u^{-}_{y}}{\partial y} \right) \right]\nonumber\\
&+\frac{\mu}{4}\left[ \left( \frac{\partial u^{+}_{x}}{\partial y} \right)^{2} + \left( \frac{\partial u^{-}_{x}}{\partial y} \right)^{2} + \left( \frac{\partial u^{+}_{y}}{\partial x} \right)^{2} + \left( \frac{\partial u^{-}_{y}}{\partial x} \right)^{2} +2\left( \frac{\partial u^{+}_{x}}{\partial y}\frac{\partial u^{+}_{y}}{\partial x}+\frac{\partial u^{-}_{x}}{\partial y}\frac{\partial u^{-}_{y}}{\partial x} \right) \right], \nonumber\\
U^{2}_{E}=&\int d^{2}\textbf{r}\frac{\lambda+\mu}{8}\left[  \frac{\partial u^{+}_{x}}{\partial x} \left( \frac{\partial h^{+}}{\partial x}\right)^2 +  \frac{\partial u^{+}_{x}}{\partial x} \left( \frac{\partial h^{-}}{\partial x}\right)^2 + 2 \frac{\partial u^{-}_{x}}{\partial x}  \frac{\partial h^{+}}{\partial x}\frac{\partial h^{-}}{\partial x} +  \frac{\partial u^{+}_{y}}{\partial y} \left( \frac{\partial h^{+}}{\partial y}\right)^2 +  \frac{\partial u^{+}_{y}}{\partial y} \left( \frac{\partial h^{-}}{\partial y}\right)^2 + 2 \frac{\partial u^{-}_{y}}{\partial y}  \frac{\partial h^{+}}{\partial y}\frac{\partial h^{-}}{\partial y}\right.\nonumber \\
&\left.  +\frac{\partial u^{+}_{y}}{\partial y} \left( \frac{\partial h^{+}}{\partial x}\right)^2 +  \frac{\partial u^{+}_{y}}{\partial y} \left( \frac{\partial h^{-}}{\partial x}\right)^2 + 2 \frac{\partial u^{-}_{y}}{\partial y}  \frac{\partial h^{+}}{\partial x}\frac{\partial h^{-}}{\partial x} +  \frac{\partial u^{+}_{x}}{\partial x} \left( \frac{\partial h^{+}}{\partial y}\right)^2 +  \frac{\partial u^{+}_{x}}{\partial x} \left( \frac{\partial h^{-}}{\partial y}\right)^2 + 2 \frac{\partial u^{-}_{x}}{\partial x}  \frac{\partial h^{+}}{\partial y}\frac{\partial h^{-}}{\partial y} \right]\nonumber\\
&+\frac{\mu}{8}\left[  \frac{\partial u^{+}_{x}}{\partial x} \left( \frac{\partial h^{+}}{\partial x}\right)^2 +  \frac{\partial u^{+}_{x}}{\partial x} \left( \frac{\partial h^{-}}{\partial x}\right)^2 + 2 \frac{\partial u^{-}_{x}}{\partial x}  \frac{\partial h^{+}}{\partial x}\frac{\partial h^{-}}{\partial x} +  \frac{\partial u^{+}_{y}}{\partial y} \left( \frac{\partial h^{+}}{\partial y}\right)^2 +  \frac{\partial u^{+}_{y}}{\partial y} \left( \frac{\partial h^{-}}{\partial y}\right)^2 + 2 \frac{\partial u^{-}_{y}}{\partial y}  \frac{\partial h^{+}}{\partial y}\frac{\partial h^{-}}{\partial y}\right. \nonumber\\
&\left.  -\frac{\partial u^{+}_{y}}{\partial y} \left( \frac{\partial h^{+}}{\partial x}\right)^2 -  \frac{\partial u^{+}_{y}}{\partial y} \left( \frac{\partial h^{-}}{\partial x}\right)^2 - 2 \frac{\partial u^{-}_{y}}{\partial y}  \frac{\partial h^{+}}{\partial x}\frac{\partial h^{-}}{\partial x} -  \frac{\partial u^{+}_{x}}{\partial x} \left( \frac{\partial h^{+}}{\partial y}\right)^2 -  \frac{\partial u^{+}_{x}}{\partial x} \left( \frac{\partial h^{-}}{\partial y}\right)^2 - 2 \frac{\partial u^{-}_{x}}{\partial x}  \frac{\partial h^{+}}{\partial y}\frac{\partial h^{-}}{\partial y} \right]\nonumber\\
&+\frac{\mu}{8}2\left[  \frac{\partial u^{+}_{x}}{\partial y}  \frac{\partial h^{+}}{\partial x}\frac{\partial h^{+}}{\partial y} +  \frac{\partial u^{+}_{x}}{\partial y}  \frac{\partial h^{-}}{\partial x}\frac{\partial h^{-}}{\partial y}  +  \frac{\partial u^{-}_{x}}{\partial y}  \frac{\partial h^{+}}{\partial x}\frac{\partial h^{-}}{\partial y} +  \frac{\partial u^{-}_{x}}{\partial y}  \frac{\partial h^{-}}{\partial x}\frac{\partial h^{+}}{\partial y} \right. \nonumber\\
&\left.  \frac{\partial u^{+}_{y}}{\partial x} \frac{\partial h^{+}}{\partial x}\frac{\partial h^{+}}{\partial y} +  \frac{\partial u^{+}_{y}}{\partial x} \frac{\partial h^{-}}{\partial x}\frac{\partial h^{-}}{\partial y}+  \frac{\partial u^{-}_{y}}{\partial x} \frac{\partial h^{+}}{\partial x}\frac{\partial h^{-}}{\partial y} +  \frac{\partial u^{-}_{y}}{\partial x}  \frac{\partial h^{-}}{\partial x}\frac{\partial h^{+}}{\partial y}  \right],\nonumber\\
U^{3}_{E}=&\int d^{2}\textbf{r}\frac{\lambda+\mu}{16}\left\{ \frac{1}{4}\left[ \left(\frac{\partial h^{+}}{\partial x}\right)^{4} +6\left(\frac{\partial h^{+}}{\partial x}\right)^{2}\left(\frac{\partial h^{-}}{\partial x}\right)^{2} +\left(\frac{\partial h^{-}}{\partial x}\right)^{4} \right] + \frac{1}{4}\left[ \left(\frac{\partial h^{+}}{\partial y}\right)^{4} +6\left(\frac{\partial h^{+}}{\partial y}\right)^{2}\left(\frac{\partial h^{-}}{\partial y}\right)^{2} +\left(\frac{\partial h^{-}}{\partial y}\right)^{4} \right]  \right.\nonumber\\
& +\frac{1}{2}\left[ \left(\frac{\partial h^{-}}{\partial x}\right)^{2}\left(\frac{\partial h^{-}}{\partial y}\right)^{2} + \left(\frac{\partial h^{+}}{\partial x}\right)^{2}\left(\frac{\partial h^{-}}{\partial y}\right)^{2} + \left(\frac{\partial h^{-}}{\partial x}\right)^{2}\left(\frac{\partial h^{+}}{\partial y}\right)^{2} + \left(\frac{\partial h^{+}}{\partial x}\right)^{2}\left(\frac{\partial h^{+}}{\partial y}\right)^{2} \right.\nonumber \\
&\left.\left. +4 \left(\frac{\partial h^{-}}{\partial x}\right)\left(\frac{\partial h^{+}}{\partial x}\right) \left(\frac{\partial h^{-}}{\partial y}\right)\left(\frac{\partial h^{+}}{\partial y}\right)\right] \right\}\nonumber \\
&+\frac{\lambda+\mu}{16}\left\{ \frac{1}{4}\left[ \left(\frac{\partial h^{+}}{\partial x}\right)^{4} +6\left(\frac{\partial h^{+}}{\partial x}\right)^{2}\left(\frac{\partial h^{-}}{\partial x}\right)^{2} +\left(\frac{\partial h^{-}}{\partial x}\right)^{4} \right] + \frac{1}{4}\left[ \left(\frac{\partial h^{+}}{\partial y}\right)^{4} +6\left(\frac{\partial h^{+}}{\partial y}\right)^{2}\left(\frac{\partial h^{-}}{\partial y}\right)^{2} +\left(\frac{\partial h^{-}}{\partial y}\right)^{4} \right]  \right.\nonumber\\
& -\frac{1}{2}\left[ \left(\frac{\partial h^{-}}{\partial x}\right)^{2}\left(\frac{\partial h^{-}}{\partial y}\right)^{2} + \left(\frac{\partial h^{+}}{\partial x}\right)^{2}\left(\frac{\partial h^{-}}{\partial y}\right)^{2} + \left(\frac{\partial h^{-}}{\partial x}\right)^{2}\left(\frac{\partial h^{+}}{\partial y}\right)^{2} + \left(\frac{\partial h^{+}}{\partial x}\right)^{2}\left(\frac{\partial h^{+}}{\partial y}\right)^{2} \right. \nonumber\\
&\left.\left. +4 \left(\frac{\partial h^{-}}{\partial x}\right)\left(\frac{\partial h^{+}}{\partial x}\right) \left(\frac{\partial h^{-}}{\partial y}\right)\left(\frac{\partial h^{+}}{\partial y}\right)\right] \right\}\nonumber \\
&+\frac{\mu}{16}\left[ \left(\frac{\partial h^{-}}{\partial x}\right)^{2}\left(\frac{\partial h^{-}}{\partial y}\right)^{2} + \left(\frac{\partial h^{+}}{\partial x}\right)^{2}\left(\frac{\partial h^{-}}{\partial y}\right)^{2} + \left(\frac{\partial h^{-}}{\partial x}\right)^{2}\left(\frac{\partial h^{+}}{\partial y}\right)^{2} + \left(\frac{\partial h^{+}}{\partial x}\right)^{2}\left(\frac{\partial h^{+}}{\partial y}\right)^{2} \right.\nonumber \\
&\left. +4 \left(\frac{\partial h^{-}}{\partial x}\right)\left(\frac{\partial h^{+}}{\partial x}\right) \left(\frac{\partial h^{-}}{\partial y}\right)\left(\frac{\partial h^{+}}{\partial y}\right) \right].
\end{align}
We introduce a the time-dependent perturbative vibration to the elastic energy. Besides, we assume that the center-of-mass component of the out-of-plane distortion vanished, i.e. $h^{+}(\textbf{r})=0$. Then, we expand the elastic energy to the second order of $\delta\textbf{u}^{\pm}(\textbf{r},t)$ and $\delta h^{\pm}(\textbf{r},t)$:
\begin{equation}
U^{1}_{E}=U^{1,(0)}_{E}+U^{1,(2)}_{E}, U^{2}_{E}=U^{2,(0)}_{E}+U^{2,(2)}_{E},
\end{equation} 
where
\allowdisplaybreaks
\begin{align}
U^{1,(2)}_{E}= &\int d^{2}\textbf{r} \frac{\kappa}{4}\left\{\left[ \left( \frac{\partial^{2} }{\partial^{2} x}+\frac{\partial^{2} }{\partial^{2} y} \right) \delta h^{+} \right]^{2} + \left[ \left( \frac{\partial^{2} }{\partial^{2} x}+\frac{\partial^{2} }{\partial^{2} y} \right) \delta h^{-} \right]^{2}\right\}\nonumber\\
&+\frac{\lambda+\mu}{4}\left[ \left( \frac{\partial \delta u^{+}_{x}}{\partial x} \right)^{2} + \left( \frac{\partial \delta u^{-}_{x}}{\partial x} \right)^{2} + \left( \frac{\partial \delta u^{+}_{y}}{\partial y} \right)^{2} + \left( \frac{\partial \delta u^{-}_{y}}{\partial y} \right)^{2} +2\left( \frac{\partial \delta u^{+}_{x}}{\partial x}\frac{\partial \delta u^{+}_{y}}{\partial y}+\frac{\partial \delta u^{-}_{x}}{\partial x}\frac{\partial \delta u^{-}_{y}}{\partial y} \right) \right]\nonumber\\
&+\frac{\mu}{4}\left[ \left( \frac{\partial \delta u^{+}_{x}}{\partial x} \right)^{2} + \left( \frac{\partial \delta u^{-}_{x}}{\partial x} \right)^{2} + \left( \frac{\partial \delta u^{+}_{y}}{\partial y} \right)^{2} + \left( \frac{\partial \delta u^{-}_{y}}{\partial y} \right)^{2} -2\left( \frac{\partial \delta u^{+}_{x}}{\partial x}\frac{\partial \delta u^{+}_{y}}{\partial y}+\frac{\partial \delta u^{-}_{x}}{\partial x}\frac{\partial \delta u^{-}_{y}}{\partial y} \right) \right]\nonumber\\
&+\frac{\mu}{4}\left[ \left( \frac{\partial \delta u^{+}_{x}}{\partial y} \right)^{2} + \left( \frac{\partial \delta u^{-}_{x}}{\partial y} \right)^{2} + \left( \frac{\partial \delta u^{+}_{y}}{\partial x} \right)^{2} + \left( \frac{\partial \delta u^{-}_{y}}{\partial x} \right)^{2} +2\left( \frac{\partial \delta u^{+}_{x}}{\partial y}\frac{\partial \delta u^{+}_{y}}{\partial x}+\frac{\partial \delta u^{-}_{x}}{\partial y}\frac{\partial \delta u^{-}_{y}}{\partial x} \right) \right] ,\nonumber\\
U^{2,(2)}_{E}=&\int d^{2}\textbf{r}\frac{\lambda+\mu}{8}\left[ \frac{\partial u^{+}_{c,x}}{\partial x} \left( \frac{\partial \delta h^{+}}{\partial x}\right)^2 +2\frac{\partial \delta u^{+}_{x}}{\partial x}  \frac{\partial h^{-}_{c}}{\partial x}\frac{\partial \delta h^{-}}{\partial x}+  \frac{\partial u^{+}_{c,x}}{\partial x} \left( \frac{\partial \delta h^{-}}{\partial x}\right)^2 +2\left( \frac{\partial \delta u^{-}_{x}}{\partial x}  \frac{\partial h^{-}_{c}}{\partial x}\frac{\partial \delta h^{+}}{\partial x} +\frac{\partial u^{-}_{c,x}}{\partial x}  \frac{\partial\delta  h^{-}}{\partial x}\frac{\partial \delta h^{+}}{\partial x} \right)  \right. \nonumber\\
&\frac{\partial u^{+}_{c,y}}{\partial y} \left( \frac{\partial \delta h^{+}}{\partial y}\right)^2 +2\frac{\partial \delta u^{+}_{y}}{\partial y}  \frac{\partial h^{-}_{c}}{\partial y}\frac{\partial \delta h^{-}}{\partial y}+  \frac{\partial u^{+}_{c,y}}{\partial y} \left( \frac{\partial \delta h^{-}}{\partial y}\right)^2 +2\left( \frac{\partial \delta u^{-}_{y}}{\partial y}  \frac{\partial h^{-}_{c}}{\partial y}\frac{\partial \delta h^{+}}{\partial y} +\frac{\partial u^{-}_{c,y}}{\partial y}  \frac{\partial\delta  h^{-}}{\partial y}\frac{\partial \delta h^{+}}{\partial y} \right) +\nonumber\\
&\frac{\partial u^{+}_{c,y}}{\partial y} \left( \frac{\partial \delta h^{+}}{\partial x}\right)^2 +2\frac{\partial \delta u^{+}_{y}}{\partial y}  \frac{\partial h^{-}_{c}}{\partial x}\frac{\partial \delta h^{-}}{\partial x}+  \frac{\partial u^{+}_{c,y}}{\partial y} \left( \frac{\partial \delta h^{-}}{\partial x}\right)^2 +2\left( \frac{\partial \delta u^{-}_{y}}{\partial y}  \frac{\partial h^{-}_{c}}{\partial x}\frac{\partial \delta h^{+}}{\partial x} +\frac{\partial u^{-}_{c,y}}{\partial y}  \frac{\partial\delta  h^{-}}{\partial x}\frac{\partial \delta h^{+}}{\partial x} \right) +\nonumber\\
&\left. \frac{\partial u^{+}_{c,x}}{\partial x} \left( \frac{\partial \delta h^{+}}{\partial y}\right)^2 +2\frac{\partial \delta u^{+}_{x}}{\partial x}  \frac{\partial h^{-}_{c}}{\partial y}\frac{\partial \delta h^{-}}{\partial y}+  \frac{\partial u^{+}_{c,x}}{\partial x} \left( \frac{\partial \delta h^{-}}{\partial y}\right)^2 +2\left( \frac{\partial \delta u^{-}_{x}}{\partial x}  \frac{\partial h^{-}_{c}}{\partial y}\frac{\partial \delta h^{+}}{\partial y} +\frac{\partial u^{-}_{c,x}}{\partial x}  \frac{\partial\delta  h^{-}}{\partial y}\frac{\partial \delta h^{+}}{\partial y} \right) \right]+\nonumber\\
&\frac{\mu}{8}\left[ \frac{\partial u^{+}_{c,x}}{\partial x} \left( \frac{\partial \delta h^{+}}{\partial x}\right)^2 +2\frac{\partial \delta u^{+}_{x}}{\partial x}  \frac{\partial h^{-}_{c}}{\partial x}\frac{\partial \delta h^{-}}{\partial x}+  \frac{\partial u^{+}_{c,x}}{\partial x} \left( \frac{\partial \delta h^{-}}{\partial x}\right)^2 +2\left( \frac{\partial \delta u^{-}_{x}}{\partial x}  \frac{\partial h^{-}_{c}}{\partial x}\frac{\partial \delta h^{+}}{\partial x} +\frac{\partial u^{-}_{c,x}}{\partial x}  \frac{\partial\delta  h^{-}}{\partial x}\frac{\partial \delta h^{+}}{\partial x} \right)  \right.\nonumber \\
&\frac{\partial u^{+}_{c,y}}{\partial y} \left( \frac{\partial \delta h^{+}}{\partial y}\right)^2 +2\frac{\partial \delta u^{+}_{y}}{\partial y}  \frac{\partial h^{-}_{c}}{\partial y}\frac{\partial \delta h^{-}}{\partial y}+  \frac{\partial u^{+}_{c,y}}{\partial y} \left( \frac{\partial \delta h^{-}}{\partial y}\right)^2 +2\left( \frac{\partial \delta u^{-}_{y}}{\partial y}  \frac{\partial h^{-}_{c}}{\partial y}\frac{\partial \delta h^{+}}{\partial y} +\frac{\partial u^{-}_{c,y}}{\partial y}  \frac{\partial\delta  h^{-}}{\partial y}\frac{\partial \delta h^{+}}{\partial y} \right) +\nonumber\\
&-\frac{\partial u^{+}_{c,y}}{\partial y} \left( \frac{\partial \delta h^{+}}{\partial x}\right)^2 -2\frac{\partial \delta u^{+}_{y}}{\partial y}  \frac{\partial h^{-}_{c}}{\partial x}\frac{\partial \delta h^{-}}{\partial x}-  \frac{\partial u^{+}_{c,y}}{\partial y} \left( \frac{\partial \delta h^{-}}{\partial x}\right)^2 -2\left( \frac{\partial \delta u^{-}_{y}}{\partial y}  \frac{\partial h^{-}_{0}}{\partial x}\frac{\partial \delta h^{+}}{\partial x} +\frac{\partial u^{-}_{c,y}}{\partial y}  \frac{\partial\delta  h^{-}}{\partial x}\frac{\partial \delta h^{+}}{\partial x} \right) +\nonumber\\
&\left. -\frac{\partial u^{+}_{c,x}}{\partial x} \left( \frac{\partial \delta h^{+}}{\partial y}\right)^2 -2\frac{\partial \delta u^{+}_{x}}{\partial x}  \frac{\partial h^{-}_{c}}{\partial y}\frac{\partial \delta h^{-}}{\partial y}+  \frac{\partial u^{+}_{c,x}}{\partial x} \left( \frac{\partial \delta h^{-}}{\partial y}\right)^2 -2\left( \frac{\partial \delta u^{-}_{x}}{\partial x}  \frac{\partial h^{-}_{c}}{\partial y}\frac{\partial \delta h^{+}}{\partial y} +\frac{\partial u^{-}_{c,x}}{\partial x}  \frac{\partial\delta  h^{-}}{\partial y}\frac{\partial \delta h^{+}}{\partial y} \right) \right]+\nonumber\\
&+\frac{\mu}{8}2\left[  \frac{\partial u^{+}_{c,x}}{\partial y}  \frac{\partial \delta h^{+}}{\partial x}\frac{\partial \delta h^{+}}{\partial y} +  \frac{\partial \delta u^{+}_{x}}{\partial y}  \frac{\partial \delta h^{-}}{\partial x}\frac{\partial h^{-}_{c}}{\partial y}  +  \frac{\partial \delta u^{+}_{x}}{\partial y}  \frac{\partial h^{-}_{c}}{\partial x}\frac{\partial \delta h^{-}}{\partial y} +  \frac{\partial u^{+}_{c,x}}{\partial y}  \frac{\partial \delta h^{-}}{\partial x}\frac{\partial \delta h^{-}}{\partial y}+ \right.\nonumber \\
&\left.  \frac{\partial \delta u^{-}_{x}}{\partial y}  \frac{\partial \delta h^{+}}{\partial x}\frac{\partial h^{-}_{c}}{\partial y} +  \frac{\partial u^{-}_{c,x}}{\partial y}  \frac{\partial \delta h^{+}}{\partial x}\frac{\partial \delta h^{-}}{\partial y}  +  \frac{\partial \delta u^{-}_{x}}{\partial y}  \frac{\partial h^{-}_{c}}{\partial x}\frac{\partial \delta h^{+}}{\partial y} +  \frac{\partial u^{-}_{c,x}}{\partial y}  \frac{\partial \delta h^{-}}{\partial x}\frac{\partial \delta h^{+}}{\partial y}+  \right.\nonumber\\
&\left.  \frac{\partial u^{+}_{c,y}}{\partial x}  \frac{\partial \delta h^{+}}{\partial x}\frac{\partial \delta h^{+}}{\partial y} +  \frac{\partial \delta u^{+}_{y}}{\partial x}  \frac{\partial \delta h^{-}}{\partial x}\frac{\partial h^{-}_{c}}{\partial y}  +  \frac{\partial \delta u^{+}_{y}}{\partial x}  \frac{\partial h^{-}_{c}}{\partial x}\frac{\partial \delta h^{-}}{\partial y} +  \frac{\partial u^{+}_{c,y}}{\partial x}  \frac{\partial \delta h^{-}}{\partial x}\frac{\partial \delta h^{-}}{\partial y}+ \right. \nonumber\\
&\left.  \frac{\partial \delta u^{-}_{y}}{\partial x}  \frac{\partial \delta h^{+}}{\partial x}\frac{\partial h^{-}_{c}}{\partial y} +  \frac{\partial u^{-}_{c,y}}{\partial x}  \frac{\partial \delta h^{+}}{\partial x}\frac{\partial \delta h^{-}}{\partial y}  +  \frac{\partial \delta u^{-}_{y}}{\partial x}  \frac{\partial h^{-}_{c}}{\partial x}\frac{\partial \delta h^{+}}{\partial y} +  \frac{\partial u^{-}_{c,y}}{\partial x}  \frac{\partial \delta h^{-}}{\partial x}\frac{\partial \delta h^{+}}{\partial y}+  \right].
\end{align}
We neglect $U^{3}_{E}$ terms in further calculation, since it is at least $100$ times smaller than other terms. 
Then, we express the force constant contributed by the elastic energy in reciprocal space:
\allowdisplaybreaks
\begin{align}
U^{1,(2)}_{E}=&\sum_{\textbf{G},\textbf{q}}\frac{\kappa}{4}[(G_{x}+q_{x})^{2}+(G_{y}+q_{y})^{2}]^{2}\delta h^{+*}_{\textbf{G}+\textbf{q}}\delta h^{+}_{\textbf{G}+\textbf{q}}+\frac{\kappa}{4}[(G_{x}+q_{x})^{2}+(G_{y}+q_{y})^{2}]^{2}\delta h^{-*}_{\textbf{G}+\textbf{q}}\delta h^{-}_{\textbf{G}+\textbf{q}}\nonumber\\
&+\sum_{\textbf{G}}\sum_{\textbf{q}}\frac{\lambda+\mu}{4}\left[ (G_{x}+q_{x})^{2}\delta u^{+*}_{\textbf{G}+\textbf{q},x}\delta u^{+}_{\textbf{G}+\textbf{q},x} + (G_{y}+q_{y})^{2}\delta u^{+*}_{\textbf{G}+\textbf{q},y}\delta u^{+}_{\textbf{G}+\textbf{q},y} \right.\nonumber\\
&+(G_{x}+q_{x})(G_{y}+q_{y})(\delta u^{+*}_{\textbf{G}+\textbf{q},x}\delta u^{+}_{\textbf{G}+\textbf{q},y}+\delta u^{+*}_{\textbf{G}+\textbf{q},y}\delta u^{+}_{\textbf{G}+\textbf{q},x}) \nonumber \\
&+(G_{x}+q_{x})^{2}\delta u^{-*}_{\textbf{G}+\textbf{q},x}\delta u^{-}_{\textbf{G}+\textbf{q},x} + (G_{y}+q_{y})^{2}\delta u^{-*}_{\textbf{G}+\textbf{q},y}\delta u^{-}_{\textbf{G}+\textbf{q},y}\nonumber \\
&+\left.(G_{x}+q_{x})(G_{y}+q_{y})(\delta u^{-*}_{\textbf{G}+\textbf{q},x}\delta u^{-}_{\textbf{G}+\textbf{q},y}+\delta u^{-*}_{\textbf{G}+\textbf{q},y}\delta u^{-}_{\textbf{G}+\textbf{q},x})  \right]\nonumber\\
&+\sum_{\textbf{G}}\sum_{\textbf{q}}\frac{\mu}{4}\left[ (G_{x}+q_{x})^{2}\delta u^{+*}_{\textbf{G}+\textbf{q},x}\delta u^{+}_{\textbf{G}+\textbf{q},x} + (G_{y}+q_{y})^{2}\delta u^{+*}_{\textbf{G}+\textbf{q},y}\delta u^{+}_{\textbf{G}+\textbf{q},y} \right.\nonumber\\
&-(G_{x}+q_{x})(G_{y}+q_{y})(\delta u^{+*}_{\textbf{G}+\textbf{q},x}\delta u^{+}_{\textbf{G}+\textbf{q},y}+\delta u^{+*}_{\textbf{G}+\textbf{q},y}\delta u^{+}_{\textbf{G}+\textbf{q},x})  \nonumber\\
&+(G_{x}+q_{x})^{2}\delta u^{-*}_{\textbf{G}+\textbf{q},x}\delta u^{-}_{\textbf{G}+\textbf{q},x} + (G_{y}+q_{y})^{2}\delta u^{-*}_{\textbf{G}+\textbf{q},y}\delta u^{-}_{\textbf{G}+\textbf{q},y} \nonumber\\
&-\left.(G_{x}+q_{x})(G_{y}+q_{y})(\delta u^{-*}_{\textbf{G}+\textbf{q},x}\delta u^{-}_{\textbf{G}+\textbf{q},y}+\delta u^{-*}_{\textbf{G}+\textbf{q},y}\delta u^{-}_{\textbf{G}+\textbf{q},x})  \right]\nonumber\\
&+\frac{\mu}{4}\left[ (G_{y}+q_{y})^{2}\delta u^{+*}_{\textbf{G}+\textbf{q},x}\delta u^{+}_{\textbf{G}+\textbf{q},x} + (G_{x}+q_{x})^{2}\delta u^{+*}_{\textbf{G}+\textbf{q},y}\delta u^{+}_{\textbf{G}+\textbf{q},y} \right.\nonumber\\
&+(G_{x}+q_{x})(G_{y}+q_{y})(\delta u^{+*}_{\textbf{G}+\textbf{q},x}\delta u^{+}_{\textbf{G}+\textbf{q},y}+\delta u^{+*}_{\textbf{G}+\textbf{q},y}\delta u^{+}_{\textbf{G}+\textbf{q},x})  \nonumber\\
&+(G_{y}+q_{y})^{2}\delta u^{-*}_{\textbf{G}+\textbf{q},x}\delta u^{-}_{\textbf{G}+\textbf{q},x} + (G_{x}+q_{x})^{2}\delta u^{-*}_{\textbf{G}+\textbf{q},y}\delta u^{-}_{\textbf{G}+\textbf{q},y} \nonumber\\
&+\left.(G_{x}+q_{x})(G_{y}+q_{y})(\delta u^{-*}_{\textbf{G}+\textbf{q},x}\delta u^{-}_{\textbf{G}+\textbf{q},y}+\delta u^{-*}_{\textbf{G}+\textbf{q},y}\delta u^{-}_{\textbf{G}+\textbf{q},x})  \right]
\end{align}
$U^{1,(2)}_{E}$ is the dominant contribution by the elastic energy, which can be written into a block diagonal matrix:
\allowdisplaybreaks
\begin{align}
&\tilde{\textbf{U}}^{1,(2)}_{E,\textbf{G},\textbf{q}}=\frac{1}{4}\left[\begin{array}{cccccc}
				(\lambda+2\mu)\tilde{G}^{2}_{x}+\mu \tilde{G}^{2}_{y} & (\lambda+\mu)\tilde{G}_{x}\tilde{G}_{y}&0&0&0&0 \\
				(\lambda+\mu)\tilde{G}_{x}\tilde{G}_{y} & (\lambda+2\mu)\tilde{G}^{2}_{y}+\mu \tilde{G}^{2}_{x}&0&0&0&0 \\
				0&0&\kappa(\tilde{G}^{2}_{x}+\tilde{G}^{2}_{y})^{2}&0&0&0\\
				0&0&0&(\lambda+2\mu)\tilde{G}^{2}_{x}+\mu \tilde{G}^{2}_{y} & (\lambda+\mu)\tilde{G}_{x}\tilde{G}_{y}&0\\
				0&0&0&(\lambda+\mu)\tilde{G}_{x}\tilde{G}_{y} & (\lambda+2\mu)\tilde{G}^{2}_{y}+\mu \tilde{G}^{2}_{x}&0\\
				0&0&0&0&0&\kappa(\tilde{G}^{2}_{x}+\tilde{G}^{2}_{y})^{2}
\end{array}\right],
\end{align}
where $\tilde{\textbf{G}}\!=\!\textbf{G}+\textbf{q}$. We can also express $U^{2,(2)}_{E}$ in reciprocal space. For simplicity, we denote $(G_{\alpha}-G'_{\alpha})(G_{\beta}+q_{\beta})(G'_{\gamma}+q_{\gamma})$ as $(\alpha,\beta,\gamma)$ with $\alpha,\beta,\gamma\!=\!x,y$. Then, $U^{2,(2)}_{E}$ is given by:
\allowdisplaybreaks
\begin{align}
U^{2,(2)}_{E}=&\sum_{\textbf{G},\textbf{G}'}\sum_{\textbf{q}}\frac{\lambda+\mu}{8}\left\{\delta h^{+*}_{\textbf{G}+\textbf{q}}i\left[(x,x,x)u^{+}_{c,\textbf{G}-\textbf{G}',x}+(y,y,y)u^{+}_{c,\textbf{G}-\textbf{G}',y}\right.\right. +\left.(y,x,x)u^{+}_{c,\textbf{G}-\textbf{G}',y}+(x,y,y)u^{+}_{c,\textbf{G}-\textbf{G}',x}\right]\delta h^{+}_{\textbf{G}'+\textbf{q}} \nonumber\\
& +h^{-}_{c,\textbf{G}-\textbf{G}'}i\left[ \delta u^{+*}_{\textbf{G}+\textbf{q},x}(x,x,x)\delta h^{-}_{\textbf{G}'+\textbf{q}}+\delta h^{-*}_{\textbf{G}+\textbf{q}}(x,x,x)\delta u^{+}_{\textbf{G}'+\textbf{q},x}+ \right. \delta u^{+*}_{\textbf{G}+\textbf{q},y}(y,y,y)\delta h^{-}_{\textbf{G}'+\textbf{q}}+\delta h^{-*}_{\textbf{G}+\textbf{q}}(y,y,y)\delta u^{+}_{\textbf{G}'+\textbf{q},y}\nonumber \\
&+\delta u^{+*}_{\textbf{G}+\textbf{q},y}(x,y,x)\delta h^{-}_{\textbf{G}'+\textbf{q}}+\delta h^{-*}_{\textbf{G}+\textbf{q}}(x,x,y)\delta u^{+}_{\textbf{G}'+\textbf{q},y}+\left.\delta u^{+*}_{\textbf{G}+\textbf{q},x}(y,x,y)\delta h^{-}_{\textbf{G}'+\textbf{q}}+\delta h^{-*}_{\textbf{G}+\textbf{q}}(y,y,x)\delta u^{+}_{\textbf{G}'+\textbf{q},x}\right]\nonumber\\
&+\delta h^{-*}_{\textbf{G}+\textbf{q}}i\left[ (x,x,x)u^{+}_{c,\textbf{G}-\textbf{G}',x}+(y,y,y)u^{+}_{c,\textbf{G}-\textbf{G}',y}+ (y,x,x)u^{+}_{c,\textbf{G}-\textbf{G}',y}+(x,y,y)u^{+}_{c,\textbf{G}-\textbf{G}',x} \right]\delta h^{-}_{\textbf{G}'+\textbf{q}}\nonumber\\
& +h^{-}_{c,\textbf{G}-\textbf{G}'}i\left[ \delta u^{-*}_{\textbf{G}+\textbf{q},x}(x,x,x)\delta h^{+}_{\textbf{G}'+\textbf{q}}+\delta h^{+*}_{\textbf{G}+\textbf{q}}(x,x,x)\delta u^{-}_{\textbf{G}'+\textbf{q},x}+ \right. \delta u^{-*}_{\textbf{G}+\textbf{q},y}(y,y,y)\delta h^{+}_{\textbf{G}'+\textbf{q}}+\delta h^{+*}_{\textbf{G}+\textbf{q}}(y,y,y)\delta u^{-}_{\textbf{G}'+\textbf{q},y} \nonumber\\
&+\delta u^{-*}_{\textbf{G}+\textbf{q},y}(x,y,x)\delta h^{+}_{\textbf{G}'+\textbf{q}}+\delta h^{+*}_{\textbf{G}+\textbf{q}}(x,x,y)\delta u^{-}_{\textbf{G}'+\textbf{q},y}+\left.\delta u^{-*}_{\textbf{G}+\textbf{q},x}(y,x,y)\delta h^{+}_{\textbf{G}'+\textbf{q}}+\delta h^{+*}_{\textbf{G}+\textbf{q}}(y,y,x)\delta u^{-}_{\textbf{G}'+\textbf{q},x}\right]\nonumber\\
&+\delta h^{-*}_{\textbf{G}+\textbf{q}}i\left[ (x,x,x)u^{-}_{c,\textbf{G}-\textbf{G}',x}+(y,y,y)u^{-}_{c,\textbf{G}-\textbf{G}',y}+ (y,x,x)u^{-}_{c,\textbf{G}-\textbf{G}',y}+(x,y,y)u^{-}_{c,\textbf{G}-\textbf{G}',x}\right]\delta h^{+}_{\textbf{G}'+\textbf{q}}\nonumber\\
&+\delta h^{+*}_{\textbf{G}+\textbf{q}}i\left[ (x,x,x)u^{-}_{c,\textbf{G}-\textbf{G}',x}+(y,y,y)(G'_{y}+q_{y})u^{-}_{c,\textbf{G}-\textbf{G}',y}+ \left.(y,x,x)u^{-}_{c,\textbf{G}-\textbf{G}',y}+(x,y,y)u^{-}_{c,\textbf{G}-\textbf{G}',x}\right]\delta h^{-}_{\textbf{G}'+\textbf{q}}\right\}\nonumber\\
&+\sum_{\textbf{G},\textbf{G}'}\sum_{\textbf{q}}\frac{\mu}{8}\left\{\delta h^{+*}_{\textbf{G}+\textbf{q}}i\left[(x,x,x)u^{+}_{c,\textbf{G}-\textbf{G}',x}+(y,y,y)u^{+}_{c,\textbf{G}-\textbf{G}',y} -(y,x,x)u^{+}_{c,\textbf{G}-\textbf{G}',y}-(x,y,y)u^{+}_{c,\textbf{G}-\textbf{G}',x}\right]\delta h^{+}_{\textbf{G}'+\textbf{q}}\right. \nonumber\\
& +h^{-}_{c,\textbf{G}-\textbf{G}'}i\left[ \delta u^{+*}_{\textbf{G}+\textbf{q},x}(x,x,x)\delta h^{-}_{\textbf{G}'+\textbf{q}}+\delta h^{-*}_{\textbf{G}+\textbf{q}}(x,x,x)\delta u^{+}_{\textbf{G}'+\textbf{q},x}+ \right. \delta u^{+*}_{\textbf{G}+\textbf{q},y}(y,y,y)\delta h^{-}_{\textbf{G}'+\textbf{q}}+\delta h^{-*}_{\textbf{G}+\textbf{q}}(y,y,y)\delta u^{+}_{\textbf{G}'+\textbf{q},y}\nonumber \\
&-\delta u^{+*}_{\textbf{G}+\textbf{q},y}(x,y,x)\delta h^{-}_{\textbf{G}'+\textbf{q}}-\delta h^{-*}_{\textbf{G}+\textbf{q}}(x,x,y)\delta u^{+}_{\textbf{G}'+\textbf{q},y}-\left.\delta u^{+*}_{\textbf{G}+\textbf{q},x}(y,x,y)\delta h^{-}_{\textbf{G}'+\textbf{q}}-\delta h^{-*}_{\textbf{G}+\textbf{q}}(y,y,x)\delta u^{+}_{\textbf{G}'+\textbf{q},x}\right]\nonumber\\
&+\delta h^{-*}_{\textbf{G}+\textbf{q}}i\left[ (x,x,x)u^{+}_{c,\textbf{G}-\textbf{G}',x}+(y,y,y)u^{+}_{c,\textbf{G}-\textbf{G}',y}-(y,x,x)u^{+}_{c,\textbf{G}-\textbf{G}',y}-(x,y,y)u^{+}_{c,\textbf{G}-\textbf{G}',x} \right]\delta h^{-}_{\textbf{G}'+\textbf{q}}\nonumber\\
& +h^{-}_{c,\textbf{G}-\textbf{G}'}i\left[ \delta u^{-*}_{\textbf{G}+\textbf{q},x}(x,x,x)\delta h^{+}_{\textbf{G}'+\textbf{q}}+\delta h^{+*}_{\textbf{G}+\textbf{q}}(x,x,x)\delta u^{-}_{\textbf{G}'+\textbf{q},x}+ \right. \delta u^{-*}_{\textbf{G}+\textbf{q},y}(y,y,y)\delta h^{+}_{\textbf{G}'+\textbf{q}}+\delta h^{+*}_{\textbf{G}+\textbf{q}}(y,y,y)\delta u^{-}_{\textbf{G}'+\textbf{q},y}\nonumber \\
&-\delta u^{-*}_{\textbf{G}+\textbf{q},y}(x,y,x)\delta h^{+}_{\textbf{G}'+\textbf{q}}-\delta h^{+*}_{\textbf{G}+\textbf{q}}(x,x,y)\delta u^{-}_{\textbf{G}'+\textbf{q},y}-\left.\delta u^{-*}_{\textbf{G}+\textbf{q},x}(y,x,y)\delta h^{+}_{\textbf{G}'+\textbf{q}}-\delta h^{+*}_{\textbf{G}+\textbf{q}}(y,y,x)\delta u^{-}_{\textbf{G}'+\textbf{q},x}\right]\nonumber\\
&+\delta h^{-*}_{\textbf{G}+\textbf{q}}i\left[ (x,x,x)u^{-}_{c,\textbf{G}-\textbf{G}',x}+(y,y,y)u^{-}_{c,\textbf{G}-\textbf{G}',y}- (y,x,x)u^{-}_{c,\textbf{G}-\textbf{G}',y}-(x,y,y)u^{-}_{c,\textbf{G}-\textbf{G}',x}\right]\delta h^{+}_{\textbf{G}'+\textbf{q}}\nonumber\\
&+\delta h^{+*}_{\textbf{G}+\textbf{q}}i\left[ (x,x,x)u^{-}_{c,\textbf{G}-\textbf{G}',x}+(y,y,y)(G'_{y}+q_{y})u^{-}_{c,\textbf{G}-\textbf{G}',y}- \left.(y,x,x)u^{-}_{c,\textbf{G}-\textbf{G}',y}-(x,y,y)u^{-}_{c,\textbf{G}-\textbf{G}',x}\right]\delta h^{-}_{\textbf{G}'+\textbf{q}}\right\}\nonumber\\
&+\frac{\mu}{8}\left\{\delta h^{+*}_{\textbf{G}+\textbf{q}}i\left[(y,x,y)u^{+}_{c,\textbf{G}-\textbf{G}',x}+(y,y,x)u^{+}_{c,\textbf{G}-\textbf{G}',x}+(x,x,y)u^{+}_{c,\textbf{G}-\textbf{G}',y}+(x,y,x)u^{+}_{c,\textbf{G}-\textbf{G}',y}\right]\delta h^{+}_{\textbf{G}'+\textbf{q}}+\right.\nonumber\\
&i*h^{-}_{c,\textbf{G}-\textbf{G}'}\left[ \delta u^{+*}_{\textbf{G}+\textbf{q},x}(y,y,x)\delta h^{-}_{\textbf{G}'+\textbf{q}} +\delta h^{-*}_{\textbf{G}'+\textbf{q}}(y,x,y)\delta u^{+}_{\textbf{G}+\textbf{q},x}+ \right.\nonumber\\
&\delta u^{+*}_{\textbf{G}+\textbf{q},y}(y,x,x)\delta h^{-}_{\textbf{G}'+\textbf{q}} +\delta h^{-*}_{\textbf{G}'+\textbf{q}}(y,x,x)\delta u^{+}_{\textbf{G}+\textbf{q},y}+\nonumber\\
&\delta u^{+*}_{\textbf{G}+\textbf{q},x}(x,y,y)\delta h^{-}_{\textbf{G}'+\textbf{q}} +\delta h^{-*}_{\textbf{G}'+\textbf{q}}(x,y,y)\delta u^{+}_{\textbf{G}+\textbf{q},x}+\nonumber\\
&\left.\delta u^{+*}_{\textbf{G}+\textbf{q},y}(x,x,y)\delta h^{-}_{\textbf{G}'+\textbf{q}} +\delta h^{-*}_{\textbf{G}'+\textbf{q}}(x,y,x)\delta u^{+}_{\textbf{G}+\textbf{q},y}\right]+\nonumber\\
&\delta h^{-*}_{\textbf{G}+\textbf{q}}i\left[(y,x,y)u^{+}_{c,\textbf{G}-\textbf{G}',x}+(y,y,x)u^{+}_{c,\textbf{G}-\textbf{G}',x}+(x,x,y)u^{+}_{c,\textbf{G}-\textbf{G}',y}+(x,y,x)u^{+}_{c,\textbf{G}-\textbf{G}',y}\right]\delta h^{-}_{\textbf{G}'+\textbf{q}}+\nonumber\\
&i*h^{-}_{c,\textbf{G}-\textbf{G}'}\left[ \delta u^{-*}_{\textbf{G}+\textbf{q},x}(y,y,x)\delta h^{+}_{\textbf{G}'+\textbf{q}} +\delta h^{+*}_{\textbf{G}'+\textbf{q}}(y,x,y)\delta u^{-}_{\textbf{G}+\textbf{q},x}+ \right.\nonumber\\
&\delta u^{-*}_{\textbf{G}+\textbf{q},y}(y,x,x)\delta h^{+}_{\textbf{G}'+\textbf{q}} +\delta h^{+*}_{\textbf{G}'+\textbf{q}}(y,x,x)\delta u^{-}_{\textbf{G}+\textbf{q},y}+\nonumber\\
&\delta u^{-*}_{\textbf{G}+\textbf{q},x}(x,y,y)\delta h^{+}_{\textbf{G}'+\textbf{q}} +\delta h^{+*}_{\textbf{G}'+\textbf{q}}(x,y,y)\delta u^{-}_{\textbf{G}+\textbf{q},x}+\nonumber\\
&\left.\delta u^{-*}_{\textbf{G}+\textbf{q},y}(x,x,y)\delta h^{+}_{\textbf{G}'+\textbf{q}} +\delta h^{+*}_{\textbf{G}'+\textbf{q}}(x,y,x)\delta u^{-}_{\textbf{G}+\textbf{q},y}\right]+\nonumber\\
&\delta h^{+*}_{\textbf{G}+\textbf{q}}i(y,x,y)u^{-}_{c,\textbf{G}-\textbf{G}',x}\delta h^{-}_{\textbf{G}'+\textbf{q}}+\delta h^{-*}_{\textbf{G}+\textbf{q}}i(y,y,x)u^{-}_{c,\textbf{G}-\textbf{G}',x}\delta h^{+}_{\textbf{G}'+\textbf{q}}+\nonumber\\
&\delta h^{+*}_{\textbf{G}+\textbf{q}}i(x,x,y)u^{-}_{c,\textbf{G}-\textbf{G}',y}\delta h^{-}_{\textbf{G}'+\textbf{q}}+\delta h^{-*}_{\textbf{G}+\textbf{q}}i(x,y,x)u^{-}_{c,\textbf{G}-\textbf{G}',y}\delta h^{+}_{\textbf{G}'+\textbf{q}}+\nonumber\\
&\delta h^{-*}_{\textbf{G}+\textbf{q}}i(y,x,y)u^{-}_{c,\textbf{G}-\textbf{G}',x}\delta h^{+}_{\textbf{G}'+\textbf{q}}+\delta h^{+*}_{\textbf{G}+\textbf{q}}i(y,y,x)u^{-}_{c,\textbf{G}-\textbf{G}',x}\delta h^{-}_{\textbf{G}'+\textbf{q}}+\nonumber\\
&\left.\delta h^{-}_{\textbf{G}+\textbf{q}}i(x,x,y)u^{-}_{c,\textbf{G}-\textbf{G}',y}\delta h^{+}_{\textbf{G}'+\textbf{q}}+\delta h^{-}_{\textbf{G}+\textbf{q}}i(x,y,x)u^{-}_{c,\textbf{G}-\textbf{G}',y}\delta h^{+}_{\textbf{G}'+\textbf{q}}\right\}
\end{align}
We can write the high order terms in a matrix form: $U^{2,(2)}_{E}=\sum_{\textbf{G},\textbf{G}',\textbf{q}}\delta\tilde{\textbf{u}}^{\dagger}_{\textbf{G}'+\textbf{q}}\tilde{\textbf{U}}^{2,(2)}_{E,\textbf{G},\textbf{G}',\textbf{q}}\delta\tilde{\textbf{u}}_{\textbf{G}+\textbf{q}}$. Despite its higher-order nature, the $\tilde{\textbf{U}}^{2,(2)}_{E,\textbf{G},\textbf{G}',\textbf{q}}$ terms couples all the components of the displacement fields together, giving rise to fruitful phononic properties in TBG. At last, we expand the kinetic energy near the equilibrium position:
\begin{equation}
\begin{split}
T&=\int d^{2}\textbf{r}\frac{\rho}{4}\left[ (\dot{u}^{+}_{x})^{2}+ (\dot{u}^{+}_{y})^{2}+ (\dot{u}^{-}_{x})^{2}+ (\dot{u}^{-}_{y})^{2}+(\dot{h}^{+})^{2}+(\dot{h}^{-})^{2} \right]\\
&=T^{(0)}+\frac{\rho\omega^{2}}{4}\sum_{\textbf{G}}\sum_{\textbf{q}}(\delta u^{+*}_{\textbf{G}+\textbf{q},x}\delta u^{+}_{\textbf{G}+\textbf{q},x}+\delta u^{+*}_{\textbf{G}+\textbf{q},y}\delta u^{+}_{\textbf{G}+\textbf{q},y}+\delta u^{-*}_{\textbf{G}+\textbf{q},x}\delta u^{-}_{\textbf{G}'+\textbf{q},x}+\delta u^{-*}_{\textbf{G}+\textbf{q},y}\delta u^{-}_{\textbf{G}+\textbf{q},y}\\
&\ \ \ \ \ \ \ \ \ \ \ \ \ \ \ \ \ \ \ \ \ \ \  +\delta h^{+*}_{\textbf{G}+\textbf{q}}\delta h^{+}_{\textbf{G}+\textbf{q}}+\delta h^{-*}_{\textbf{G}+\textbf{q}}\delta h^{-}_{\textbf{G}+\textbf{q}})\\
&=T^{(0)}+\sum_{\textbf{G},\textbf{q}}\frac{\rho\omega^{2}}{4}\delta\tilde{\textbf{u}}^{\dagger}_{\textbf{G}+\textbf{q}}\delta\tilde{\textbf{u}}_{\textbf{G}+\textbf{q}}, 
\end{split}
\end{equation}
where $\rho\!=\!7.61\times 10^{-7}\,$kg/$m^{2}$ is the mass per area of the single layer graphene, $\omega$ is the phonon frequency. 

After evaluating the second-order functional derivatives of the total energy with respect to the displacement fields, we obtain the equation of motion in reciprocal space:
\begin{equation}
\frac{\rho\omega^{2}}{4}\delta\tilde{u}_{\textbf{G}+\textbf{q}}=\tilde{\textbf{U}}^{1,(2)}_{E,\textbf{G},\textbf{q}}\delta\tilde{u}_{\textbf{G}+\textbf{q}}+\sum_{\textbf{G}'}\tilde{\textbf{U}}^{2,(2)}_{E,\textbf{G},\textbf{G}',\textbf{q}}\delta\tilde{u}_{\textbf{G}'+\textbf{q}}+\sum_{\textbf{G}'}\tilde{\textbf{U}}^{(2)}_{B,\textbf{G},\textbf{G}'}\delta\tilde{u}_{\textbf{G}'+\textbf{q}}
\end{equation}

\vspace{12pt}
\begin{center}
\textbf{\large \V.  Details in the formalism for the effective electronic continuum model}
\end{center}


In Sec.~\ref{sec:bands}, we study the influence of the lattice distortions on the electronic band structure. We derive the interlayer hopping amplitudes under lattice distortions in Eq.~(\ref{eq:intert0t1t2}). Here we present the details in the calculations of the coefficients. We first consider $n_{h,1}+n'_{h,1}=0, n_{h,2}+n'_{h,2}=0, \dots$, i.e. the effects from the out-of-plane distortions are neglected. Taking the approximation $\textbf{Q}_{\parallel}\approx\textbf{Q}_{j}$:
\begin{equation}
\begin{split}
&-\frac{d_{0}}{2\pi}\int\mathrm{d}p_{z}\,t(\textbf{Q}_{j}+p_{z}\textbf{e}_{z})e^{i\,p_{z}d_{0}}\\
=&-\frac{d_{0}}{2\pi}\int\mathrm{d}p_{z}\int\mathrm{d}^{3}r\frac{1}{S_{0}d_{0}} T(\textbf{r}+d'\textbf{e}_{z})e^{-i(\textbf{Q}_{j}+p_{z}\textbf{e}_{z})\cdot(\textbf{r}+d'\textbf{e}_{z})}e^{i\,p_{z}d_{0}}\\
=&-\frac{1}{S_{0}}\int\mathrm{d}^{2}r\ T(\textbf{r}+d_{0}\textbf{e}_{z})e^{-i\textbf{Q}_{j}\cdot\textbf{r}}\\
\approx &0.101\,\textrm{eV},
\end{split}
\end{equation}
where $d_{0}=3.3869\angstrom$ is the average distance of the fully relaxed magic-angle TBG. Then, we consider the first order effect from the out-of-plane distortions. Specifically, for certain moir\'e reciprocal lattice vector $\textbf{G}^{h}_{m_{1}}$, we have $n_{h,m_{1}}+n'_{h,m_{1}}\!=\!1$ and $n_{h,m}+n'_{h,m}\!=\!0$ with $m\!\neq\!m_{1}$. The integration over $p_z$ can also be done analytically:
\begin{equation}
\begin{split}
&-\frac{d_{0}}{2\pi}\int\mathrm{d}p_{z}\,t(\textbf{Q}_{j}+p_{z}\textbf{e}_{z})e^{i\,p_{z}d_{0}}\times\left[i\,p_{z}h^{-}_{\textbf{G}^{h}_{m_{1}}}\right]\\
=&-\frac{d_{0}}{2\pi}\int\mathrm{d}p_{z}\,t(\textbf{Q}_{j}+p_{z}\textbf{e}_{z})\left.\frac{\mathrm{d}}{\mathrm{d}d'}\left(e^{i\,p_{z}d'}\right)\right|_{d'=d_{0}}h^{-}_{\textbf{G}^{h}_{m_{1}}}\\
=&-\frac{d_{0}}{2\pi}\int\mathrm{d}p_{z}\left.\frac{\mathrm{d}}{\mathrm{d}d'}\left(\int\mathrm{d}^{3}r\frac{1}{S_{0}d_{0}} T(\textbf{r}+d''\textbf{e}_{z})e^{-i(\textbf{Q}_{j}+p_{z}\textbf{e}_{z})\cdot(\textbf{r}+d''\textbf{e}_{z})}e^{i\,p_{z}d'}\right)\right|_{d'=d_{0}}h^{-}_{\textbf{G}^{h}_{m_{1}}}\\
=&-\frac{1}{S_{0}}\int\mathrm{d}^{2}r\left.\frac{\mathrm{d}}{\mathrm{d}d'}\left(T(\textbf{r}+d'\textbf{e}_{z})\right)\right|_{d'=d_{0}}e^{-i\textbf{Q}_{j}\cdot\textbf{r}}h^{-}_{\textbf{G}^{h}_{m_{1}}}\\
=&-\frac{1}{S_{0}}\int\mathrm{d}^{2}r\left[(V_{p}-V_{\sigma})\left(\frac{-2d_{0}}{\left|\textbf{r}+d_{0}\textbf{e}_{z}\right|^{2}}+\frac{2d^{3}_{0}}{\left|\textbf{r}+d_{0}\textbf{e}_{z}\right|^{4}}\right)+T(\textbf{r}+d_{0}\textbf{e}_{z})\frac{-d_{0}/r_{0}}{\left|\textbf{r}+d_{0}\textbf{e}_{z}\right|}\right]e^{-i\textbf{Q}_{j}\cdot\textbf{r}}h^{-}_{\textbf{G}^{h}_{m_{1}}}\\
\approx&-0.248\,h^{-}_{\textbf{G}^{h}_{m_{1}}}\,\textrm{eV}
\end{split}
\end{equation}
We also consider the second-order effects to the interlayer hopping from out-of-plane lattice distortions. For some moir\'e reciprocal lattice vector $\textbf{G}^{h}_{m_{1}}$ and $\textbf{G}^{h}_{m_{2}}$, we have $n_{h,m_{1}}+n'_{h,m_{1}}\!=\!1$,$n_{h,m_{2}}+n'_{h,m_{2}}\!=\!1$ and $n_{h,m}+n'_{h,m}\!=\!0$ with $m\!\neq\!m_{1}, m_{2}$., which leads to
\begin{equation}
\begin{split}
\left.\frac{\mathrm{d}^{2}}{\mathrm{d}d'^{2}}T(\textbf{r}+d'\textbf{e}_{z})\right|_{d'=d_{0}}=&(V_{P}-V_{\sigma})\left(\frac{-r_{z}}{r_{0}r_{m}}\right)*\left(\frac{-2r_{z}}{r_{m}^{2}}+\frac{2r_{z}^{3}}{r_{m}^{4}}\right)\\
&+(V_{P}-V_{\sigma})*\left(\frac{-2}{r_{m}^{2}}+\frac{4r_{z}^{2}}{r_{m}^{4}}+\frac{6r_{z}^{2}}{r_{m}^{4}}-\frac{8r_{z}^{4}}{r_{m}^{6}}\right)\\
&+V_{p}\left(\frac{-r_{z}}{r_{0}r_{m}}\right)^{2}\left(1-\left(\frac{r_{z}}{r_{m}}\right)^{2}\right)+V_{p}\left(\frac{-1}{r_{0}r_{m}}\right)\left(1-\left(\frac{r_{z}}{r_{m}}\right)^{2}\right)\\
&+V_{p}\left(\frac{-r_{z}}{r_{0}r_{m}}\right)\left(\frac{-2d_{0}}{\left|\textbf{r}+d_{0}\textbf{e}_{z}\right|^{2}}+\frac{2d^{3}_{0}}{\left|\textbf{r}+d_{0}\textbf{e}_{z}\right|^{4}}\right)\\
&+V_{\sigma}\left(\frac{-r_{z}}{r_{0}r_{m}}\right)^{2}\left(\frac{r_{z}}{r_{m}}\right)^{2}+V_{\sigma}\left(\frac{-1}{r_{0}r_{m}}\right)\left(\frac{r_{z}}{r_{m}}\right)^{2}\\
&+V_{\sigma}\left(\frac{-r_{z}}{r_{0}r_{m}}\right)\left(\frac{2d_{0}}{\left|\textbf{r}+d_{0}\textbf{e}_{z}\right|^{2}}-\frac{2d^{3}_{0}}{\left|\textbf{r}+d_{0}\textbf{e}_{z}\right|^{4}}\right).
\end{split}
\end{equation}
The integration over $p_z$ can be done analytically:
\begin{equation}
\begin{split}
-\frac{1}{S_{0}}\int\mathrm{d}^{2}r\left.\frac{\mathrm{d}^{2}}{\mathrm{d}d'^{2}}T(\textbf{r}+d'\textbf{e}_{z})\right|_{d'=d_{0}}e^{-i\textbf{Q}_{j}\cdot\textbf{r}}h_{\textbf{G}^{h}_{m_{1}}}h_{\textbf{G}^{h}_{m_{2}}}\approx0.508\ h_{\textbf{G}^{h}_{m_{1}}}h_{\textbf{G}^{h}_{m_{2}}}\,\textrm{eV}
\end{split}
\end{equation}
We have derived the coefficients in Eq.~\ref{eq:intert0t1t2}, with the value of $t_{0}(\textbf{Q}_{j})\!\approx\!0.101\,$eV, $t_{1}(\textbf{Q}_{j})\!\approx\!0.248\,$eV$/\angstrom$ and $t_{2}(\textbf{Q}_{j})\!\approx\!0.508\,$eV$/\angstrom^{2}$.

The lattice distortions also has influence on the interlayer hopping terms of TBG. To study this influence, we start with a monolayer graphene with lattice vector: $\textbf{a}_{1}\!=\!(a,0,0)$, and $\textbf{a}_{2}\!=\!a(\frac{1}{2},\frac{\sqrt{3}}{2},0)$. $\bm{\tau}_{A}\!=\!(0,0,0)$ and $\bm{\tau}_{B}\!=\!\frac{a}{\sqrt{3}}(0,-1,0)$ represent the relative position of two sublattices. Consider the hopping from B sublattice to the nearest neighbor  A sublattice, the vectors connecting two sites are given by $\textbf{r}^{0}_{1}\!=\!(0,a_{0},0)$, $\textbf{r}^{0}_{2}\!=\!(\frac{\sqrt{3}}{2},-\frac{1}{2},0)a_{0}$ and $\textbf{r}^{0}_{3}\!=\!(-\frac{\sqrt{3}}{2},-\frac{1}{2},0)a_{0}$ with $a_{0}\!=\!a/\sqrt{3}$. In order to treat both in-plane and out-of-plane lattice distortions on equal footing, we introduce the three dimensional lattice strain $\hat{u}$, which is expressed as
\begin{align}\label{strain}
\begin{split}
\hat{u}=\left[\begin{array}{ccc}
				u_{xx} & u_{xy} & u_{hx} \\
				u_{xy} & u_{yy} & u_{hy} \\
				u_{hx} & u_{hy} & u_{hh}
\end{array}\right],\\
\end{split}
\end{align}
 where $u_{\alpha\beta}\!=\!(\partial u_{\alpha}/\partial r_{\beta}+ \partial u_{\beta}/\partial r_{\alpha})/2$, $\alpha,\beta\!=\!x, y$, and $u_{h\alpha}\!=\!\partial h/\partial r_{\alpha}$, $\alpha\!=\!x, y$. Three vectors from $B$ sublattice to $A$ sublattice undergoes small shifts from $\textbf{r}^{0}_{i}$ to $\textbf{r}_{i}\!=\!(1+\hat{u})\textbf{r}^{0}_{i}$, $i=1,2,3$. As a result, the hopping amplitudes change from $t(\textbf{r}^{0}_{i})$ to $t(\textbf{r}_{i})$. For $i\!=\!1$, we can expand the hopping amplitude to second order term of lattice strain.
\allowdisplaybreaks
\begin{align}
t(\textbf{r}_{1})&\approx V^{0}_{\pi}\left[  1-\beta\left(\frac{\left|\textbf{r}_{1}\right|}{a_{0}}-1\right)+\frac{\beta^{2}}{2}\left(\frac{\left|\textbf{r}_{1}\right|}{a_{0}}-1\right)^{2} \right]\left(1-u^{2}_{hy}\right)+V^{0}_{\sigma}e^{-(a_{0}-d_{0})/r_{0}}u^{2}_{hy}, \nonumber\\
&=V^{0}_{\pi}\left[1-\beta\left(u_{yy}+\frac{1}{2}u^{2}_{xy}+\frac{1}{2}u^{2}_{hy}\right)+\frac{\beta^{2}}{2}\left(u_{yy}+\frac{1}{2}u^{2}_{xy}+\frac{1}{2}u^{2}_{hy}\right)^{2}\right]\left(1-u^{2}_{hy}\right)+V^{0}_{\sigma}e^{-(a_{0}-d_{0})/r_{0}}u^{2}_{hy}\nonumber\\
& \approx V^{0}_{\pi}\left[1-\beta\left(u_{yy}+\frac{1}{2}u^{2}_{xy}+\frac{1}{2}u^{2}_{hy}\right)+\frac{\beta^{2}}{2}u_{yy}^{2}\right]\left(1-u^{2}_{hy}\right)+V^{0}_{\sigma}e^{-(a_{0}-d_{0})/r_{0}}u^{2}_{hy}\nonumber\\
&\approx V^{0}_{\pi}\left[1-\beta\left(u_{yy}-\frac{\beta}{2}u^{2}_{yy}+\frac{1}{2}u^{2}_{xy}+\frac{1}{2}u^{2}_{hy}\right)-u^{2}_{hy}\right]+V^{0}_{\sigma}e^{-(a_{0}-d_{0})/r_{0}}u^{2}_{hy}.
\end{align}
Then, the differences of the hopping amplitudes is given by:
\begin{align}
\delta t(\textbf{r}_{1})=-V^{0}_{\pi}\beta\left(u_{yy}-\frac{\beta}{2}u^{2}_{yy}+\frac{1}{2}u^{2}_{xy}+\frac{1}{2}u^{2}_{hy}\right)-V^{0}_{\pi}u^{2}_{hy}+V^{0}_{\sigma}e^{-(a_{0}-d_{0})/r_{0}}u^{2}_{hy},
\end{align}
which $\beta\!=\!a_{0}/r_{0}$. Similarly, we evaluate the difference of hopping amplitudes under strain for $i=2$ and $i=3$.
\allowdisplaybreaks
\begin{align}
\delta t(\textbf{r}_{2})&=-V^{0}_{\pi}\beta\left[\frac{3}{4}u_{xx}-\frac{\sqrt{3}}{2}u_{xy}+\frac{1}{4}u_{yy}\right.\nonumber\\
&+\left(\frac{3}{32}-\frac{9\beta}{32}\right)u^{2}_{xx}+\left(\frac{3}{32}-\frac{\beta}{32}\right)u^{2}_{yy}+\left(\frac{1}{8}-\frac{3\beta}{8}\right)u^{2}_{xy}\nonumber\\
&+\left(\frac{\sqrt{3}}{8}+\frac{3\sqrt{3}\beta}{8}\right)u_{xx}u_{xy}-\left(\frac{\sqrt{3}}{8}-\frac{\sqrt{3}\beta}{8}\right)u_{yy}u_{xy}-\left(\frac{3}{16}+\frac{3\beta}{16}\right)u_{xx}u_{yy}\nonumber\\
&+\left.\frac{3}{8}u^{2}_{hx}-\frac{\sqrt{3}}{4}u_{hx}u_{hy}+\frac{1}{8}u^{2}_{yy}\right]-V^{0}_{\pi}\left(\frac{\sqrt{3}}{2}u_{hx}-\frac{1}{2}u_{hy}\right)^{2}+V^{0}_{\sigma}e^{-(a_{0}-d_{0})/r_{0}}\left(\frac{\sqrt{3}}{2}u_{hx}-\frac{1}{2}u_{hy}\right)^{2}\nonumber\\
\delta t(\textbf{r}_{3})&=-V^{0}_{\pi}\beta\left[\frac{3}{4}u_{xx}+\frac{\sqrt{3}}{2}u_{xy}+\frac{1}{4}u_{yy}\right.\nonumber\\
&+\left(\frac{3}{32}-\frac{9\beta}{32}\right)u^{2}_{xx}+\left(\frac{3}{32}-\frac{\beta}{32}\right)u^{2}_{yy}+\left(\frac{1}{8}-\frac{3\beta}{8}\right)u^{2}_{xy}\nonumber\\
&-\left(\frac{\sqrt{3}}{8}+\frac{3\sqrt{3}\beta}{8}\right)u_{xx}u_{xy}+\left(\frac{\sqrt{3}}{8}-\frac{\sqrt{3}\beta}{8}\right)u_{yy}u_{xy}-\left(\frac{3}{16}+\frac{3\beta}{16}\right)u_{xx}u_{yy}\nonumber\\
&+\left.\frac{3}{8}u^{2}_{hx}+\frac{\sqrt{3}}{4}u_{hx}u_{hy}+\frac{1}{8}u^{2}_{yy}\right]-V^{0}_{\pi}\left(\frac{\sqrt{3}}{2}u_{hx}+\frac{1}{2}u_{hy}\right)^{2}+V^{0}_{\sigma}e^{-(a_{0}-d_{0})/r_{0}}\left(\frac{\sqrt{3}}{2}u_{hx}+\frac{1}{2}u_{hy}\right)^{2}
\end{align}
Then, we consider the monolayer graphene Hamiltonian in reciprocal space:
\begin{align}\label{strain}
\begin{split}
\hat{u}=\left[\begin{array}{cc}
				0 & g(\textbf{k},\hat{u})  \\
				g^{*}(\textbf{k},\hat{u}) & 0
\end{array}\right],\\
\end{split}
\end{align}
where $g(\textbf{k},\hat{u})\!=\!\sum^{3}_{i=1}t(\textbf{r}_{i})e^{i\textbf{k}\cdot\textbf{r}_{i}}$. We can expand $g(\textbf{k})$ near the $\mathbf{K}^{\mu}$ valley, with $\mathbf{K}^{\mu}\!=\!\mu(-4\pi/(3a),0,0)$. Then, we have 
\begin{equation}
g(\bar{\textbf{k}},\hat{u})=-\hbar v_{F}[\mu(\bar{k}_{x}+\mu A_{x})-i(\bar{k}_{y}+\mu A_{y})]\;, 
\end{equation}
with $\bar{\textbf{k}}\!=\!\textbf{k}-\textbf{K}^{\mu}$.
Compared with the monolayer graphene Hamiltonian, the pseudo vector potential induced by strain in the $l$th layer  graphene is given by:
\allowdisplaybreaks
\begin{align}
A_{x}^{(l)}=&\frac{\beta\gamma_{0}}{ev}\left[\frac{3}{4}(\frac{\partial u^{(l)}_{x}}{\partial x}-\frac{\partial u^{(l)}_{y}}{\partial y})+\frac{3}{32}\left((1-3\beta)\left(\frac{\partial u^{(l)}_{x}}{\partial x}\right)^{2}-(4+4\beta)\left(\frac{\partial u^{(l)}_{x}}{\partial y}+\frac{\partial u^{(l)}_{y}}{\partial x}\right)^{2}\right.\right.\nonumber\\
&\left.\left.-(2+2\beta)\frac{\partial u^{(l)}_{x}}{\partial x}\frac{\partial u^{(l)}_{y}}{\partial y}+(1+5\beta)\left(\frac{\partial u^{(l)}_{y}}{\partial y}\right)^{2}\right)\right]\nonumber+\left(\mu\frac{\beta\gamma_{0}}{ev}\frac{3}{8}+\mu\frac{M}{ev}\frac{3}{4}\right)\left[\left(\frac{\partial h^{(l)}}{\partial x}\right)^{2}-\left(\frac{\partial h^{(l)}}{\partial y}\right)^{2}\right],\nonumber\\
A_{y}^{(l)}=&\frac{\beta\gamma_{0}}{ev}\left[-\frac{3}{4}\left(\frac{\partial u^{(l)}_{x}}{\partial y}+\frac{\partial u^{(l)}_{y}}{\partial x}\right)+\frac{3}{16}\left((1+3\beta)\frac{\partial u^{(l)}_{x}}{\partial x}\left(\frac{\partial u^{(l)}_{x}}{\partial y}+\frac{\partial u^{(l)}_{y}}{\partial x}\right)-(1-\beta)\left(\frac{\partial u^{(l)}_{x}}{\partial y}+\frac{\partial u^{(l)}_{y}}{\partial x}\right)\frac{\partial u^{(l)}_{y}}{\partial y}\right)\right]\nonumber\\
&-\left(\mu\frac{\beta\gamma_{0}}{ev}\frac{3}{4}+\mu\frac{M}{ev}\frac{3}{2}\right)\frac{\partial h^{(l)}}{\partial x}\frac{\partial h^{(l)}}{\partial y},
\end{align}
where $M\!=\!-V^{0}_{\pi}+V^{0}_{\sigma}e^{-(a_{0}-d_{0})/r_{0}}\!\approx\!36\,$eV, $\beta\!=\!a_{0}/r_{0}\!\approx\!3.14$ and $\gamma_{0}=\left|V^{0}_{\pi}\right|\!=\!2.7\,$ eV. $v$ in the above equation is just the Fermi velocity $v_F$.
The pseudo vector potential in reciprocal space is expressed as:
\allowdisplaybreaks
\begin{align}
A^{(l)}_{x}=&\mu\frac{\beta\gamma_{0}}{ev}\left[\frac{3}{4}\sum_{\textbf{G}}(i\,G_{x}u^{(l)}_{\textbf{G},x}-i\,G_{y}u^{(l)}_{\textbf{G},y})e^{i\textbf{G}\cdot\textbf{r}}+\frac{3}{32}\sum_{\textbf{G},\textbf{G}'}\left[(1-3\beta)i\,G_{x}i\,G'_{x}u^{(l)}_{\textbf{G},x}u^{(l)}_{\textbf{G}',x}\right.\right.\nonumber\\
&-(\frac{4+4\beta}{4})(i\,G_{y}i\,G'_{y}u^{(l)}_{\textbf{G},x}u^{(l)}_{\textbf{G}',x}+2i\,G_{y}i\,G'_{x}u^{(l)}_{\textbf{G},x}u^{(l)}_{\textbf{G}',y}+i\,G_{x}i\,G'_{x}u^{(l)}_{\textbf{G},y}u^{(l)}_{\textbf{G}',y})\nonumber\\
&-\left.\left.(2+2\beta)i\,G_{x}i\,G'_{y}u^{(l)}_{\textbf{G},x}u^{(l)}_{\textbf{G}',y}+(1+5\beta)i\,G_{y}i\,G'_{y}u^{(l)}_{\textbf{G},y}u^{(l)}_{\textbf{G}',y}\right]e^{i(\textbf{G}+\textbf{G}')\cdot\textbf{r}}\right]\nonumber\\
&+\mu\frac{\beta\gamma_{0}}{ev}\left(\frac{3}{8}+\frac{3M}{4\beta\gamma_{0}}\right)(i\,G_{x}i\,G'_{x}h^{(l)}_{\textbf{G}}h^{(l)}_{\textbf{G}'}-i\,G_{y}i\,G'_{y}h^{(l)}_{\textbf{G}}h^{(l)}_{\textbf{G}'})e^{i(\textbf{G}+\textbf{G}')\cdot\textbf{r}},\nonumber\\
A^{(l)}_{y}=&\mu\frac{\beta\gamma_{0}}{ev}\left[\frac{3}{4}\sum_{\textbf{G}}(-i\,G_{y}u^{(l)}_{\textbf{G},x}-i\,G_{x}u^{(l)}_{\textbf{G},y})e^{i\textbf{G}\cdot\textbf{r}}+\frac{3}{8}\sum_{\textbf{G},\textbf{G}'}\left[\frac{1+3\beta}{2}(i\,G_{x}i\,G'_{y}u^{(l)}_{\textbf{G},x}u^{(l)}_{\textbf{G}',x}\right.\right.\nonumber\\
&+\left.\left.i\,G_{x}i\,G'_{x}u^{(l)}_{\textbf{G},x}u^{(l)}_{\textbf{G}',y})-\frac{1-\beta}{2}(i\,G_{y}i\,G'_{y}u^{(l)}_{\textbf{G},x}u^{(l)}_{\textbf{G}',y}+i\,G_{x}i\,G'_{y}u^{(l)}_{\textbf{G},y}u^{(l)}_{\textbf{G}',y})\right]e^{i(\textbf{G}+\textbf{G}')\cdot\textbf{r}}\right]\nonumber\\
&+\mu\frac{\beta\gamma_{0}}{ev}\left(\frac{3}{4}+\frac{3M}{2\beta\gamma_{0}}\right)(-i\,G_{x}i\,G'_{y}h^{(l)}_{\textbf{G}}h^{(l)}_{\textbf{G}'})e^{i(\textbf{G}+\textbf{G}')\cdot\textbf{r}},
\end{align}
Thus, the intralayer Hamiltonian of valley $\mu$ for the $l$th layer is given by:
\begin{equation}
H^{\mu,(l)}(\textbf{k})=-\hbar v_{F}\left[\left(\textbf{k}+\mu\frac{e}{\hbar}\textbf{A}^{(l)}-\textbf{K}^{\mu,(l)}\right)\right]\cdot(\mu\sigma_{x},\sigma_{y}),
\end{equation}
where $\sigma_{x}$ and $\sigma_{y}$ are Pauli matrices defined in the sublattice space and $\textbf{K}^{\mu,(l)}$ is the Dirac point in the $l$th layer from valley $\mu$.

\end{widetext}

\bibliography{tmg}

\end{document}